%% file: marquez.tex
\documentclass{aa}
\usepackage{morefloats}
\usepackage{epsfig}
\usepackage{natbib}
\begin{document}
\def\kms{km~s$^{-1}${}}

\title{Rotation curves and metallicity gradients from HII regions in spiral
   galaxies
\thanks{Based on data obtained Asiago/Ekar Observatory. Also
based on observations made with INT operated on the
island of La Palma by ING in the
Spanish Observatorio del Roque de Los Muchachos of the
Instituto de Astrofísica de Canarias.}
}
\author{I. M\'arquez \inst{1}\thanks{Visiting Astronomer, German-Spanish
Astronomical Center, Calar Alto, operated by the Max-Planck-Institut
fur Astronomie jointly with the Spanish National Commission for
Astronomy}
          \and
  J.~Masegosa \inst{1}
\and
  M.~Moles$^{**}$ \inst{2}
\and
  J. Varela \inst{2}
\and
  D. Bettoni \inst{3}
\and
  G. Galletta \inst{4}
}

   \offprints{I. M\'arquez (\sl{isabel@iaa.es})}

   \institute{Instituto de Astrof\'{\i}sica de Andaluc\'{\i}a (CSIC), Apartado 3004 , E-18080
   Granada, Spain
         \and
         Instituto de Matem\'aticas y F\'{\i}sica Fundamental (CSIC), C) Serrano 113B, E-28006 Madrid, Spain.
           \and
           Osservatorio Astronomico di Padova, Vicolo Osservatorio 5, 35122 Padova, Italy
\and
Dipartimento di Astronomia, Universit\`a di Padova, Vicolo Osservatorio 2, 35122 Padova, Italy
             }

   \date{Received: ; accepted: } \authorrunning{M\'arquez et al.}

\titlerunning{Rotation curves and metallicities of Spiral Galaxies}

\abstract {In this paper we study long slit spectra in the region of
H$\alpha$ emission line of a sample of 111 spiral galaxies 
with recognizable and well defined spiral morphology and with a well
determined environmental status, ranging from isolation to
non-disruptive interaction with satellites or companions.  The form
and properties of the rotation curves are considered as a function of
the isolation degree, morphological type and luminosity. The line
ratios are used to estimate the metallicity of all the detected HII
regions, thus producing a composite metallicity profile for different
types of spirals. We have found that isolated galaxies tend to be of
later types and lower luminosity than the interacting galaxies. The
outer parts of the rotation curves of isolated galaxies tend to be
flatter than in interacting galaxies, but they show similar relations
between global parameters. The scatter of the Tully-Fisher relation
defined by isolated galaxies is significantly lower than that of
interacting galaxies. The [NII]/H$\alpha$ ratios, used as metallicity
indicator, show a clear trend between Z and morphological type, t,
with earlier spirals showing larger ratios; this trend is tighter when
instead of t the gradient of the inner rotation curve, G, is used; no
trend is found with the interaction status. The Z-gradient of the
disks depends on the type, being almost flat for early spirals, and
increasing for later types. The [NII]/H$\alpha$ ratios measured for
disk HII regions of interacting galaxies are higher than
for normal/isolated objects, even if all the galaxy families present
similar distributions of H$\alpha$ Equivalent Width.  
 } 
\maketitle

%

\section{Introduction}

The analysis of the rotation curves of disk galaxies is the most direct way to obtain
information on the mass distribution of spiral galaxies. The ionized gas has been
used for long as a tracer of their kinematics. During the 80's, Rubin and
collaborators started a systematic effort to obtain accurate rotation curves of
spiral galaxies of all morphological types and luminosity (Rubin et al. 1991, and
references therein). The accumulation of data from different sources helped to get an
overall picture of the form of the rotation curve of spirals, and its relation with
other galactic properties. It is now recognized that the maximum rotational velocity,
V$_{m}$, is related with the total mass (and luminosity) of the galaxy, with the
optical scale radius of the disk and with the morphological type (see Persic et al.
1996). The flatness of the outer rotation curve in most cases also led to accept the
existence of massive dark halos in spiral galaxies (see Rubin et al. 1991;
Sofue 1998).

Most of those analysis were based on data sets assembled with no completeness
criteria. In particular, the galaxies were considered as field or cluster objects,
and no further attention was payed to their environmental status, in spite of the
expected influence of even small companions on the mass distribution, and star
formation history of a given galaxy. Spiral galaxies in very close isolated
pairs were studied by Keel (1993, 1996). Trying to
understand the effects of the interaction on the dynamics of
disk galaxies, M\'arquez \& Moles (1996; hereafter Paper I) studied
the properties of isolated (see below for the definition of isolation)
spiral galaxies, to set a zero-point for the effects of the
interaction; see also Mathewson et al. 1992, and Courteau 1997, for an
analysis of field spirals). M\'arquez \& Moles (1999; hereafter Paper
II), studied also the properties of spirals in isolated pairs, and
compared them to those of the isolated galaxies in Paper I. It was
found that the main differences
is the presence of type II disk profiles in interacting systems (but not in isolated
galaxies), and a flatter outer rotation curve in isolated galaxies. No distorted
curves were found among isolated disk galaxies.

More recently, 2D Fabry-Perot rotation curves have been obtained for a
number of cluster spirals in order to determine the environmental
effects in such large aggregates. The results show a complex pattern
(Amram et al. 1996). Barton et al.  (2000, 2001) have also analyzed
the rotation curves of spiral galaxies in close pairs and in the
general field in order to put some constraints on the cluster effects
on the kinematical properties of galaxies, and the consequences in
their use for distance estimation by means of the Tully-Fisher
relation. Their results do confirm the earlier results in Paper II, in
the sense of a more scattered T-F relation for non isolated objects.
Similarly, galaxies with interacting companions in the recent
analysis by Kannappan et al. (2002) fall on the high luminosity/low
velocity width side of the TF and show more scatter.

We emphasize that the so called field galaxies should be carefully
investigated since some of them could still be perturbed by small
companions or satellites, that could produce significant effects
(Athanassoula 1984; Conselice \& Gallagher 1999; Conselice et
al. 2000). In Paper I and II a quantitative criterion of isolation was
given, trying to identify truly isolated objects to build up a
reference for the analysis of the effects of gravitational
interaction. We will use a similar approach here.

The same long slit spectroscopic data used for the study of the gas
kinematics can also be used, through the flux ratios of the observed
emission lines, to trace the metallicity, Z, along the disk, and to
determine the existence of Z-gradients. The existing analysis point
out that the global metallicity is related to the mass (hence, to
V$_{m}$), and that Z decreases gently outwards (see
the review by Henry \& Worthey 1999). Ferguson et al. (1998) have
extended the analysis towards the extreme outer regions of disks,
finding that Z drops there abruptly, but keeping values far from
pristine.

In the present paper, we will study a sample of 111 spiral
galaxies with a well studied environmental status, ranging from
isolation to mild interaction with satellites or companions. In all
cases however the interaction is non disruptive (they have been
selected to have recognizable and well defined spiral morphology).
The data comprise new long slit spectroscopy for 85 spiral
galaxies. The remaining data are from Paper I. The form and
properties of the rotation curves are considered as a function of the
isolation degree, morphological type and luminosity.  The line ratios
are used to estimate the metallicity of all the detected HII regions,
thus producing a composite metallicity profile for different types of
spirals.

Section 2 is devoted to the description of the sample and the
determination of the interaction status. In Section 3, the
observations and data reduction procedures are presented. Section 4
and 5 deal with the rotation curves and the Tully-Fisher relation,
respectively. In Section 6 the properties of nuclear and extranuclear
HII regions are described. The summary and conclusions are given in
Section 7.

\section{The sample}

The 85 galaxies
with new long slit data come from three sources:

\begin{itemize}
\item{Automatically selected isolated galaxies from the CfA catalogue
(Huchra et al.  1999), among those brighter than m$_B$ = 13 (and not
bigger than 4\arcmin~in diameter, for practical reasons), with an
inclination between 32\degr and 73\degr, to minimize the uncertainties
with the correction in both, photometric and kinematic parameters. The
isolation criterium for automatic selection was that they had no CfA
companions (i. e., with known red-shift) within a projected distance
of 0.5~Mpc, or with a red-shift difference greater than 500~km/s (see
M\'arquez 1994; Paper I). A total of 43 galaxies were selected
in that way}

\item{An additional sample of 29 spiral galaxies also from the CfA
catalogue, also imposing the automatical isolation criteria and with
$|b| \geq$ 40\degr but with no restrictions in apparent magnitude,
size or inclination}

\item{Thirteen spiral galaxies in 7 pairs taken from the catalogue of
isolated pairs by Karachentsev (1972). We applied the automathic 
isolation criterium to the pair as a whole. In addition to this, 
only those galaxies with still
recognizable and well defined morphology were retained. Therefore, we
implicitly avoid strongly interacting galaxies, with about 40 \% of
pairs being widely separated, in contrast with previous studies of
pairs of spirals, where mostly close pairs were selected (Keel 1993,
1996; Barton et al. 2001).}

\end{itemize}

We have also included in our final sample a total of 26 spiral
galaxies (17 isolated
and 9 spirals in 5 S-S pairs) from Paper I\footnote{Isolated
galaxies: NGC 818, UGC 3511, UGC 3804, NGC 2532, NGC 2712, IC 529, NGC
3294, NGC 3549, NGC 3811, NGC 3835, NGC 4162, NGC 4290, NGC 4525, NGC
5962, NGC 6155, NGC 6395, NGC 6643; spirals in Karachentsev's pairs:
NGC 2798/99, NGC 4567/68, NGC 3646, NGC 7469/IC 5283, NGC 7537/41}, so
the analysis refers to the whole sample of 111 spiral galaxies. Notice
that since we imposed that the candidates would have a well defined
spiral morphology, and no bright (CfA) companions, our sample does not
include cases of very strong and/or disruptive
interaction. Nevertheless, even in those mildly interacting systems
the perturbations can be recognizable depending, other than on the
properties of the galaxy itself, on the mass, size, distance and
relative velocity of the perturbing agent.

\subsection{The isolation status revisited}

As a second step for defining a sample of isolated galaxies,
the previously selected as isolated were investigated for the presence
of perturbing companions. It is well known that the influence of
small, close companions can produce secular alterations on the
dynamics of the primary system (Athanassoula 1984; Sundelius et
al. 1987; Byrd \& Howard 1992). Moreover, the effects can manifest
themselves on very different time scales. To cope with all those
aspects, we define a galaxy as isolated when the possible past
perturbations by neighboring objects, if any, have been completely
erased by now. Accepting that typical time scales for the decay of the
perturbation effects are not longer than a few times 10$^9$ years, a
criterion as given before can be defined.

Given the incompleteness of the CfA catalog, and the lack of red-shift
information for many of the possible companions, we have adopted the
parameter $f$, defined as $f = 3 \times log (r/D) + 0.4 \times (m -
m_p)$ \footnote{$D$ is the distance from the target galaxy to the
perturber, $r$ is the radius of the target galaxy, $m$ is the
magnitude of the target galaxy and $m_p$ is the magnitude of the
perturber.}  to describe the environmental status of our sample
galaxies.  Numerical results (Athanassoula 1984) indicate that
ponderable effects are expected for $f$ larger than $-$2. To take into
account the possible past effects, we have adopted as truly isolated
those galaxies with $f \leq -$4 (the details and justification of the
adopted criterion will be given in Varela et al., in preparation). The
search for companions was done with a catalog complete to m$_B$ = 18,
kindly made available to us by G. Paturel. The results were visually
inspected to discard any candidate that is not a galaxy. As a result,
we have an indication of the status of each galaxy just given by the
maximum of the $f$ values corresponding to all its (candidate)
companions.

According to that status we have classified our sample galaxies
into 3 groups.  The first one includes 24 truly isolated
galaxies in the sense defined before (interaction class INT = 1); the
second, with 44 objects, includes possibly interacting galaxies,
with a companion (with log f $> -$4) of unknown red-shift (INT = 2);
it will most probably contain both truly isolated and truly
interacting galaxies and will therefore be considered separately from
the others. Group 3, with 43 objects, is for interacting
galaxies, including the 22 galaxies in Karachentsev's pairs and the
21 isolated spirals from the authomatic rearch that resulted to
have confirmed companion(s).  The type and absolute magnitude
distributions of the two subgroups included in the last Interaction
Grop are similar. We have performed the Kolmogorov-Smirnov test and
found that they are not significantly different, and therefore can be
considered as defining the same group.

\subsection{Characterization of the sample}

\begin{figure}
\psfig{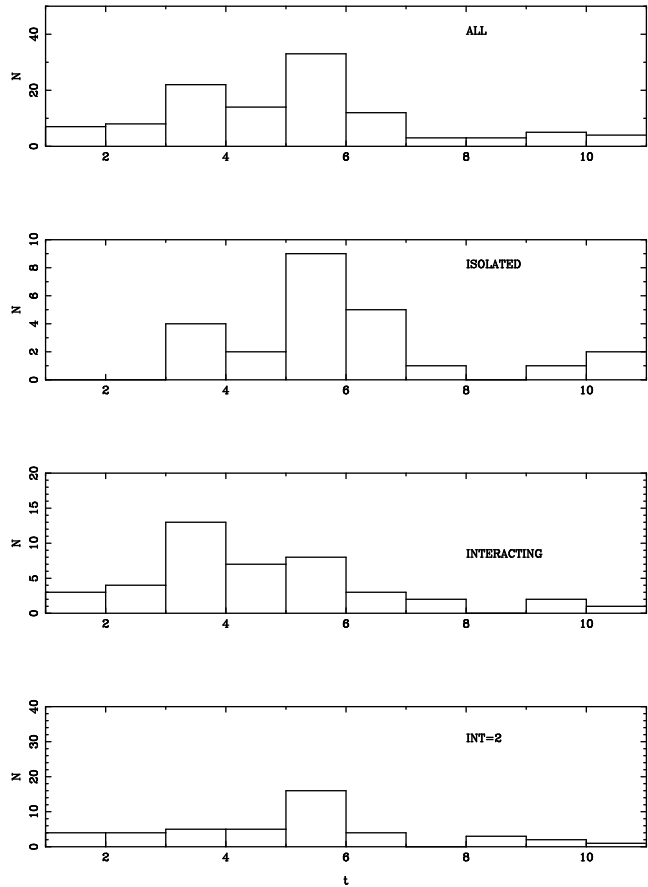}
\caption{Distribution of morphological types in our sample.}
\label{type}
\end{figure}

We have considered the morphology and luminosity of the whole sample
galaxies to see if the selection criteria would have resulted in
differences among the galaxies with different status. The type and
magnitude are from the RC3 catalogue. The distribution of the types is
shown in Fig.~\ref{type}a. It appears that 37 galaxies have t
$<$ 4, and 47 have 4 $\leq$ t $<$ 6). The remaining 27 galaxies
with t $\geq$ 6 (24\%) (12 galaxies with t $=$ 6) present different
morphological peculiarities, but in all cases an underlying disk 
does exist.  


We have compared the morphology of the confirmed isolated galaxies
(Interacting Type 1, 24 objects) with that of interacting galaxies
(Interacting Type 3, 43 objects).  As can be seen in
Fig.~\ref{type}, both t-distributions look different, in particular
due to the lack of galaxies with t $<$ 3, plus the scarcity of t = 3
galaxies (only 4) among the isolated objects in our sample. To
quantify the differences we performed a Kolmogorov-Smirnov test,
finding that, at the 95\% level, the difference is statistically
significant. The same kind of result is found when only galaxies with
t $<$ 7 are considered.

Regarding the presence of bars for the whole sample, 32 are
barred systems (SB code in the RC3), 24 are of intermediate type
(SX), and 34 have been classified as unbarred systems (SA). We
notice that these numbers are similar to those reported in studies on
the fraction of barred spirals based on optical studies (see for
instance Moles et al. 1995; Ho et al. 1997). For the other 21
galaxies there is no information about the presence of a bar in the
RC3. The same percentage of isolated objects (33\%) is found
among barred and unbarred galaxies.

\begin{figure}
\psfig{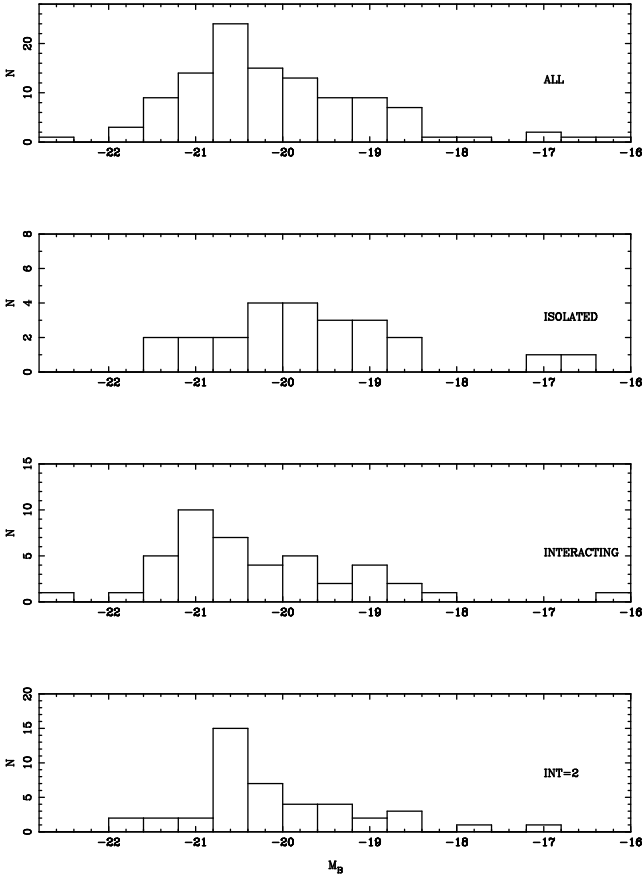}
\caption{Distribution of B absolute magnitudes.}
\label{histmb}
\end{figure}

The distribution of absolute magnitudes for the whole sample is given in
Fig.~\ref{histmb}. The absolute magnitude
ranges from $-$16.31 to $-$22.43, the low luminosity tail
corresponding to late type galaxies with morphological classification
later than t = 6.
For those late types alone, the median value is $-$18.63, whereas it
reaches $-$20.42 for the other types. Looking at those galaxies with t
$\leq$ 6, we find that the interacting objects tend to be brighter,
M$_B = -$20.69 mag, {\sl versus} M$_B = -$20.28 mag for isolated
objects, with a dispersion of 0.62 mag in both cases.

\begin{figure}
\psfig{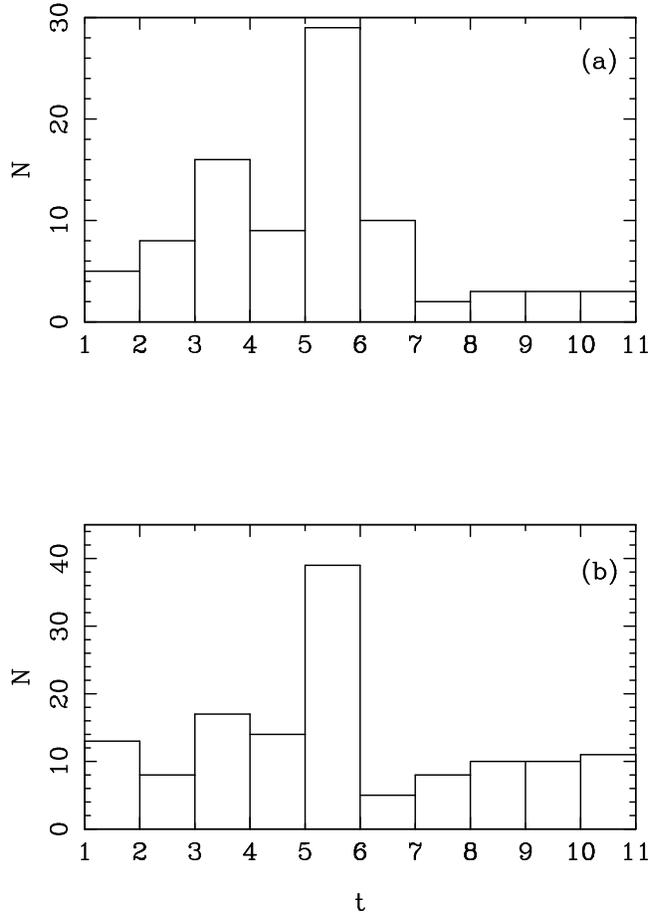}
\caption{Comparison of the distribution of morphological types for  
authomatically selected isolated galaxies in our sample (a) and in Jansen's 
sample (b).}
\label{comp_tJansen}
\end{figure}

Both the morphological type and absolute magnitude distributions for
the whole sample are similar to those of the spiral galaxies analyzed
by de Jong \& van der Kruit (1994), who also selected non-disrupted
spiral galaxies. We have also compared with the sample by Jansen et
al. (2000):
considering galaxies defined as isolated with our first criteria,
i. e., with no companions within a projected distance of 0.5~Mpc, or
with a red-shift difference greater than 500~km/s, the t-distribution
is similar to that of our galaxies, excepting that their selection
method produces more t$\geq$7 galaxies and (by construction) a much
flatter distribution of absolute magnitudes (see
Fig.~\ref{comp_tJansen}).  Indeed, the Kolmogorov-Smirnov test
for galaxies earlier than t = 7 (92 and 79 galaxies in Jansen et
al. sample and in our sample, respectively) gives the result that both
distributions are not statistically different.
We also note that only one galaxy
among the 8 secondary members of the isolated pairs has t=3, the rest
been later types. Therefore, we are confident that our sample galaxies
is representative of normal spirals and it is not biased to any
particular property, and the differences between isolated and
interacting galaxies seem to be real.

\section{Observations and Data Reductions}

The spectroscopic data were collected with three different
instruments: the Cassegrain Spectrograph attached to the 2.2m
telescope in Calar Alto (CAHA, Spain), the Intermediate Dispersion
Spectrograph at the 2.5m Isaac Newton telescope in La Palma (Spain),
and the Boller \& Chivens spectrograph attached to the 1.82m
Asiago-Ekar telescope at the Asiago Observatory (Italy). The setup and
main characteristics of the observations are given in Table~\ref{obs}.

\begin{table*}[h]
\caption[]{Observations}
\label{obs}
\begin{flushleft}
\begin{scriptsize}
\begin{tabular}{l  c c c c c c c}
\hline
\noalign{\smallskip}
\multicolumn {1}{l}{Telescope}
& \multicolumn {1}{c}{Date}
& \multicolumn {1}{c}{Instrument}
& \multicolumn {1}{c}{Spectral res.}
& \multicolumn {1}{c}{Range}
& \multicolumn {1}{c}{Slit width}
& \multicolumn {1}{c}{Spatial res.}
& \multicolumn {1}{c}{Average seeing}\\
 & & & (\AA/pix) & (\AA) & (arcsec) & (arcsec/pix) & (arcsec)\\
\noalign{\smallskip}
\hline\noalign{\smallskip}
2.2m CAHA & June 1991  & Boller\& Chivens & 1.31 & 5926 - 7151& 1.5 & 1.35 & 1.2 - 1.5\\
2.2m CAHA & March 1993 & Boller\& Chivens & 1.36 & 5736 - 7138& 1.5 & 1.69 & 1.1 - 1.4\\
INT       & Febr. 1993 & IDS              & 0.79 & 6000 - 7000& 1.5 & 0.65 & 1.1\\
ASIAGO    & 1996, 1997 & Boller\& Chivens & 0.96 & 5000 - 7000& 2.5 & 1.16 & 1.4 - 2.4\\
\hline
\end{tabular}
\end{scriptsize}
\end{flushleft}
\end{table*}

All the galaxies were observed with the slit oriented along the galaxy
major axis.  We determined the PA of the axis from our own broad band
images (Paper I), or from DSS images. In most cases the values we
measured are very close to that catalogued in the RC3. In general the
differences are within 15\degr. Only in three cases
we found a big discrepancy, namely for N3769a (40\degr), N3044
(100\degr), and U11577 (85\degr). In those cases we adopted our own
values for the PA of the major axis. The PA are given in Table
\ref{obsspec}, together with the exposure time. For 22 galaxies we
also obtained spectra along one additional PA, corresponding in 16
cases to the minor axis.

The data obtained in Calar Alto and La Palma were reduced in a similar
way, using the appropriate routines in FIGARO. After bias subtraction,
and flat field correction, the data were wavelength calibrated using
He-Ar comparison spectra, that were observed before and after each
object exposure. The resulting accuracy was tested using the sky lines
[OI]$\lambda$6300, and [OI]$\lambda$6364, as references. The centroids
of the lines were measured section by section. The average value was
then compared with the reference wavelengths to fix the zero-point
offset of each spectrum, and the rms error as an indication of the
error in the resulting velocity distribution induced by the residual
distortion (see Table \ref{obsspec}).

\begin{table*}[h]
\caption[]{Detailed log of the spectra}
\label{obsspec}

\begin{flushleft}
\begin{scriptsize}
\begin{tabular}{l|  c r r   l | l |  c r r   l }
\hline
\noalign{\smallskip}
\multicolumn {1}{l|}{Galaxy}
& \multicolumn{1}{c}{Date}
& \multicolumn{1}{r}{PA}
& \multicolumn{1}{c}{Time}
& \multicolumn{1}{c|}{RMS}
&\multicolumn {1}{l|}{Galaxy}
& \multicolumn{1}{c}{Date}
& \multicolumn{1}{r}{PA}
& \multicolumn{1}{c}{Time}
& \multicolumn{1}{c}{RMS}\\
\noalign{\smallskip}
\hline\noalign{\smallskip}
NGC 828   & 02/16/93&100 &3600  & 1.4 &NGC 493   &29/10/97& 58   &3600& 1.5\\
NGC 2460  & 02/16/93& 40 &3600  & 2.0 &NGC 658   &28/10/97& 20   &2983& 2.1\\
NGC 2543  & 02/12/93& 45 &3600  & 4.9 &NGC 864   &05/10/96& 20   &3600& 1.9\\
          & 02/17/93&145 &3600  & 3.1 &NGC 1036  &30/12/97&  5   &1803& 3.0\\
NGC 2552  & 02/17/93& 45 &3600  & 1.4 &NGC 1137  &05/10/96& 20   &1463& 2.8\\
NGC 2608  & 03/27/93& 60 &3000  & 1.0 &NGC 1507  &28/12/97& 11   &3600& 1.6\\
NGC 2633  & 03/26/93&175 &3000  & 0.6 &NGC 1530  &27/12/97& 23   &3600& 0.9\\
NGC 2701  & 03/28/93& 23 &3600  & 1.3 &          &27/12/97& 113  &3600& 1.3\\
NGC 2748  & 03/29/93& 38 &3000  & 1.9 &NGC 2344  &28/12/97& 0    &3600& 2.5\\
NGC 2770  & 03/26/93&148 &3000  & 0.5 &          &28/12/97& 90   &3600& 3.0\\
NGC 2964  & 03/30/93& 97 &3000  & 0.5 &NGC 2424  &29/01/96& 81   &3600& 2.5\\
NGC 2998  & 03/29/93& 53 &3000  & 0.4 &NGC 2469  &29/01/96& 40   &3600& 3.4\\
NGC 3041  & 03/28/93& 95 &3600  & 0.4 &NGC 2545  &05/06/97& 170  &3600& 1.8\\
NGC 3183  & 03/29/93&170 &3000  & 0.3 &NGC 2628  &31/10/97& 177  &3600& 4.0\\
NGC 3320  & 02/12/93& 20 &3600  & 3.1 &          &31/10/97& 87   &2000& 2.9\\
NGC 3370  & 02/14/93&148 &3600  & 2.5 &NGC 2906  &11/04/97& 75   &3600& 2.5\\
NGC 3395  & 03/29/93& 50 &1800  & 0.7 &          &11/04/97& 120  &3600& 3.0\\
NGC 3396  & 03/28/93&100 &3600  & 0.5 &NGC 3044  &04/03/97& 114  &3600& 3.1\\
NGC 3471  & 02/13/93& 14 &3600  & 4.4 &NGC 3055  &12/04/97& 63   &3428& 2.5\\
NGC 3501  & 03/30/93& 25 &3000  & 1.0 &          &12/04/97& 83   &3600& 1.9\\
NGC 3507  & 03/27/93&110 &3000  & 0.4 &NGC 3526  &05/03/97& 55   &3600& 0.9\\
NGC 3689  & 02/13/93& 97 &3600  & 3.0 &NGC 4455  &05/03/97& 16   &3600& 1.6\\
NGC 3769a & 03/31/93&110 &3000  & 0.3 &NGC 5147  &14/04/96& 0    &3600& 2.6\\
NGC 3769  & 03/31/93&152 &3000  & 1.0 &NGC 5375  &12/04/97& 0    &3600& 0.6\\
NGC 3976  & 03/27/93& 53 &3000  & 1.0 &NGC 5894  &12/04/97& 13   &3600& 3.0\\
NGC 4047  & 02/16/93&105 &3600  & 3.5 &NGC 5908  &05/03/97& 154  &2339& 2.1\\
NGC 4284  & 03/29/93&102 &3600  & 0.5 &IC 391    &28/12/97& 0    &3600& 0.8\\
NGC 4389  & 03/28/93&105 &3000  & 0.5 &IC 396    &29/12/97&  85  &3600& 1.5\\
NGC 4496a & 02/17/93& 70 &3200  & 10  &          &29/12/97& 175  &3600& 2.0\\
NGC 4496b & 03/27/93&115 &3000  & 1.0 &UGC 1155  &28/10/97& 165  &3600& 0.8\\
NGC 4793  & 02/13/93& 50 &3600  & 1.7 &UGC 3580  &29/12/97& 3    &3600& 3.0\\
NGC 4800  & 03/27/93& 25 &3000  & 0.5 &UGC 4107  &29/10/97& 40   &3600& 2.6\\
NGC 5012  & 06/22/91& 10 &1800  & 4.3 &          &29/10/97& 130  &3600& 1.9\\
          & 06/17/91&170 &1800  & 6.5 &UGC 11577 &28/10/97& 90   &3600& 3.0\\
NGC 5172  & 06/19/91&103 &2500  & 3.3 &UGC 12178 &29/10/97& 20   &3600& 1.9\\
          & 06/19/91& 13 &1000  & 3.1 &UGC 12857 &05/10/96& 34   &3600& 2.4\\
NGC 5351  & 06/20/91&150 &2000  & 4.2 &&&&&\\
          & 06/22/91& 10 &1000  & 6.0 &&&&&\\
NGC 5394  & 06/18/91&145 &1000  & 19  &&&&&\\
          & 06/23/91& 55 &1200  & 20  &&&&&\\
NGC 5395  & 06/19/91&170 &2000  & 3.9 &&&&&\\
NGC 5480  & 06/22/91&177 &1500  & 5.5 &&&&&\\
          & 06/22/91& 87 &1000  & 6.5 &&&&&\\
NGC 5533  & 06/21/91& 30 &1800  & 4.5 &&&&&\\
NGC 5641  & 06/19/91&158 &2000  & 4.3 &&&&&\\
NGC 5656  & 06/21/91& 50 &2000  & 3.3 &&&&&\\
          & 06/22/91&140 &1000  & 4.9 &&&&&\\
NGC 5678  & 06/21/91&  5 &1800  & 2.9 &&&&&\\
NGC 5740  & 06/21/91&160 &2000  & 2.2 &&&&&\\
NGC 5774  & 06/20/91&142 &2000  & 4.5 &&&&&\\
NGC 5775  & 06/20/91&146 &1800  & 4.2 &&&&&\\
NGC 5899  & 06/18/91& 18 &2000  & 10  &&&&&\\
          & 06/17/91&162 &2500  & 5.1 &&&&&\\
NGC 5963  & 06/19/91& 45 &2000  & 4.9 &&&&&\\
          & 06/22/91&135 &1200  & 7.1 &&&&&\\
NGC 5970  & 06/21/91& 88 &1800  & 5.9 &&&&&\\
          & 06/23/91&178 & 953  & 7.5 &&&&&\\
NGC 6070  & 06/17/91& 62 &2500  & 9.3 &&&&&\\
          & 06/23/91&152 &1200  & 1.6 &&&&&\\
NGC 6106  & 06/20/91&140 &1800  & 6.7 &&&&&\\
          & 06/22/91& 50 &1000  & 4.5 &&&&&\\
NGC 6181  & 06/20/91&175 &2000  & 5.8 &&&&&\\
NGC 6207  & 06/21/91& 22 &1402  & 9.0 &&&&&\\
NGC 6239  & 06/18/91&118 &2000  & 6.0 &&&&&\\
NGC 7177  & 06/18/91& 90 &2000  & 4.5 &&&&&\\
          & 06/23/91&  0 &1200  & 6.7 &&&&&\\
NGC 7217  & 06/24/91&100 &1200  & 3.7 &&&&&\\
          & 06/22/91& 10 &1200  & 3.9 &&&&&\\
NGC 7448  & 06/21/91&170 &1800  & 5.5 &&&&&\\
NGC 7479  & 06/24/91& 45 &1200  & 5.2 &&&&&\\
          & 06/24/91&135 &1200  & 6.4 &&&&&\\
\noalign{\smallskip}
\hline
\end{tabular}
\noindent
\smallskip\\
February 1993: INT telescope; 1996, 1997: Asiago,
 otherwise: 2.2m CAHA telescope.\\
\end{scriptsize}
\end{flushleft}

\end{table*}

The sky background level was determined taking median averages over
two strips at both sides of the galaxy signal. The parameters of the
lines were measured with the program LINES kindly provided to us by
Dr. Perea. This program perform a simultaneous interactive polynomial
fitting of the continuum and Gaussian fitting of the emission selected
lines providing the Equivalent Width, central wavelength of the
Gaussian fit and the integrated fluxes of the lines. The errors
in Table \ref{parhii} have
been calculated by quadratic addition of photon counting errors 
and the errors in the continuum level determination. 

We used cross-correlation technics as described in paper I to extract
the kinematic information. The spatial section in the 2D spectrum with
the highest S/N ratio was used as a template for the
cross-correlation. The errors are referred to the determination of the
velocity shift with respect to the template spectrum. They are shown
as the error bars in all the velocity distributions (see
Figs. \ref{vel} and \ref{jesus}).

\begin{figure*}[htp]
\psfig{figure=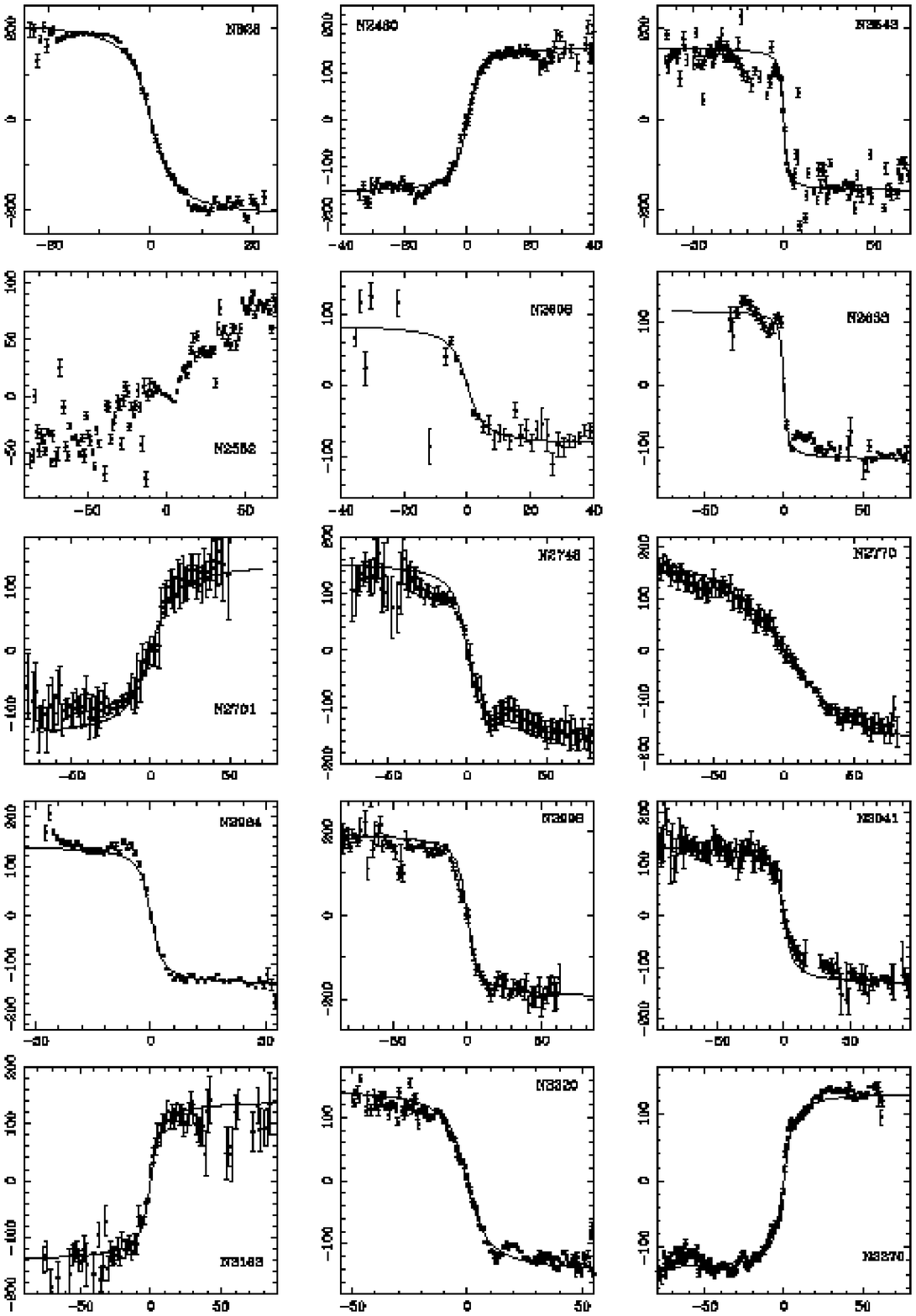,width=17truecm}
\caption{Rotation curves from INT and CAHA spectra. Velocities are in km/s 
(y axis) and distances to the center in arcseconds.}
\label{vel}
\end{figure*}

\begin{figure*}[htp]
\psfig{figure=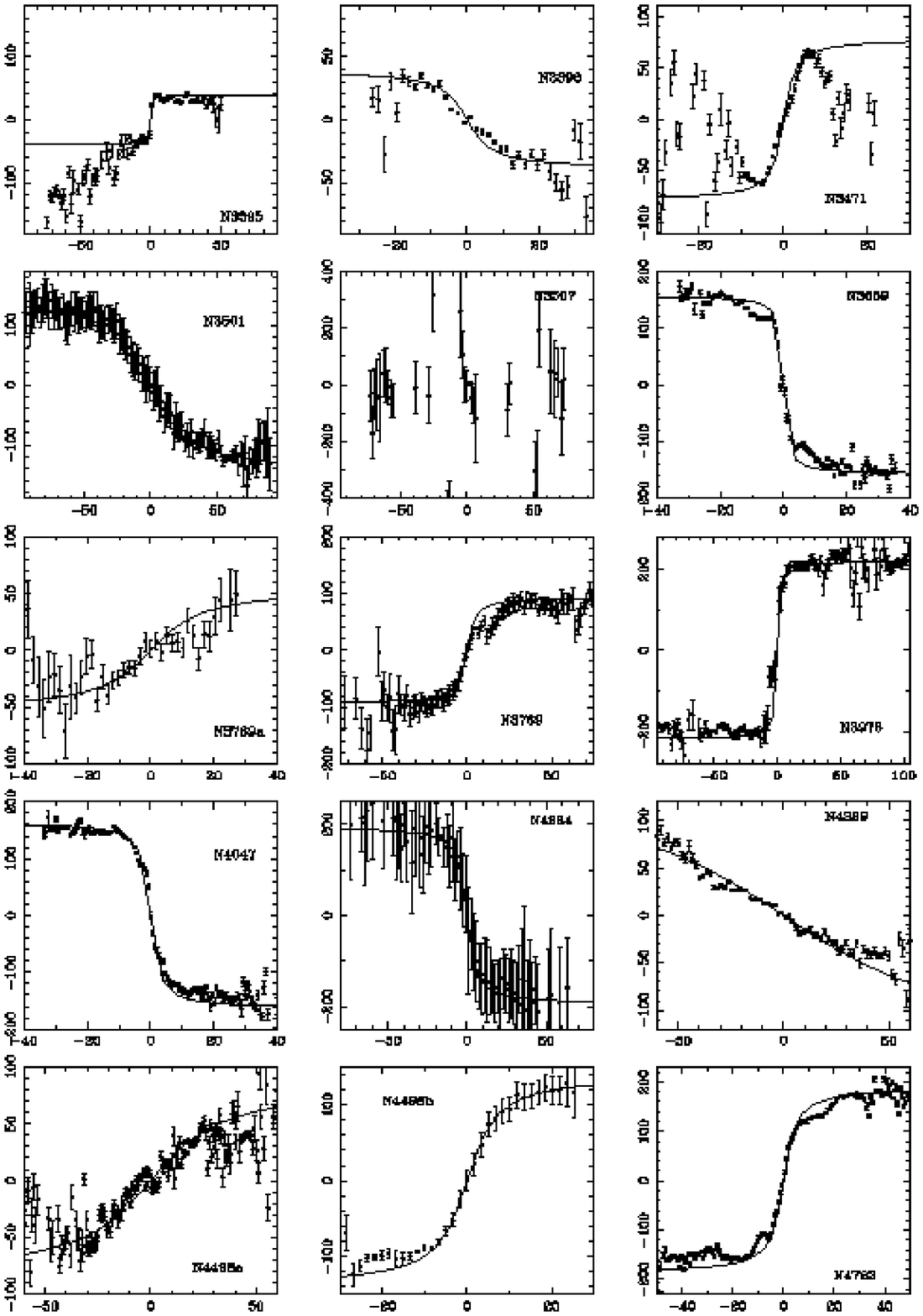,width=17truecm}
\addtocounter{figure}{-1}
\caption{Rotation curves from INT and CAHA spectra (cont.).}
\end{figure*}

\begin{figure*}[htp]
\psfig{figure=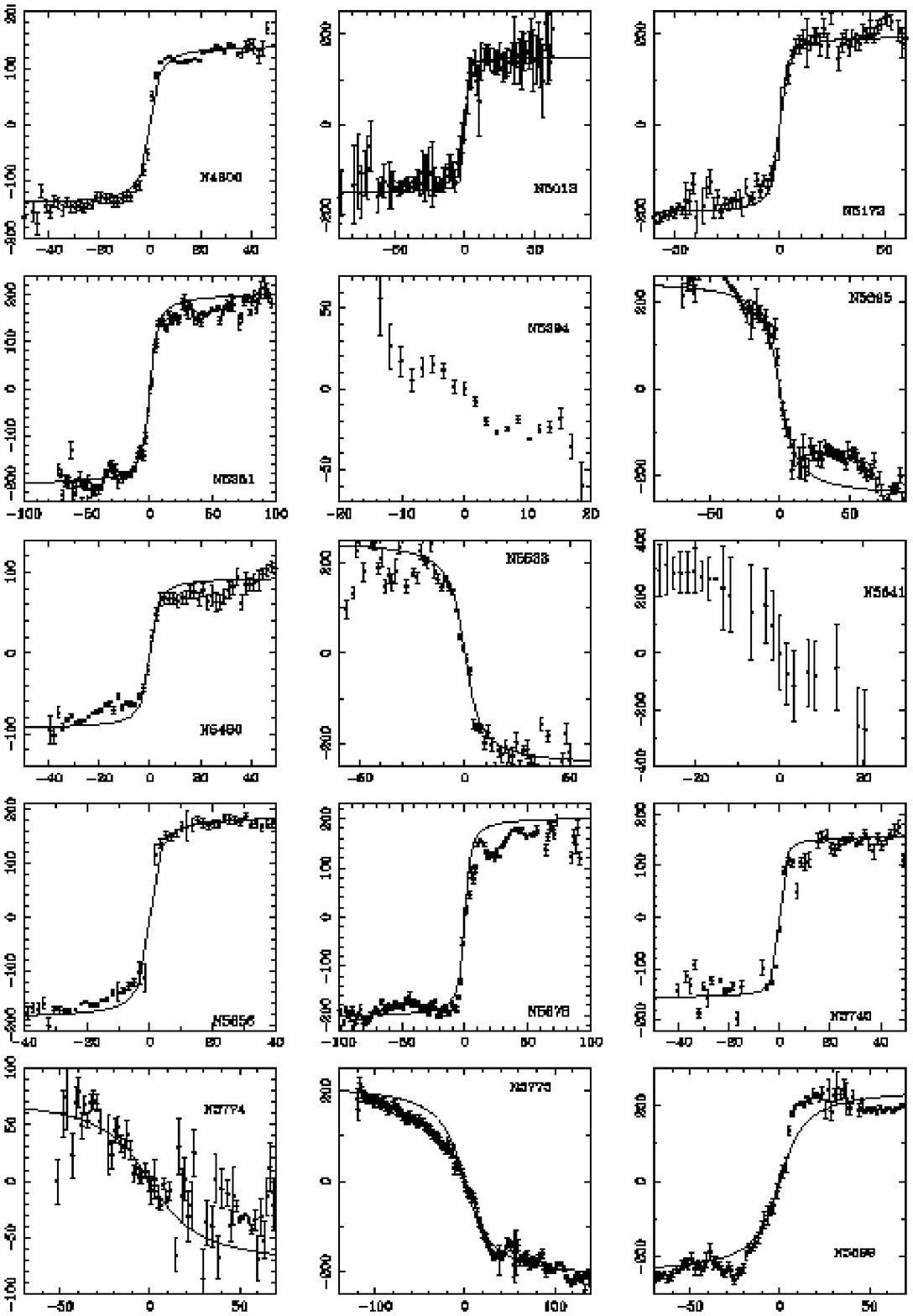,width=17truecm}
\addtocounter{figure}{-1}
\caption{Rotation curves from INT and CAHA spectra (cont.).}
\end{figure*}

\begin{figure*}[htp]
\psfig{figure=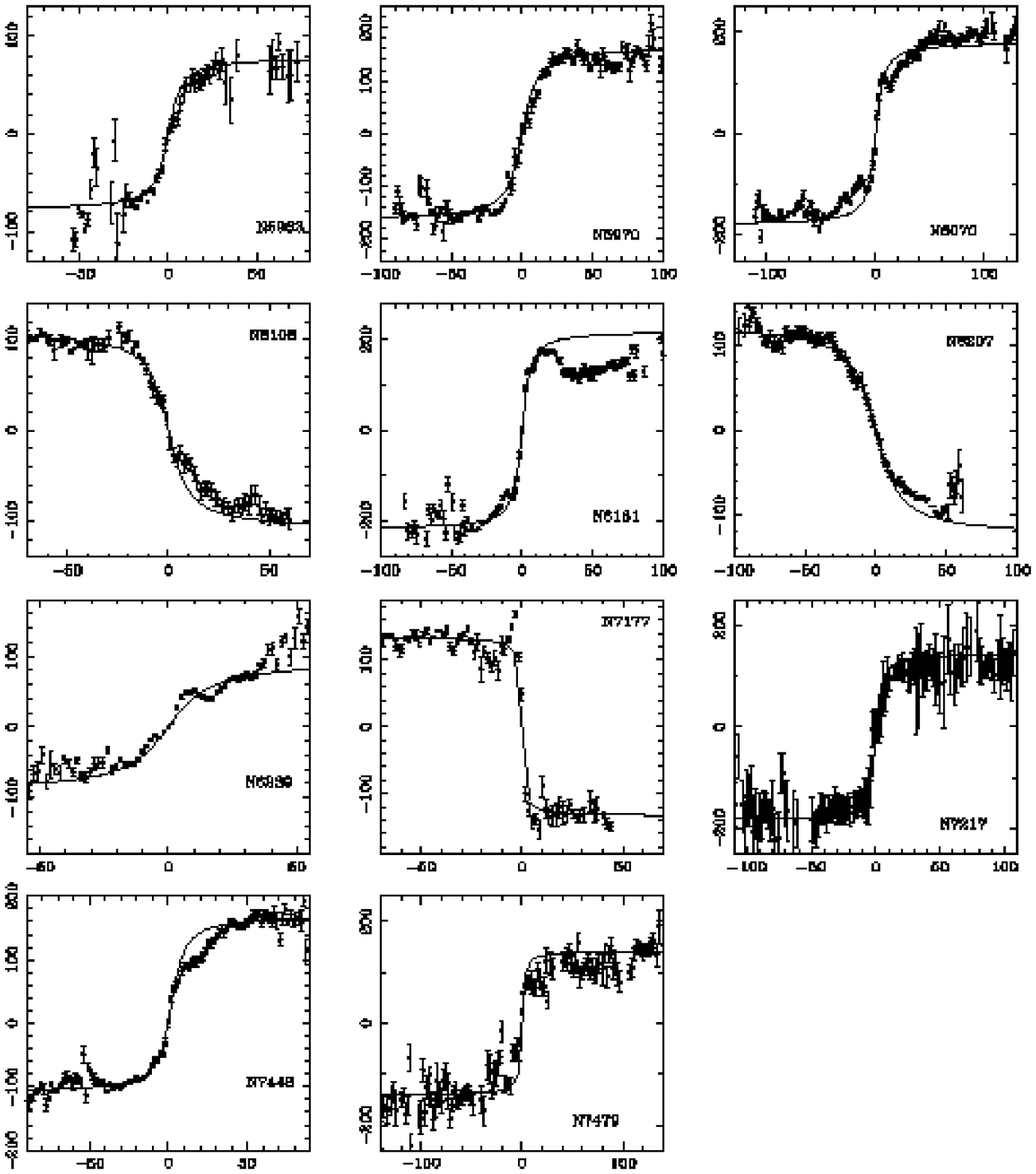,width=17truecm}
\addtocounter{figure}{-1}
\caption{Rotation curves from INT and CAHA spectra (cont.).}
\end{figure*}

\begin{figure*}[htp]
\psfig{figure=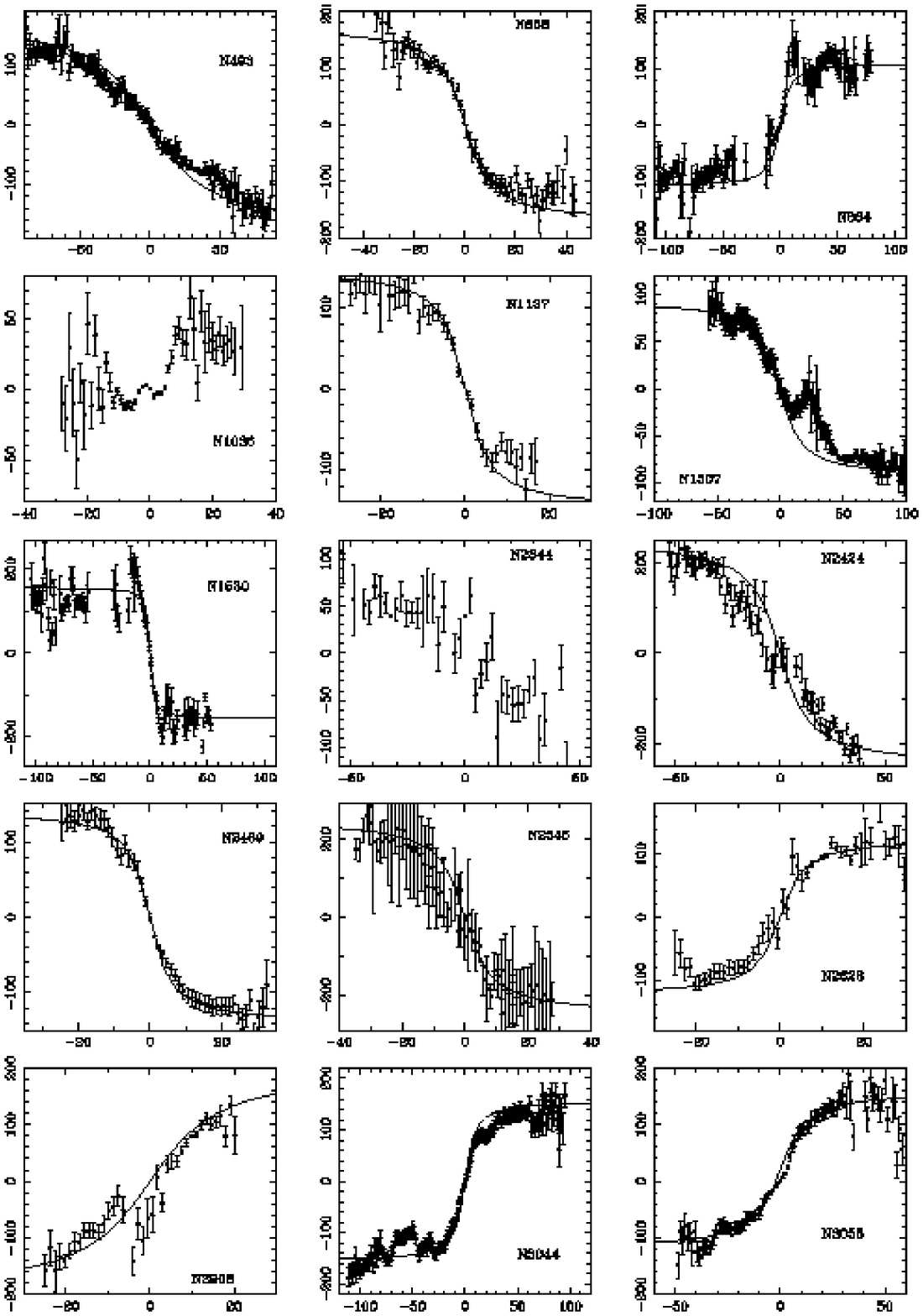,width=17truecm}
\caption{Rotation curves from Asiago spectra.}
\label{jesus}
\end{figure*}

\begin{figure*}[htp]
\psfig{figure=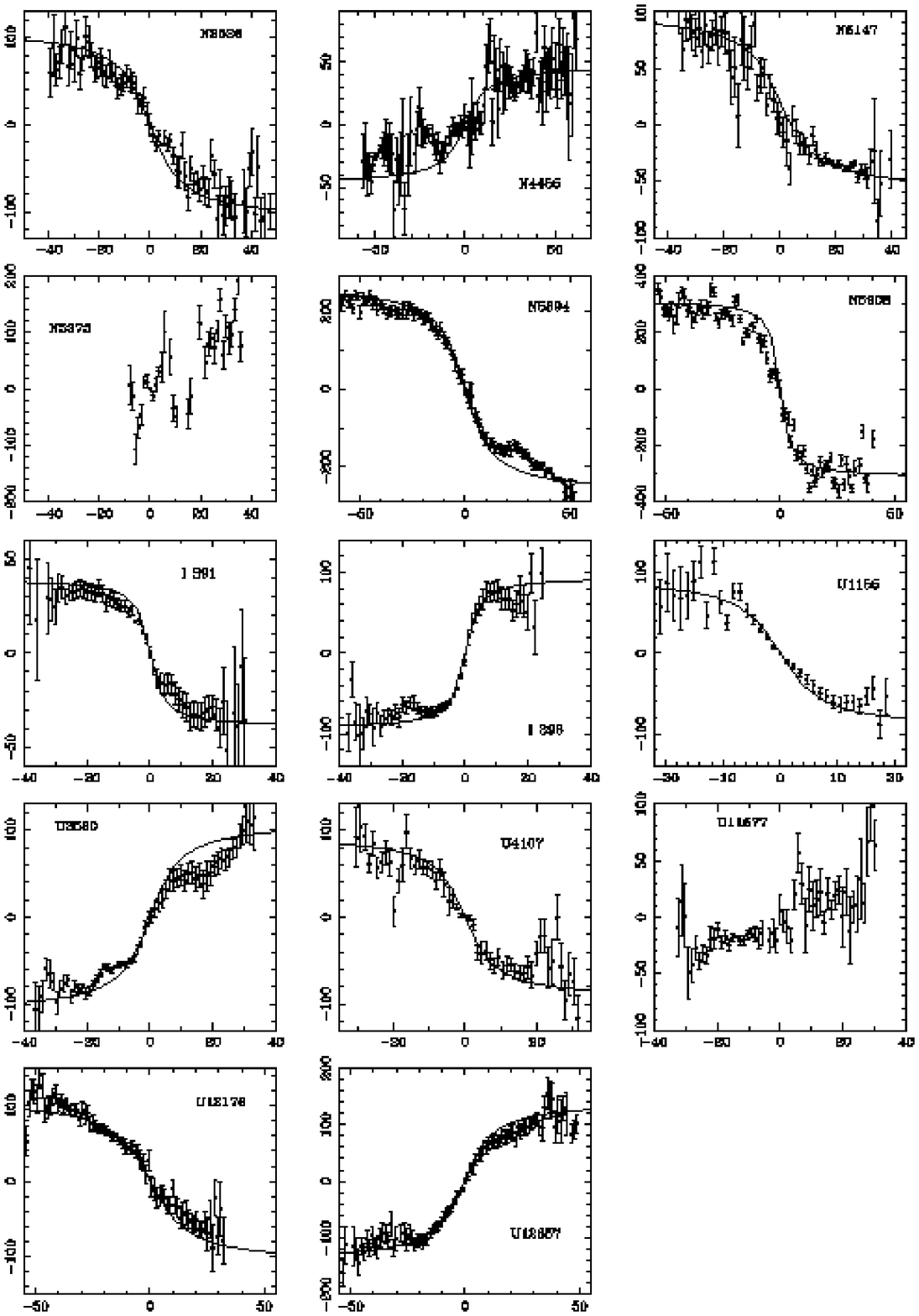,width=17truecm}
\addtocounter{figure}{-1}
\caption{Rotation curves from Asiago spectra (cont.).}
\end{figure*}

The data obtained at Asiago (a total of 29 galaxies) was reduced using IRAF
routines, with the same criteria and definitions as for the other data.

\section{The shape of the rotation curves}

The long slit spectra along the major axis we obtained contain
adequate information to elaborate the rotation curves for 78 out of
the 85 galaxies.
They are presented in Figs. \ref{vel} and \ref{jesus}.
The 7 galaxies with velocity distributions not reliable enough to
trace their rotation curves are NGC~3507, NGC~5394, NGC~5641,
NGC~2344, NGC~1036, NGC~5375 and UGC~11577.  The parameters derived
from our rotation curves (without further correction) are given in
Tables \ref{parspec1} and \ref{parspec2}.

The kinematic center was defined as that section for which the
differences between the two branches of the rotation curve (in
particular, for the most symmetric region which normally corresponds
to the central parts), are minimized. In most cases, within our
resolution, that center corresponds to the photometric center, i.e.,
the absolute maximum in the continuum distribution along the slit. The
red-shift corresponding to that kinematical center was adopted as the
red-shift of the system in all cases and is given in Tables
\ref{parspec1} and \ref{parspec2}. The derived distances were
determined allowing for correction for galactocentric motion,
following the RC2, and for virgocentric motion. Since the galaxies are
rather nearby, the last correction should be considered. It was
calculated following Kraan-Korteweg (1986; model 1). We found that
those corrections typically amount to less than 15\%.  The distances
calculated from the corrected values and H$_0$ = 75\kms~Mpc$^{-1}$ are
given in Tables \ref{parspec1} and \ref{parspec2}.

\begin{table*}[h]
\caption[]{Parameters derived from the mayor axes velocity distributions}
\label{parspec1}
\begin{flushleft}
\begin{scriptsize}
\begin{tabular}{l  c c c c c c c c c c c c r r}
\hline
\noalign{\smallskip}
\multicolumn {1}{l}{Galaxy}
& \multicolumn {1}{c}{m$_B$(RC3)}
& \multicolumn {1}{c}{t}
& \multicolumn {1}{c}{i}
& \multicolumn {1}{c}{D$_{25}$}
& \multicolumn {1}{c}{r$_G$}
& \multicolumn {1}{c}{v$_G$}
& \multicolumn {1}{c}{r$_1$}
& \multicolumn {1}{c}{v$_1$}
& \multicolumn {1}{c}{r$_m$}
& \multicolumn {1}{c}{v$_m$}
& \multicolumn {1}{c}{INT}
& \multicolumn {1}{c}{Shape}
& \multicolumn {1}{c}{cz}
& \multicolumn {1}{c}{D}\\
 & (mag) & & & & (arcsec) &(km/s) &(arcsec) &(km/s) &(arcsec) &(km/s) & & & (km/s) & (Mpc)\\
\noalign{\smallskip}
\hline\noalign{\smallskip}
NGC~828     &      12.87   &  1.0  &  0.17  &   1.46 &  0.98  &  47.0   &  10.0  &  195.0  &  23.0  &  175.0  &  2  &  1  &   5383  &  73.4\\
NGC~2460    &      12.34   &  1.0  &  0.12  &   1.39 &   2.9  &  74.0   &  14.0  &  154.0  &  28.0  &  146.0  &  3  &  2  &   1462  &  24.1\\
NGC~2543$^*$    &      12.14   &  3.0  &  0.24  &   1.37 &  0.97  &  64.0   &  5.00  &  125.0  &  75.0  &  145.0  &  2  &  4  &   2478  &  36.2\\
NGC~2552    &      12.21   &  9.0  &  0.18  &   1.54 &   5.4  &  7.00   &  16.0  &  42.00  &  69.0  &  64.00  &  2  &  5  &    519  &   6.7\\
NGC~2608    &      12.55   &  4.0  &  0.22  &   1.36 &   1.7  &  29.0   &  30.0  &  106.0  &  35.5  &  71.00  &  2  &  5  &   2167  &  31.5\\
NGC~2633    &      12.31   &  3.0  &  0.20  &   1.39 &   1.8  &  83.0   &  5.00  &  103.0  &  34.0  &  110.0  &  3  &  4  &   2199  &  34.8\\
NGC~2701    &      12.32   &  5.0  &  0.13  &   1.34 &   1.6  &  6.00   &  26.0  &  110.0  &  46.0  &  113.0  &  2  &  2  &   2334  &  35.5\\
NGC~2748    &      11.59   &  4.0  &  0.42  &   1.48 &   2.0  &  4.80   &  8.00  &  113.0  &  46.0  &  138.0  &  2  &  2  &   1499  &  25.2\\
NGC~2770    &      11.76   &  5.0  &  0.52  &   1.58 &   1.8  &  8.00   &  34.0  &  122.0  &  90.0  &  165.0  &  2  &  1  &   1988  &  29.8\\
NGC~2964    &      11.64   &  3.0  &  0.26  &   1.46 &   3.4  &  72.0   &  12.0  &  135.0  &  46.0  &  156.0  &  3  &  4  &   1336  &  21.5\\
NGC~2998    &      12.53   &  5.0  &  0.33  &   1.46 &   2.5  &  54.0   &  16.0  &  174.0  &  62.0  &  172.0  &  3  &  2  &   4805  &  67.1\\
NGC~3041    &      11.94   &  5.0  &  0.19  &   1.57 &   2.3  &  49.0   &  25.0  &  139.0  &  94.0  &  134.0  &  2  &  2  &   1418  &  21.8\\
NGC~3183    &      12.18   &  4.0  &  0.23  &   1.37 &   1.7  &  40.0   &  12.0  &  118.0  &  76.0  &  136.0  &  3  &  3  &   3122  &  45.9\\
NGC~3320    &      12.48   &  6.0  &  0.35  &   1.34 &   1.0  &  23.0   &  13.0  &  111.0  &  50.0  &  129.0  &  2  &  1  &   2333  &  35.8\\
NGC~3370    &      11.87   &  5.0  &  0.25  &   1.50 &   2.1  &  57.0   &  33.0  &  138.0  &  57.0  &  126.0  &  1  &  1  &   1275  &  22.8\\
NGC~3395    &      12.09   &  6.0  &  0.23  &   1.32 &  0.90  &  24.0   &  4.00  &  39.00  &  0.00  &  0.000  &  3  &  5  &   1698  &  26.4\\
NGC~3396    &      12.29   &  10.  &  0.42  &   1.49 &   2.6  &  7.00   &  13.0  &  36.00  &  30.0  &  13.00  &  3  &  4  &   1708  &  26.7\\
NGC~3471    &      12.87   &  1.0  &  0.32  &   1.24 &   1.4  &  14.0   &  5.50  &  65.00  &  16.0  &  18.00  &  2  &  3  &   2130  &  33.2\\
NGC~3501    &      12.27   &  5.0  &  0.88  &   1.59 &   1.7  &  15.0   &  55.0  &  125.0  &  90.0  &  126.0  &  3  &  2  &   1159  &  19.2\\
NGC~3507    &      11.63   &  3.0  &  0.07  &   1.53 &   --   &  --     &  --    &  --     &  --    &  --     &  3  &  5  &    973  &  13.1\\
NGC~3689    &      12.80   &  5.0  &  0.17  &   1.22 &   1.6  &  68.0   &  19.0  &  155.0  &  32.0  &  157.0  &  2  &  3  &   2700  &  40.0\\
NGC~3769    &      11.78   &  3.0  &  0.50  &   1.49 &   1.7  &  15.0   &  32.0  &  96.00  &  73.0  &  97.00  &  3  &  4  &    751  &  13.1\\
NGC~3769A   &      14.70   &  9.0  &  0.38  &   1.03 &   --   &  --     &  --    &  --     &  --    &  --     &  3  &  5  &    830  &  15.9\\
NGC~3976    &      11.67   &  3.0  &  0.50  &   1.58 &   3.0  &  180.   &  13.0  &  211.0  &  106.  &  210.0  &  3  &  1  &   2538  &  36.4\\
NGC~4047    &      12.83   &  3.0  &  0.07  &   1.20 &   1.3  &  67.0   &  12.0  &  150.0  &  32.0  &  147.0  &  2  &  1  &   3445  &  49.4\\
NGC~4284    &      13.75   &  4.0  &  0.33  &   1.40 &   1.1  &  52.0   &  20.0  &  186.0  &  53.0  &  192.0  &  2  &  2  &   4244  &  58.2\\
NGC~4389    &      12.23   &  4.0  &  0.29  &   1.42 &  0.90  &  2.60   &  7.50  &  16.00  &  65.0  &  107.0  &  3  &  3  &    712  &  12.2\\
NGC~4496A   &      12.00   &  9.0  &  0.10  &   1.60 &   --   &  --     &  31.0  &  55.00  &  68.0  &  70.00  &  2  &  4  &   1727  &  27.2\\
NGC~4496B   &      0.000   &  10.  &  0.03  &   1.01 &   5.1  &  71.0   &  12.0  &  105.0  &  25.0  &  115.0  &  2  &  1  &   4539  &  58.5\\
NGC~4793    &      11.86   &  5.0  &  0.27  &   1.44 &   1.3  &  44.0   &  8.00  &  120.0  &  56.0  &  155.0  &  3  &  4  &   2487  &  37.8\\
NGC~4800    &      12.21   &  3.0  &  0.13  &   1.20 &   3.0  &  85.0   &  8.00  &  124.0  &  41.0  &  140.0  &  2  &  3  &    902  &  13.3\\
NGC~5012    &      12.32   &  5.0  &  0.24  &   1.46 &   4.0  &  49.0   &  42.0  &  220.0  &  70.0  &  179.0  &  2  &  2  &   2635  &  40.7\\
NGC~5172    &      11.99   &  4.0  &  0.29  &   1.52 &   4.7  &  166.   &  40.0  &  240.0  &  53.0  &  235.0  &  2  &  1  &   4076  &  57.0\\
NGC~5351$^*$    &      12.48   &  3.0  &  0.28  &   1.47 &   2.7  &  94.0   &  14.0  &  164.0  &  53.0  &  186.0  &  3  &  2  &   3665  &  52.7\\
NGC~5394$^*$    &      13.90   &  3.0  &  0.25  &   1.24 &   --   &  --     &  --    &  --     &  --    &  --     &  3  &  5  &   3460  &  46.8\\
NGC~5395    &      12.11   &  3.0  &  0.27  &   1.46 &   2.7  &  104.   &  12.0  &  177.0  &  54.0  &  225.0  &  3  &  4  &   3496  &  51.8\\
NGC~5480$^*$    &      12.54   &  5.0  &  0.18  &   1.24 &   2.0  &  46.0   &  4.70  &  66.00  &  32.0  &  86.00  &  3  &  2  &   1887  &  30.4\\
NGC~5533    &      12.39   &  2.0  &  0.21  &   1.49 &   2.0  &  40.0   &  13.0  &  226.0  &  41.0  &  226.0  &  3  &  4  &   3856  &  55.7\\
NGC~5641    &      12.73   &  2.0  &  0.26  &   1.39 &   2.7  &  144.   &  1.70  &  144.0  &  16.0  &  282.0  &  3  &  5  &   4320  &  63.3\\
NGC~5656$^*$    &      12.59   &  2.0  &  0.13  &   1.28 &   1.4  &  114.   &  18.0  &  170.0  &  31.0  &  170.0  &  3  &  2  &   3177  &  47.2\\
NGC~5678    &      11.68   &  3.0  &  0.31  &   1.52 &   4.0  &  127.   &  10.0  &  180.0  &  84.0  &  187.0  &  3  &  4  &   1940  &  36.2\\
NGC~5740    &      12.07   &  3.0  &  0.29  &   1.47 &   3.0  &  120.   &  4.00  &  120.0  &  32.0  &  141.0  &  3  &  2  &   1579  &  24.7\\
NGC~5774    &      13.01   &  7.0  &  0.09  &   1.48 &   --   &  --     &  5.00  &  110.0  &  89.0  &  161.0  &  3  &  5  &   1544  &  24.7\\
NGC~5775    &      11.25   &  5.0  &  0.62  &   1.62 &   4.0  &  28.0   &  32.0  &  142.0  &  96.0  &  209.0  &  3  &  4  &   1707  &  24.8\\
NGC~5899$^*$    &      11.78   &  5.0  &  0.43  &   1.50 &   2.6  &  53.0   &  19.0  &  239.0  &  56.0  &  227.0  &  3  &  4  &   2621  &  39.1\\
NGC~5963$^*$    &      12.76   &  1.0  &  0.11  &   1.52 &   2.7  &  25.0   &  9.00  &  60.00  &  50.0  &  65.00  &  2  &  4  &    663  &  13.8\\
NGC~5970    &      11.85   &  5.0  &  0.17  &   1.46 &   2.6  &  41.0   &  17.0  &  149.0  &  70.0  &  148.0  &  3  &  3  &   1968  &  31.3\\
NGC~6070    &      11.58   &  6.0  &  0.30  &   1.55 &  0.85  &  35.5   &  7.00  &  110.0  &  100.  &  190.0  &  3  &  3  &   1995  &  29.9\\
NGC~6106    &      12.29   &  5.0  &  0.30  &   1.40 &   2.0  &  35.0   &  20.0  &  110.0  &  58.0  &  110.0  &  1  &  4  &   1459  &  22.9\\
NGC~6181    &      11.85   &  5.0  &  0.35  &   1.40 &   2.7  &  118.   &  19.0  &  184.0  &  53.0  &  174.0  &  2  &  4  &   2401  &  35.6\\
NGC~6207    &      11.59   &  5.0  &  0.36  &   1.47 &   5.3  &  38.0   &  24.0  &  95.00  &  54.0  &  106.0  &  2  &  2  &    835  &  16.0\\
NGC~6239    &      12.44   &  3.0  &  0.45  &   1.41 &   2.7  &  28.0   &  10.0  &  47.00  &  40.0  &  84.00  &  2  &  1  &    935  &  17.6\\
NGC~7177$^*$    &      11.50   &  3.0  &  0.19  &   1.49 &   1.8  &  102.   &  3.00  &  148.0  &  68.0  &  148.0  &  1  &  4  &   1118  &  18.5\\
NGC~7217$^*$    &      10.52   &  2.0  &  0.08  &   1.59 &   --   &  --     &  5.00  &  138.0  &  86.0  &  183.0  &  3  &  1  &    950  &  16.5\\
NGC~7448    &      11.48   &  4.0  &  0.34  &   1.43 &   2.7  &  52.0   &  31.0  &  133.0  &  68.0  &  135.0  &  1  &  4  &   2120  &  32.0\\
NGC~7479$^*$    &      11.21   &  5.0  &  0.12  &   1.61 &   1.4  &  52.0   &  39.0  &  138.0  &  107.  &  154.0  &  1  &  4  &   2334  &  33.7\\
\hline
\end{tabular}
\null
$^*$ Galaxies showing evidences of non-circular motions.
\end{scriptsize}
\end{flushleft}
\end{table*}

\begin{table*}[h]
\caption[]{Parameters derived from the mayor axes velocity distributions}
\label{parspec2}
\begin{flushleft}
\begin{scriptsize}
\begin{tabular}{l  c c c c c c c c c c c c r r}
\hline
\noalign{\smallskip}
\multicolumn {1}{l}{Galaxy}
& \multicolumn {1}{c}{m$_B$(RC3)}
& \multicolumn {1}{c}{t}
& \multicolumn {1}{c}{i}
& \multicolumn {1}{c}{D$_{25}$}
& \multicolumn {1}{c}{r$_G$}
& \multicolumn {1}{c}{v$_G$}
& \multicolumn {1}{c}{r$_1$}
& \multicolumn {1}{c}{v$_1$}
& \multicolumn {1}{c}{r$_m$}
& \multicolumn {1}{c}{v$_m$}
& \multicolumn {1}{c}{INT}
& \multicolumn {1}{c}{Shape}
& \multicolumn {1}{c}{cz}
& \multicolumn {1}{c}{D}\\
 & (mag) & & & & (arcsec) &(km/s) &(arcsec) &(km/s) &(arcsec) &(km/s) & & & (km/s) & (Mpc)\\
\noalign{\smallskip}
\hline\noalign{\smallskip}
NGC~493     &      12.01   &  6.0  &  0.62  &   1.53 &   2.0  &  11.7   &  32.0  &  94.00  &  74.0  &  138.0  &  2  &  1  &   2354  &  31.7\\
NGC~658     &      12.58   &  3.0  &  0.28  &   1.48 &   1.5  &  30.9   &  20.0  &  138.0  &  34.0  &  112.0  &  1  &  4  &   2990  &  40.8\\
NGC~864     &      11.26   &  5.0  &  0.12  &   1.67 &   1.0  &  5.90   &  12.0  &  136.0  &  91.0  &  107.0  &  2  &  5  &   1543  &  21.7\\
NGC~1036    &      13.75   &  10.  &  0.14  &   1.16 &   --   &  --     &  --    &  --     &  --    &  --     &  1  &  5  &    800  &  12.7\\
NGC~1137    &      12.61   &  3.0  &  0.21  &   1.33 &  0.50  &  7.90   &  11.0  &  126.0  &  26.0  &  150.0  &  1  &  4  &   3028  &  40.2\\
NGC~1507    &      11.82   &  9.0  &  0.62  &   1.56 &   2.5  &  14.0   &  44.5  &  84.00  &  84.0  &  87.00  &  1  &  5  &    888  &  12.0\\
NGC~1530$^*$    &      11.42   &  3.0  &  0.28  &   1.66 &   2.5  &  69.0   &  9.50  &  192.0  &  114.  &  157.0  &  1  &  4  &   2496  &  37.5\\
NGC~2344    &      12.48   &  5.0  &  0.01  &   1.23 &   --   &  --     &  --    &  --     &  --    &  --     &  2  &  4  &    990  &  17.1\\
NGC~2424    &      12.19   &  3.0  &  0.80  &   1.58 &  0.50  &  1.40   &  28.5  &  207.0  &  45.5  &  217.0  &  3  &  5  &   3365  &  45.6\\
NGC~2469    &      13.12   &  4.0  &  0.17  &   1.05 &   2.0  &  42.0   &  14.0  &  128.6  &  27.0  &  119.6  &  1  &  1  &   3505  &  48.0\\
NGC~2545    &      12.66   &  2.0  &  0.24  &   1.30 &   1.0  &  151.   &  2.00  &  196.0  &  30.0  &  183.0  &  2  &  1  &   3414  &  46.8\\
NGC~2628    &      13.85   &  5.0  &  0.02  &   1.06 &  0.50  &  29.0   &  10.5  &  99.00  &  24.5  &  105.5  &  1  &  3  &   3609  &  47.5\\
NGC~2906$^*$    &      12.97   &  6.0  &  0.23  &   1.16 &   1.5  &  56.0   &  4.50  &  98.00  &  21.5  &  95.40  &  1  &  5  &   2168  &  26.1\\
NGC~3044    &      11.13   &  5.0  &  0.85  &   1.69 &   3.0  &  22.0   &  38.0  &  138.0  &  95.0  &  179.0  &  2  &  3  &   1313  &  20.4\\
NGC~3055    &      12.18   &  5.0  &  0.21  &   1.32 &   2.0  &  11.7   &  28.0  &  148.0  &  46.0  &  143.0  &  2  &  4  &   1827  &  27.9\\
NGC~3526    &      12.79   &  5.0  &  0.65  &   1.28 &   2.0  &  27.0   &  13.0  &  72.00  &  38.0  &  111.0  &  1  &  2  &   1425  &  22.6\\
NGC~4455    &      12.12   &  7.0  &  0.53  &   1.44 &   1.0  &  11.0   &  12.0  &  48.00  &  53.0  &  60.00  &  1  &  5  &    621  &   6.6\\
NGC~5147    &      12.29   &  8.0  &  0.09  &   1.28 &   1.0  &  6.30   &  8.00  &  54.00  &  31.0  &  77.00  &  2  &  3  &   1083  &  18.5\\
NGC~5375    &      12.29   &  2.0  &  0.07  &   1.51 &   --   &  --     &  --    &  --     &  --    &  --     &  2  &  5  &   2391  &  36.7\\
NGC~5894    &      12.33   &  8.0  &  0.85  &   1.48 &   1.0  &  28.0   &  16.0  &  174.0  &  51.0  &  223.0  &  2  &  3  &   2483  &  38.6\\
NGC~5908    &      12.29   &  3.0  &  0.43  &   1.51 &   1.0  &  62.0   &  16.0  &  305.0  &  45.0  &  325.0  &  3  &  4  &   3331  &  49.2\\
IC~391      &      12.56   &  5.0  &  0.02  &   1.06 &   1.0  &  7.90   &  11.0  &  31.50  &  28.0  &  28.00  &  1  &  1  &   1587  &  26.0\\
IC~396      &      12.08   &  10.  &  0.16  &   1.32 &   1.0  &  24.3   &  8.00  &  74.70  &  30.0  &  76.00  &  1  &  2  &    882  &  14.9\\
UGC~1155    &      14.17   &  6.0  &  0.24  &  0.850 &   5.0  &  46.9   &  10.0  &  87.00  &  18.0  &  70.00  &  1  &  3  &   3188  &  44.6\\
UGC~3580    &      12.20   &  1.0  &  0.28  &   1.53 &   1.5  &  10.7   &  6.50  &  48.80  &  29.5  &  90.60  &  2  &  4  &   1198  &  21.0\\
UGC~4107    &      13.52   &  5.0  &  0.01  &   1.15 &   1.0  &  1.54   &  9.00  &  58.90  &  26.0  &  82.50  &  1  &  4  &   3486  &  47.1\\
UGC~11577   &      13.44   &  6.0  &  0.12  &   1.21 &   --   &  --     &  --    &  --     &  --    &  --     &  1  &  5  &   3767  &  52.5\\
UGC~12178   &      13.13   &  8.0  &  0.26  &   1.47 &  0.50  &  17.3   &  23.5  &  89.60  &  48.5  &  109.4  &  2  &  2  &   1936  &  28.8\\
UGC~12857   &      13.40   &  4.0  &  0.68  &   1.26 &   2.0  &  23.4   &  17.0  &  95.00  &  45.0  &  112.0  &  2  &  3  &   2474  &  35.5\\
\hline
\end{tabular}
\null
$^*$ Galaxies showing evidences of non-circular motions.
\end{scriptsize}
\end{flushleft}
\end{table*}

\begin{figure*}[htp]
\psfig{figure=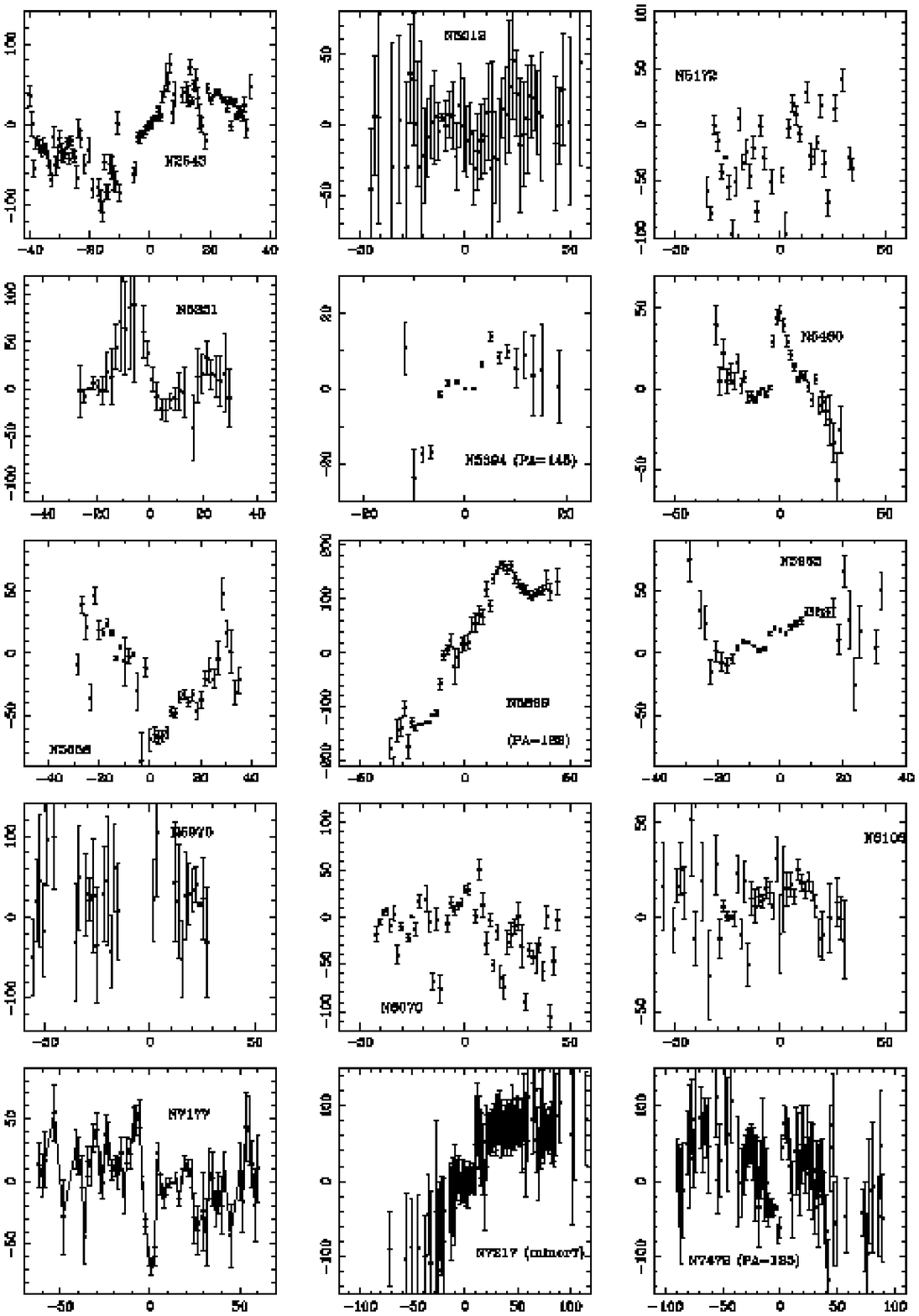,width=17truecm}
\caption{Velocity distributions for additional PAs (minor axes,
otherwise the PA is given) from INT and CAHA spectra.}
\label{minornos}
\end{figure*}

\begin{figure*}[htp]
\psfig{figure=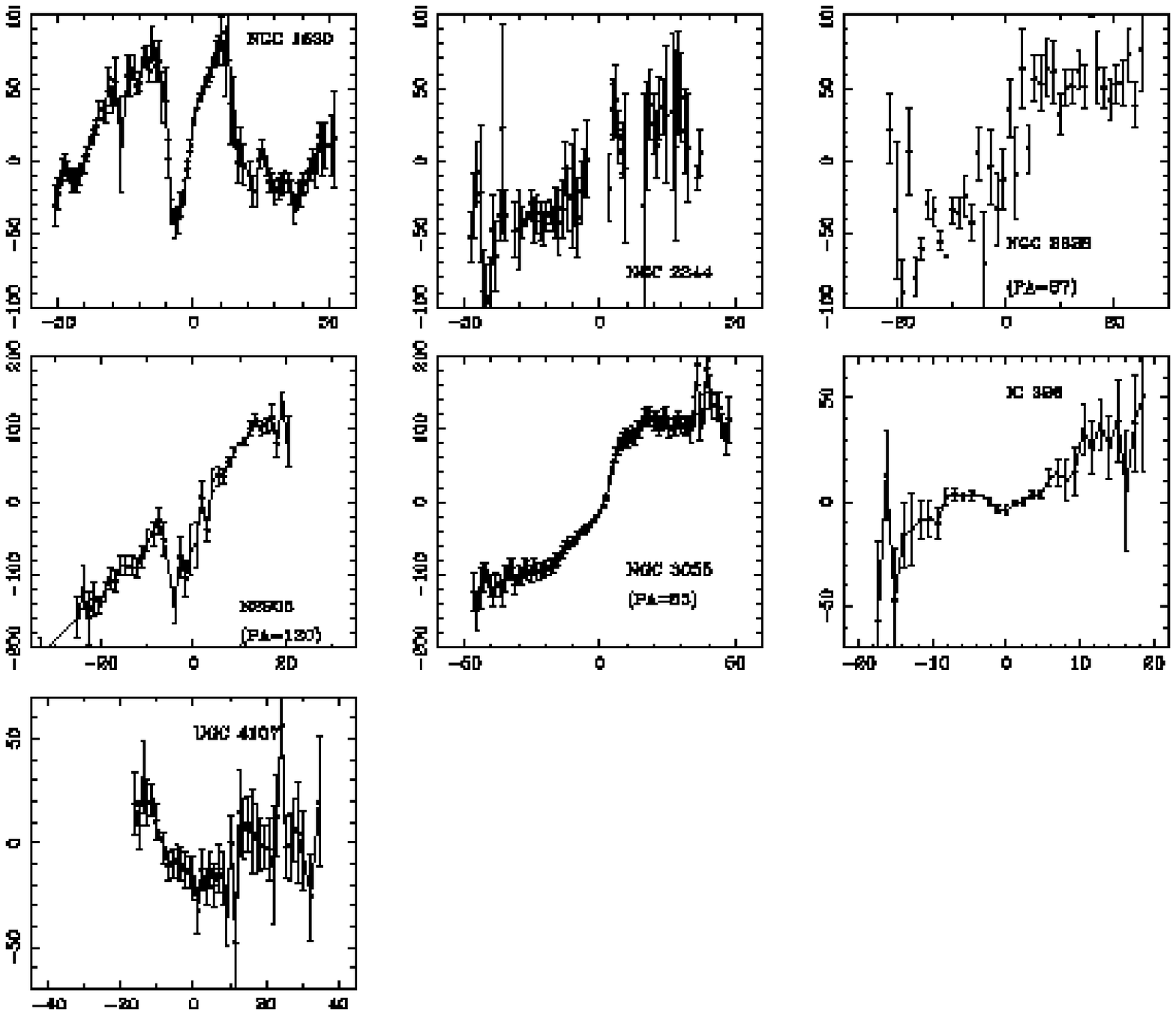,width=17truecm}
\caption{Velocity distributions for additional PAs (minor axes,
otherwise the PA is given) from Asiago spectra.}
\label{minorjesus}
\end{figure*}

We have also obtained 21 additional spectra, 16 of them along the
minor axis.  The resulting velocity distributions are shown in
Figs. \ref{minornos} and \ref{minorjesus}. We notice that for NGC~2543
and NGC~5963 the velocity distributions along the quoted minor axes
show velocity amplitudes for the outskirts of about 50 km/s, implying
that a misalignment exists between photometric and kinematical minor
axes. For NGC~7217 and NGC~7479 the resulting amplitude is much
higher, up to about 100 km/s.
The presence of non-circular motions is clearly detected in the
central regions of NGC~1530, NGC~2543, NGC~5351, NGC~5480, NGC~5656,
NGC~5963, NGC~7177, NGC~7217 and NGC~7479. We note that, excepting
NGC~5656 and NGC~5963, for which the detailed morphological
classification is missing in the RC3 catalogue, the other galaxies are
classified as having rings, and four of them are weakly (NGC~7177) or
strongly (NGC~1530, NGC~2543, NGC~7479) barred.  Non-circular motions
are detectable in 3 out of the 5 non-major-axis spectra, namely
NGC~5394, NGC~5899 and NGC~2906, in the inner 5, 10 and 7 arc-seconds,
respectively. The exact connection between morphological and
kinematical features, for which the imaging is needed, is out the
scope of this paper. We just note that the percentage of galaxies
showing noncircular motions (11\%) is in agreement with the 17\%
reported by Rubin et al. (1997).
%

\begin{figure}
\psfig{file=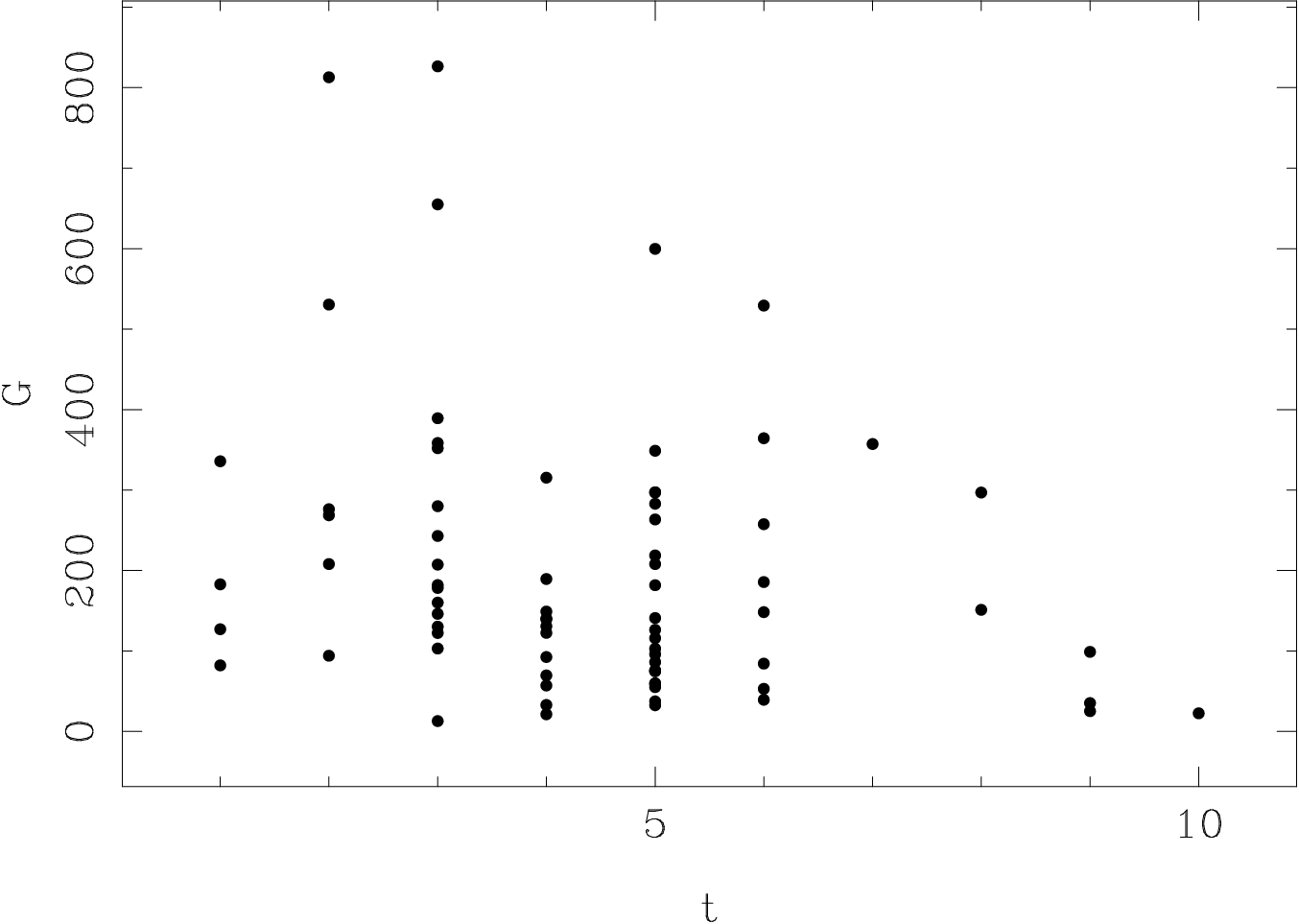,width=6truecm,angle=-90}
\caption{Distribution of G values (representing the inner solid-body
rotation gradient) as a function of the morphological type, t.}
\label{G}
\end{figure}

Different parameters are used to describe the rotation curves
depending on their final purpose. We have measured all the parameters
that could be of some interest for statistical purposes. In particular
we have measured the first local maximum, V$_{max}$, in the observed
velocity distribution, and the corresponding distance to the center,
R$_{max}$. We have also measured the slope, G (in km/s/Kpc) of the
inner rotation curve. This parameter, directly related to the
bulge-to-disk ratio, appears as one of the two main parameters
(together with the total mass or luminosity) to describe the family of
isolated spiral galaxies
and correlates with the bulge-to-disk ratio better than the
morphological type (Paper I and II and see below). Indeed, we
also measured the absolute maximum velocity, V$_m$, and the central
distance at which it is reached, R$_m$, and the maximum extent of the
observed rotation curve, R$_M$.
All those parameters show relations between them and with other
galactic properties.  Thus, as shown in Fig. \ref{G}, later type
galaxies tend to have lower G values (see also Paper II and
Baiesi-Pillastrini 1987). A better correlation was found between the
same slope G, and a quantitative descriptor of the morphological type
as the bulge to disk ratio, B/D, in particular for isolated galaxies
(Paper II).  Unfortunately, the available photometric data in the
literature is not abundant and homogeneous enough to further test that
relation here.

Regarding the outer part of the rotation curve, a descriptor was
defined (Paper I and II) for its behavior,

$$\Delta = arctan [((v_m - v_{max})/max(v_m,v_{max})) \times (r_m -
r_{max})/r_{25}]$$

\begin{figure}
\psfig{file=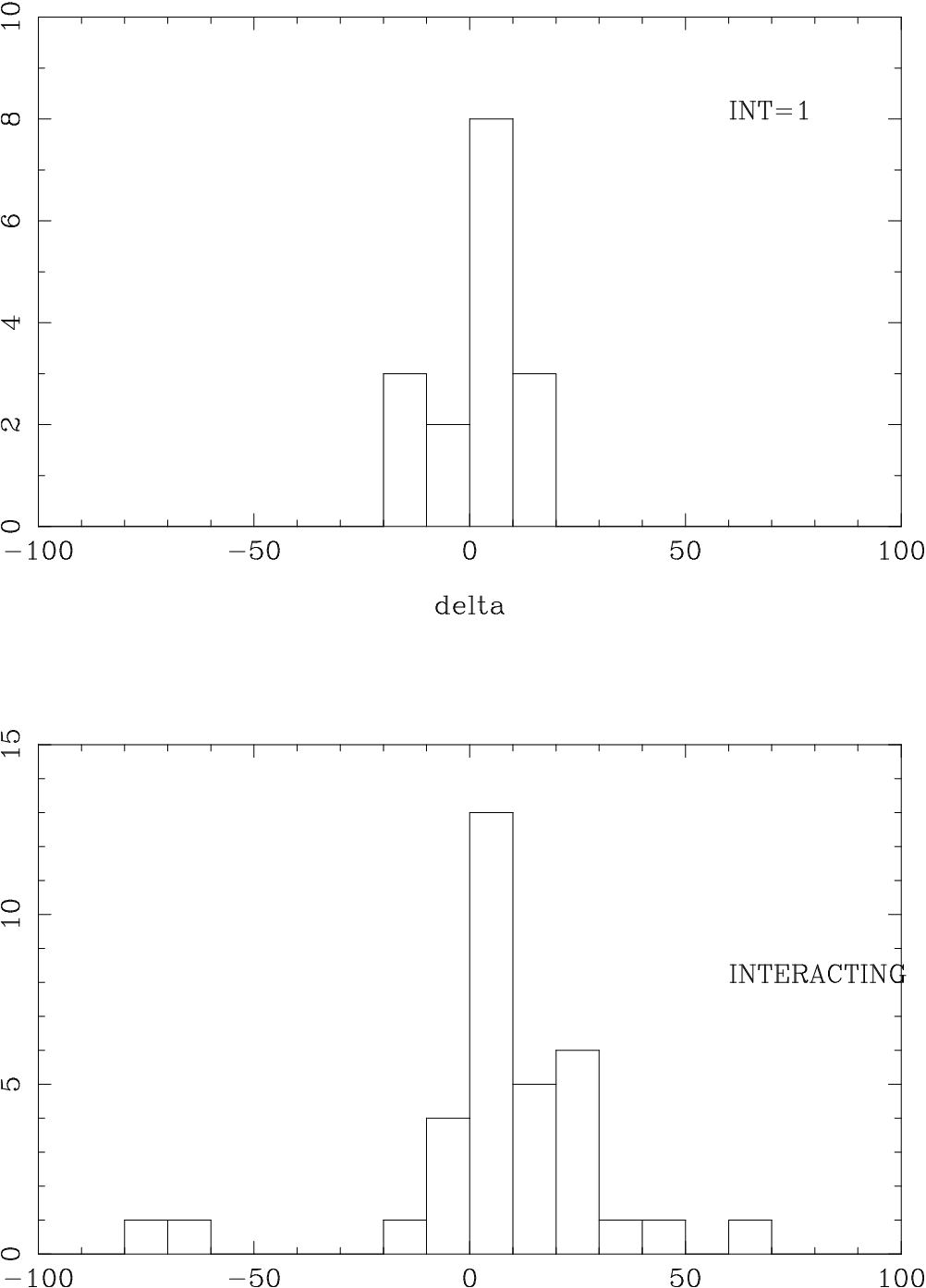,width=8.5truecm}
\caption{Distribution of $\Delta$ values (representing the shape of the outer
rotation curve.}
\label{delta}
\end{figure}

It takes values around zero for flat rotation curves, whereas it is
positive for rising curves and negative for declining curves. The
distribution of this parameter for isolated and interacting is shown
in Fig.\ref{delta}. We find that for isolated galaxies $\Delta$ =
0$\pm$ 7, and for interacting galaxies  $\Delta$ = 5
$\pm$ 16.
This result confirms and extends the earlier one reported in Paper II,
where galaxies in isolated pairs were compared to isolated galaxies.

Sometimes a different parameter is adopted to describe the gradient of
the outer rotation curve, defined as (see for instance Dale et
al. 2001),
$$ OG = [V(0.70 \times R_{25}) - V(0.35 \times R_{25})]/V(0.70 \times R_{ 25}) $$
We found OG = 7.8$\pm$6.6 for isolated galaxies and 10.4$\pm$15.5 for
interacting galaxies, showing that both parameters, $\Delta$ and OG give
similar information about the outer rotation curve.

\begin{figure}
\psfig{file=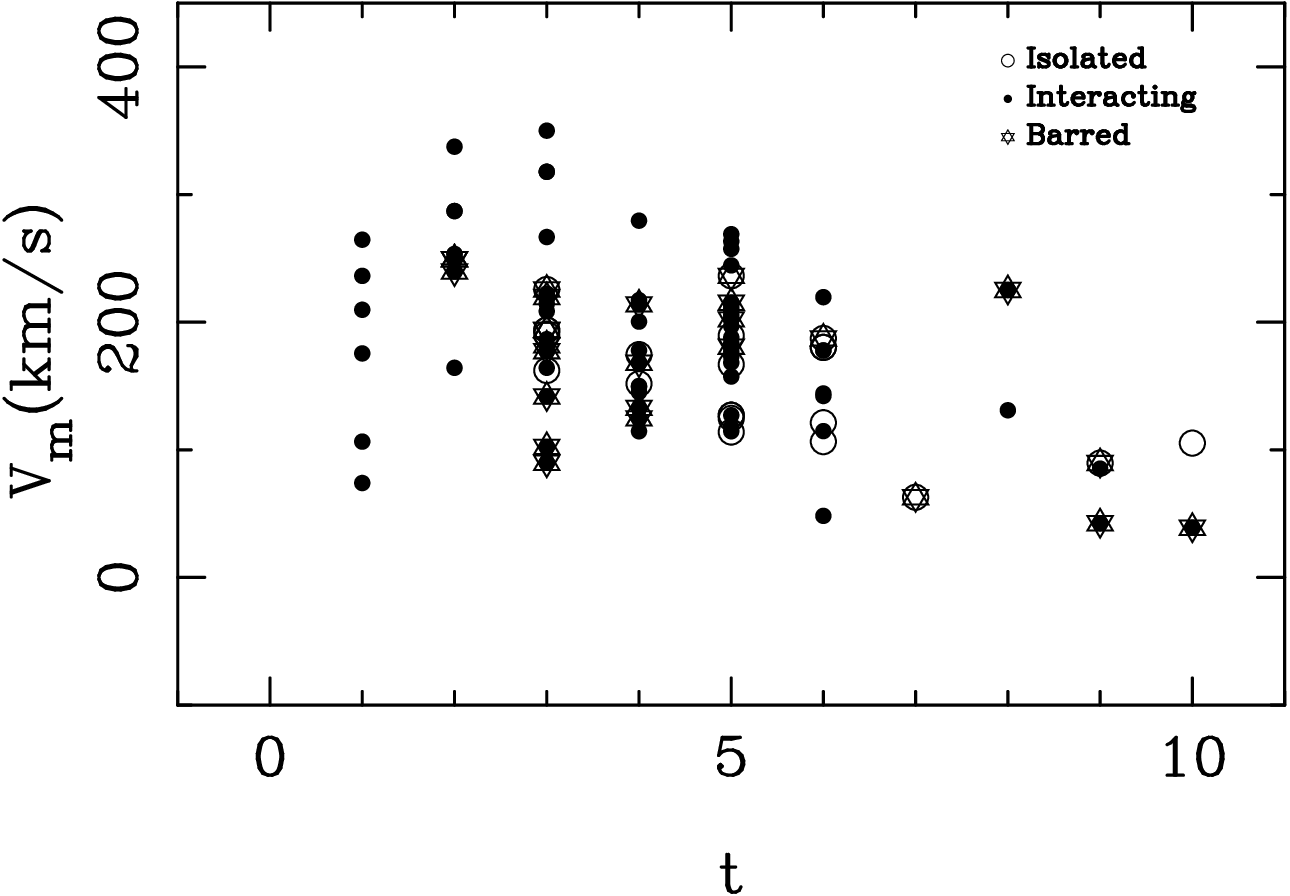,width=6truecm,angle=-90}
\caption{Absolute maximum velocity versus morphological type, $t$.}
\label{Vmax}
\end{figure}

\begin{figure}
\psfig{file=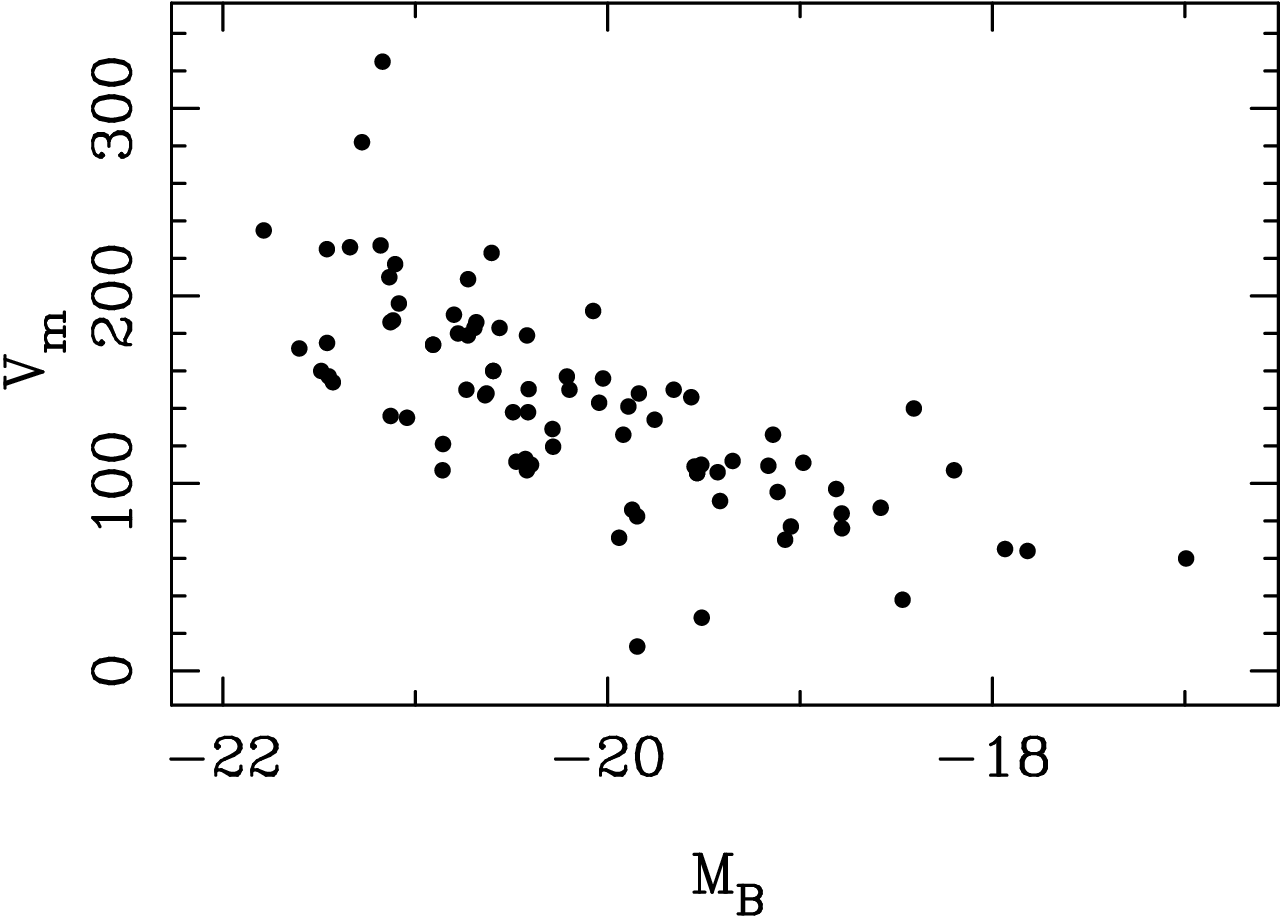,width=6truecm,angle=-90}
\caption{Absolute maximum velocity as a function of absolute B magnitude.}
\label{VMB}
\end{figure}

Regarding the maximum velocity\footnote{All the velocities have
been corrected for inclination following Peterson et al. (1978).}, V$_m$, it
has been known that it is also related with the morphological type
(Rubin et al. 1991). We find the same tendency with the present data
set (Fig. \ref{Vmax}). The median values for type 1 to 6 are
223$\pm$46, 253$\pm$29, 194$\pm$41, 160$\pm$33, 186$\pm$37, and
142$\pm$37, respectively. The same V$_m$ is related to the absolute
magnitude (Fig. \ref{VMB}), with less scatter for the isolated
objects.

\begin{figure}
\psfig{file=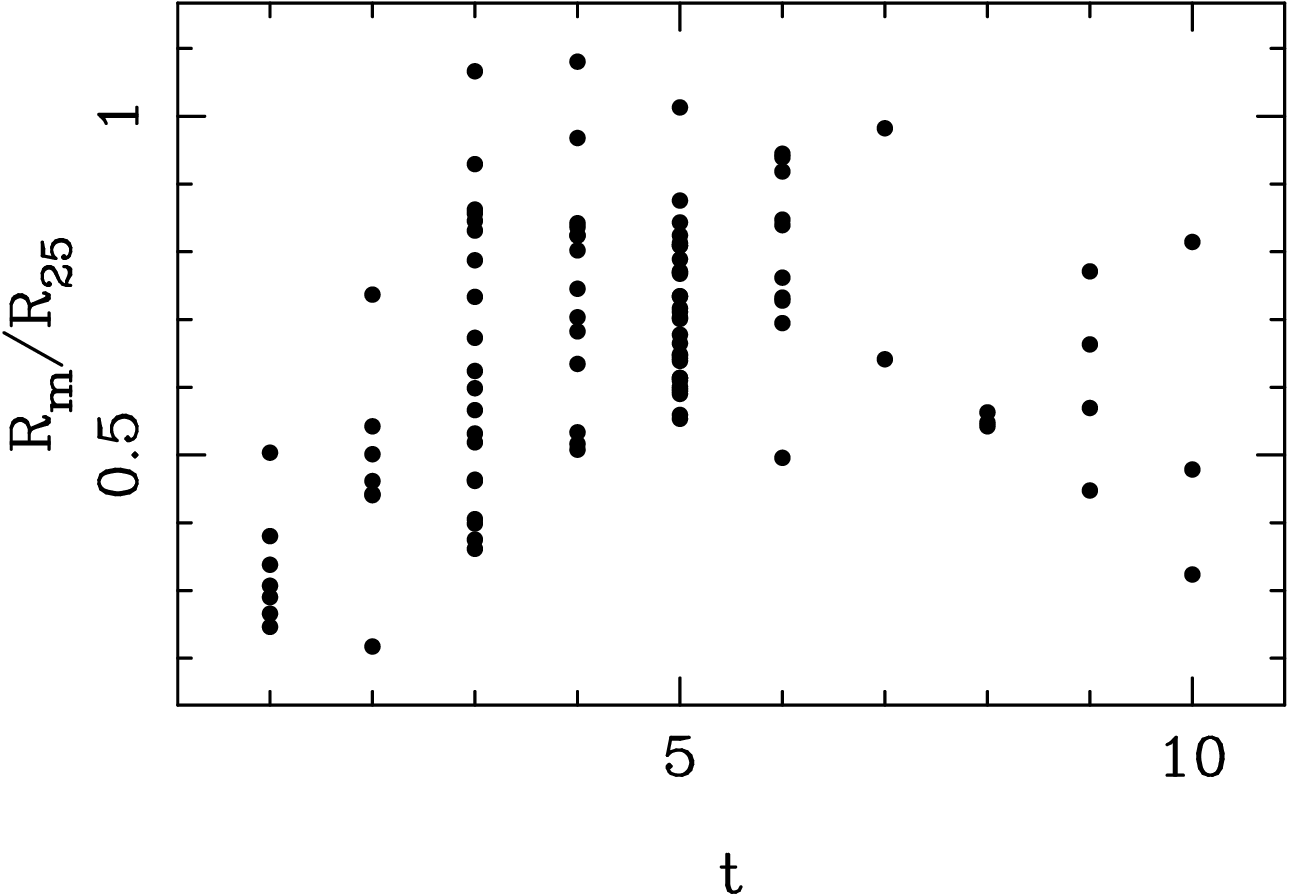,width=6truecm,angle=-90}
\caption{Maximum extension of the rotation curve, R$_M$ (in units of
R$_{25}$ {\it versus} the morphological type t.
}
\label{rcut_t}
\end{figure}

We have further investigated whether there the tendency found by Rubin
et al.  (1999) for later spiral galaxies to show more extended HII
distributions, what translates into longer rotation curves. This
tendency is confirmed, as shown in Fig. \ref{rcut_t}, where the
maximum extension of the rotation curve, R$_M$ (in units of R$_{25}$)
is plotted versus the morphological type. The increasing trend is
clear for spirals up to t = 6. No difference is found when the sample
is separated according to the interaction state.

The central mass (inside the solid-body rotation region) seems to be
slightly higher for early type galaxies, consistent with them hosting
bigger bulges. No tendency is found between central mass and
interaction class. The total masses within R$_M$ and R$_{25}$ (this
last from the RC3 catalogue), for an homogeneous and spherically
symmetric distribution (M$_T$ = 2.3265 $\times$ 10$^5$ r v$^2$(r)
M$_{\odot}$, with r in Kpc and v(r) in km/s) are also given. The M/L
relation is similar for galaxies with t$<$7, with no clear trend
neither with the morphological type, nor with the interaction class.

\section{The Tully-Fisher relation}

As shown by Courteau (1997), the best kinematical tracer is V$_{2.2}$,
which is the velocity attained at R$_{2.2} \approx$ 2.2 $r_d$
($r_d$ = disk scale length).  Instead of $r_d$ we have used
1.3$\times$ R$_{eff}$, equivalent to 0.65$\times$R$_{25}$, that results to be a
good approximation for spiral galaxies and does not require a
bulge-disk decomposition (Courteau 1997).
Since the rotation curves derived from optical emission lines do not
usually reach R$_{2.2}$, we fitted a model curve to the observed
rotation curve in the inner part of the galaxy, and extrapolate it to
the outermost regions. We have followed that procedure using models as
simple as possible. We have used the normalized arctangent rotation
curve fitting function given by $$v(R) = v_0 + 2/\Pi \times v_c \times
arctan(R)$$ with $R=(r-r_0)/r_t$, $v_0$ is the velocity center of
rotation, $r_0$ is the spatial center of the galaxy, $v_c$ is an
asymptotic velocity and $r_t$ is a radius that corresponds to the
transition region between the rising and the flat parts of the
rotation curve. This function has been shown to reproduce adequately
the shape of the rotation curves with the smallest number of arguments
(Courteau 1997) and emerges naturally from the standard
parameterization of the density profiles of dark halos (Gilmore et
al. 1990). As noted by Courteau (1997) the resulting parameters cannot
be used to describe rotation amplitudes and scales, but since it
allows to trace smoother curves and to somewhat extrapolate to
0.65$\times$R$_{25}$, it will be useful for our purposes. The model
curves are shown in Figs. \ref{vel} and \ref{jesus}.  According to this model,
we have calculated the residuals, what allows us to obtain a
quantitative estimation of the the goodness of the model: the
rotation curves are classed as regular when the differences between
the observed and the arctangent velocity distributions are smaller
then 10\%; if these differences are higher than 10\% but the shape of
the curve still follows that of the model, the curve is classified as
distorted; the rest are considered as peculiar. The curves may be also
symmetric or asymmetric in the extension of the two
branches. Therefore, the final classification is as following: 1:
regular and symmetric; 2: regular and asymmetric; 3: distorted
symmetric; 4: distorted asymmetric; 5: peculiar.

\begin{figure}
\psfig{file=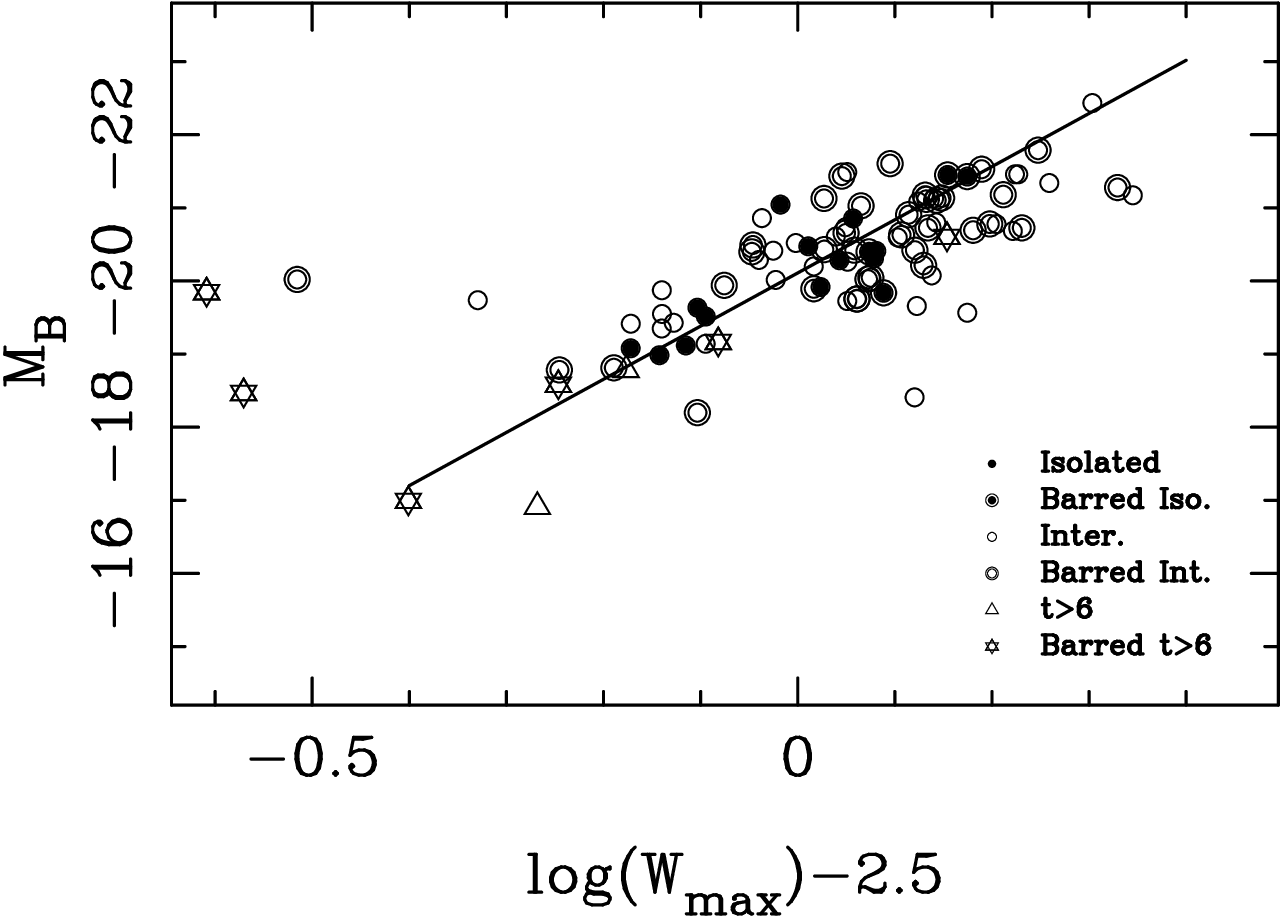,width=6truecm,angle=-90} \caption{TF relation for our sample
galaxies. The solid line represents the TF line  by Tully \& Pierce (2001).}
\label{tf}
\end{figure}

The resulting TF relation is shown in Fig. \ref{tf}, where the
different interaction classes have been marked with different
symbols. It can be seen that isolated galaxies trace the TF relation
with the smallest scatter. 
Moreover, the three outliers belong to the group of interacting pairs. We have
quantified the residuals with respect to the TF relation as given by Tully \& Pierce
(2000). We find 0.32 for isolated galaxies (INT = 1) and 0.65 for
interacting spirals (INT = 3).
We notice that the three outliers (NGC~2799, NGC~3395 and NGC~3396)
show peculiar rotation curves (only NGC~2799 is not later than t = 6,
but is a secondary member), and that their position is much
closer to the TF line when HI velocity amplitudes are considered.
The same analysis has been applied to isolated and interacting
objects.  Therefore, even if the results shown here are only
indicative due to small number statistics, we point out that a
possibility seems to exist for reducing the scatter of the TF relation
determined with optical data when using the most isolated objects.

We have also explored the eventual dependence of the departures from
the TF line as a function of the degree of peculiarity of the rotation
curves.\footnote{Notice that systematic effects in the third parameter 
analysis cannot not be addressed mainly due to the lack of accurate enough 
photometric information in our sample galaxies (see
Kannappan et al. 2002).}
The dispersion from the canonic TF line is 0.30 for galaxies with
regular rotation curves, 0.34 for galaxies with distorted rotation
curves and 0.36 for peculiar rotation curves.  This appears to be at
variance with the results reported by Barton et al. (2001) who found
that the presence of strong kinematic distortions is a significant
predictor of TF residuals.  The difference could be due to the fact
that, as already pointed out, our sample includes only isolated and
mildly interacting objects, that do not present strong
distortions. Our data are in agreement with Dale et al. (2001) who
find no differences between cluster and field spirals in the degree of
asymmetry.

\section{Nuclear and Disk spectral characteristics}

The distribution of the H$\alpha$ line emission along the slit was
traced for each 2D spectrum. The local peaks in the distribution were
taken as the centers of HII regions, that were defined to comprise all
the spatial sections within the FWHM around that peak. The resulting
1D spectra were measured as explained before, producing the data
collected in Table \ref{parhii}. 
In the following the region comprising the center of
the galaxy is referred as the Nuclear Region. Typical sizes for
the Nuclear Regions range between 0.5 and 2 kpc. The spectra are not
flux-calibrated. However, since the lines used in the analysis are
very close in wavelength, the count ratios are a good measure of the
flux ratios.

\begin{table*}[h]
\caption[]{Parameters of the HII regions}
\label{parhii}

\begin{flushleft}
\begin{scriptsize}
\begin{tabular}{l| r r r r r r r r r}
\hline
\noalign{\smallskip}
\multicolumn {1}{l|}{Galaxy}
& \multicolumn{1}{c}{R}
& \multicolumn{1}{c}{EW(H$\alpha$)}
& \multicolumn{1}{c}{[OI]/H$\alpha$}
& \multicolumn{1}{c}{err([OI]/H$\alpha$)}
& \multicolumn{1}{c}{[NII]/H$\alpha$}
& \multicolumn{1}{c}{err([NII]/H$\alpha$)}
& \multicolumn{1}{c}{[SII]/H$\alpha$}
& \multicolumn{1}{c}{err([SII]/H$\alpha$)}
& \multicolumn{1}{c}{S1/S2}\\
 & ('') & (\AA) & & & & & & &\\
\noalign{\smallskip}
\hline\noalign{\smallskip}
\input{tabla_lin1.tex}
\noalign{\smallskip}
\hline
\end{tabular}
\end{scriptsize}
\end{flushleft}
\end{table*}

\begin{table*}[h]
\addtocounter{table}{-1}
\caption[]{Parameters of the HII regions (Cont.)}

\begin{flushleft}
\begin{scriptsize}
\begin{tabular}{l| r r r r r r r r r}
\hline
\noalign{\smallskip}
\multicolumn {1}{l|}{Galaxy}
& \multicolumn{1}{c}{R}
& \multicolumn{1}{c}{EW(H$\alpha$)}
& \multicolumn{1}{c}{[OI]/H$\alpha$}
& \multicolumn{1}{c}{err([OI]/H$\alpha$)}
& \multicolumn{1}{c}{[NII]/H$\alpha$}
& \multicolumn{1}{c}{err([NII]/H$\alpha$)}
& \multicolumn{1}{c}{[SII]/H$\alpha$}
& \multicolumn{1}{c}{err([SII]/H$\alpha$)}
& \multicolumn{1}{c}{S1/S2}\\
 & ('') & (\AA) & & & & & & &\\
\noalign{\smallskip}
\hline\noalign{\smallskip}
\input{tabla_lin2.tex}
\noalign{\smallskip}
\hline
\end{tabular}
\end{scriptsize}
\end{flushleft}
\end{table*}

\begin{table*}[h]
\addtocounter{table}{-1}
\caption[]{Parameters of the HII regions (Cont.)}

\begin{flushleft}
\begin{scriptsize}
\begin{tabular}{l| r r r r r r r r r}
\hline
\noalign{\smallskip}
\multicolumn {1}{l|}{Galaxy}
& \multicolumn{1}{c}{R}
& \multicolumn{1}{c}{EW(H$\alpha$)}
& \multicolumn{1}{c}{[OI]/H$\alpha$}
& \multicolumn{1}{c}{err([OI]/H$\alpha$)}
& \multicolumn{1}{c}{[NII]/H$\alpha$}
& \multicolumn{1}{c}{err([NII]/H$\alpha$)}
& \multicolumn{1}{c}{[SII]/H$\alpha$}
& \multicolumn{1}{c}{err([SII]/H$\alpha$)}
& \multicolumn{1}{c}{S1/S2}\\
 & ('') & (\AA) & & & & & & &\\
\noalign{\smallskip}
\hline\noalign{\smallskip}
\input{tabla_lin3.tex}
\noalign{\smallskip}
\hline
\end{tabular}
\end{scriptsize}
\end{flushleft}
\end{table*}

\begin{table*}[h]
\addtocounter{table}{-1}
\caption[]{Parameters of the HII regions (Cont.)}

\begin{flushleft}
\begin{scriptsize}
\begin{tabular}{l| r r r r r r r r r}
\hline
\noalign{\smallskip}
\multicolumn {1}{l|}{Galaxy}
& \multicolumn{1}{c}{R}
& \multicolumn{1}{c}{EW(H$\alpha$)}
& \multicolumn{1}{c}{[OI]/H$\alpha$}
& \multicolumn{1}{c}{err([OI]/H$\alpha$)}
& \multicolumn{1}{c}{[NII]/H$\alpha$}
& \multicolumn{1}{c}{err([NII]/H$\alpha$)}
& \multicolumn{1}{c}{[SII]/H$\alpha$}
& \multicolumn{1}{c}{err([SII]/H$\alpha$)}
& \multicolumn{1}{c}{S1/S2}\\
 & ('') & (\AA) & & & & & & &\\
\noalign{\smallskip}
\hline\noalign{\smallskip}
\input{tabla_lin4.tex}
\noalign{\smallskip}
\hline
\end{tabular}
\end{scriptsize}
\end{flushleft}
\end{table*}

\vskip 0.5truecm
Given the
spectral coverage of the data, the standard diagnostic tools to
classify the spectra (Baldwin et al. 1981, Veilleux \&
Osterbrock 1987) cannot be used.
But, as shown by the early work by Keel (1984),
the [NII]/H$\alpha$ line ratio can be used as a rough, first order
estimator to classify the spectral nuclear types (SNT) into Active
Galactic Nuclei (regardless of the kind of activity, since into the
AGN category we have included both Seyfert and LINER-like nuclei) and
HII like objects. Accumulated evidence has shown
that that line ratio is very sensitive to the presence of any kind of
activity, thus allowing for an easy spectral classification of the
nuclei. The existence of absorption under the Balmer line could
however induce the misclassification of some objects, and special care
has to be taken.

Apart from a spectroscopic classification of the nuclear spectra, we
have also attempted to give an estimation of the metallicity of the
disk and its possible gradient,
taking the [NII]/H$\alpha$ line ratio as an estimator.  For low
metallicity objects both, Nitrogen and Oxygen are of primary origin
and their abundances correlate well (Masegosa et al. 1994). For higher
metallicities, a fraction
of the measured Nitrogen is of secondary origin, what modifies the
previous relation, even if it is still monotonic and, therefore,
useful to probe Z.
van Zee et al. (1998) have found that, for 12 + log (O/H) $<$ 9.1, a
relation does exist of the form 12 + log (O/H) = 1.02 log
([NII]/H$\alpha$) $+$ 9.36. The use of the [NII]/H$\alpha$ line ratio
to estimate Z has the advantage of being insensitive to reddening.
But, as Stasinska \& Sodr\'e (2001) have pointed out, that
calibration relation is adequate only for HII regions, and important
errors could be produced when the integrated spectra of spiral
galaxies, or a complex ISM with shocked gas is being analyzed. Based
on the [NII]/H$\alpha$ ratio Denicol\'o et al. (2002) obtained an
improved calibration on the oxygen abundances. They clearly showed the
power of this estimator when analyzing large survey data to rank their
metallicities,
even if the uncertainties on individual objects can reach up to
0.6 dex, mainly due to O/N abundance ratio and ionization degree
variations. Here we only consider it to study global trends of Z from
the spectra of HII regions or HII-like nuclei in the collected sample
of spiral galaxies.

\subsection{Nuclear spectra}

Nuclear H$\alpha$ emission has been detected in 91 out of 98 
galaxies in the sample. For the other 13, the stored data were corrupted.
The nuclear spectra are presented in
Fig. \ref{nucleo}.
Three of the galaxies without
H$\alpha$ emission (namely, NGC~3976, NGC~5641, and NGC~2424) show
[NII] emission, suggesting that shock ionization would be important in
these nuclei. They
were classified as LINERs in previous studies (Carrillo et
al. 1999), and they belong to the class of interacting
systems. For the remaining 4 galaxies (NGC~2344, NGC~2545, NGC~3835
and NGC~5147) only very faint or even absent emission has been
detected partly due to the poor S/N ratio of the spectra.
In any case the emission cannot be strong.
They do not show any other peculiarity and can be considered as normal spiral
galaxies (Jansen et al. 2000).

Given the purpose of
the work and the rather low S/N ratio in many of the spectra
we have not applied any correction for absorption. To cope with the
problem of the presence of absorption under H$\alpha$, we visually
inspected all the spectra, identifying the cases where it was
conspicuous. All those galaxies were classified as SNT = 3, i. e.,
nuclei in which the Balmer absorption is so strong that the measured
[NII]/H$\alpha$ ratio is not reliable to classify it. For the
remaining nuclei, without any appreciable absorption under H$\alpha$,
those with spectral characteristics of HII regions were classified as
SNT = 1, and those with line ratios similar to active galaxies as SNT
= 2. Indeed this is a rather crude classification but, as we will see
later, some
conclusions on the nuclei of spiral galaxies and their relation to some global
properties can be drawn.

\begin{figure}
\psfig{file=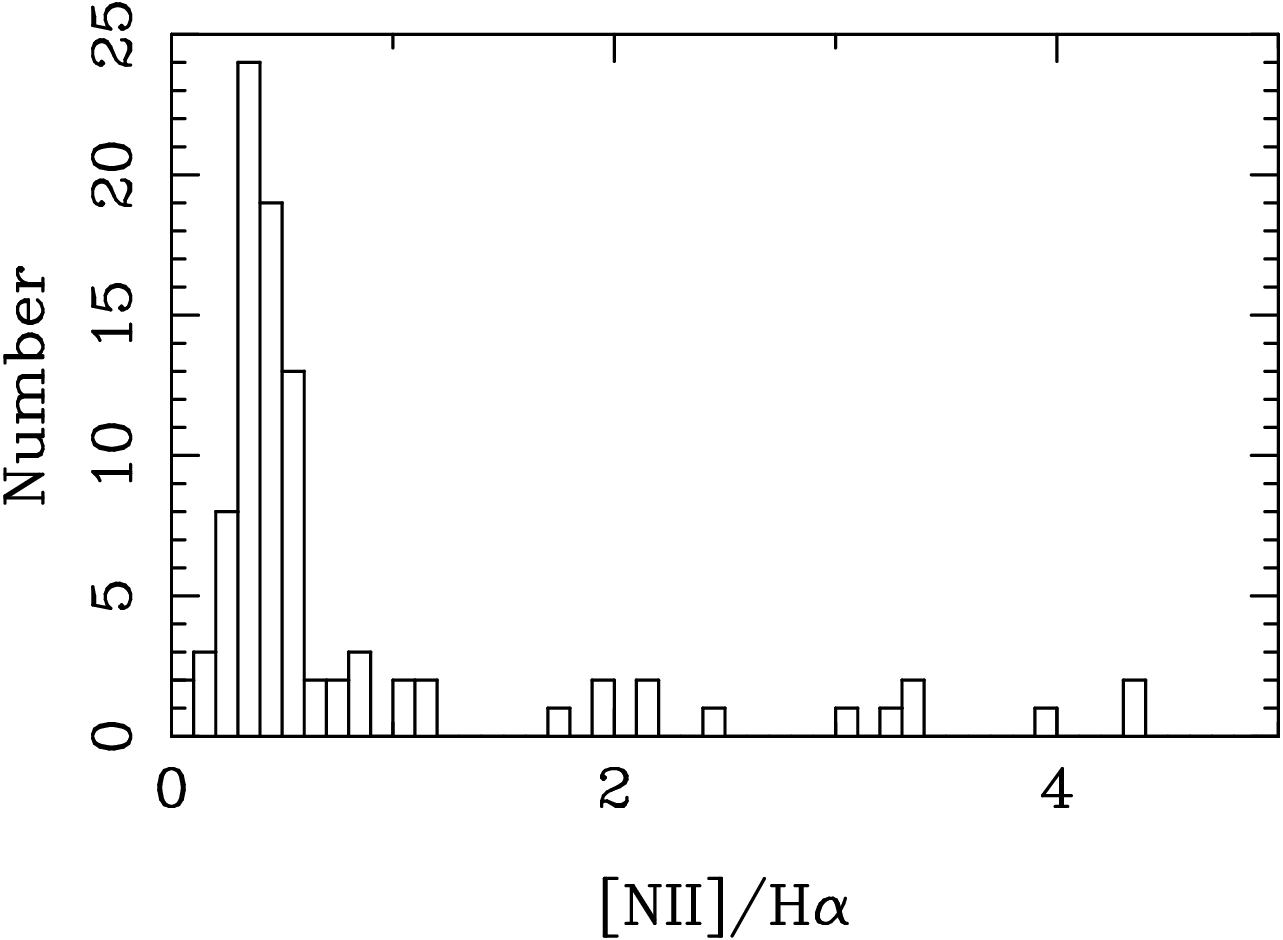,width=6truecm,angle=-90}
\caption{Distribution of nuclear [NII]/H$\alpha$ for the whole sample.
NGC 7217 has not been included due to the large measured ratio (see text).}
\label{histnii-all}
\end{figure}

The distribution of the [NII]/H$\alpha$ nuclear values shows that for
most of the galaxies it is lower than 1 (see
Fig. \ref{histnii-all}). The data are presented in
Table \ref{parhii}.  We notice that all
the galaxies with SNT = 3 have EW(H$\alpha$)$\leq$ 10, what produces
an artificially high ratio if no correction is applied to cope with
the underlying absorption and are consequently excluded hereafter from the
discussion.

%
%

Judged from the [NII]/H$\alpha$ ratio, we find 11 AGN candidate
objects (about 10\%) in our sample.
Seven of them were already observed by Ho et al. (1997), who
classified them as 6 LINERs and 1 Seyfert. For the remaining 4
galaxies we find that 2 of them are Seyfert 1 based on the
width of the H$\alpha$ line.
For the other two the information we have is not enough to classify
them as Seyfert 2 or LINER.
The largest line ratio is found for N7217,
with
[NII]/H$\alpha$ = 8.6.
It is
a known LINER (Filippenko \& Sargent 1985) frequently quoted to illustrate the
signature of strong shocks (see the models by Dopita \& Sutherland 1995).

\begin{figure}
\psfig{file=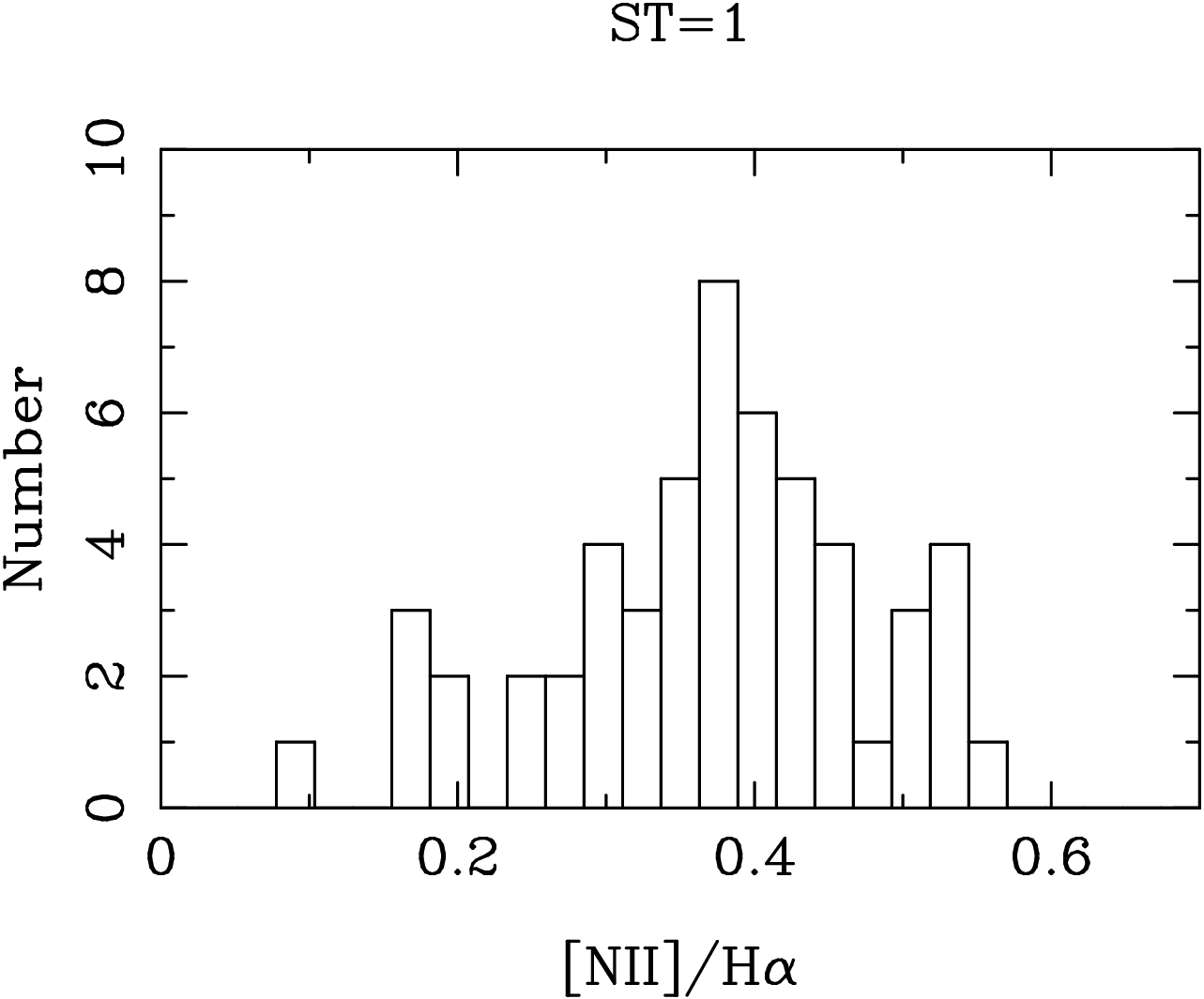,width=7truecm,angle=-90}
\caption{Distribution of nuclear [NII]/H$\alpha$ for the galaxies with
nuclear  spectral type SNT = 1.}
\label{histnii-ST1}
\end{figure}

Regarding the HII-like nuclei, we find that they define a rather
narrow distribution of the [NII]/H$\alpha$ ratio
(Fig. \ref{histnii-ST1}).  Only two galaxies (not shown in the figure)
depart from the general trend, N5172 with [NII]/H$\alpha$ = 1.92, and
N5678 with [NII]/H$\alpha$ = 1.16.  For N5172, our data are of very
poor S/N ratio and the [NII]/H$\alpha$ ratio we obtained is very
uncertain. And N5678 is a composite LINER/HII galaxy after Filho et
al. (2000).  Excluding those two objects, the remaining galaxies
present a range of values of the line ratio, corresponding to values
typical of irregular galaxies and disk HII regions (see Vila-Costas \&
Edmunds 1993 and McCall et al. 1985).  For all the data classified as
SNT = 1, (with the quoted exceptions), included the latest spirals (t
$>$ 6), the median value is 0.38, with a dispersion of 0.07.  
Excluding late type spirals it amounts to 0.39 with the same
dispersion.

\begin{figure}
\psfig{file=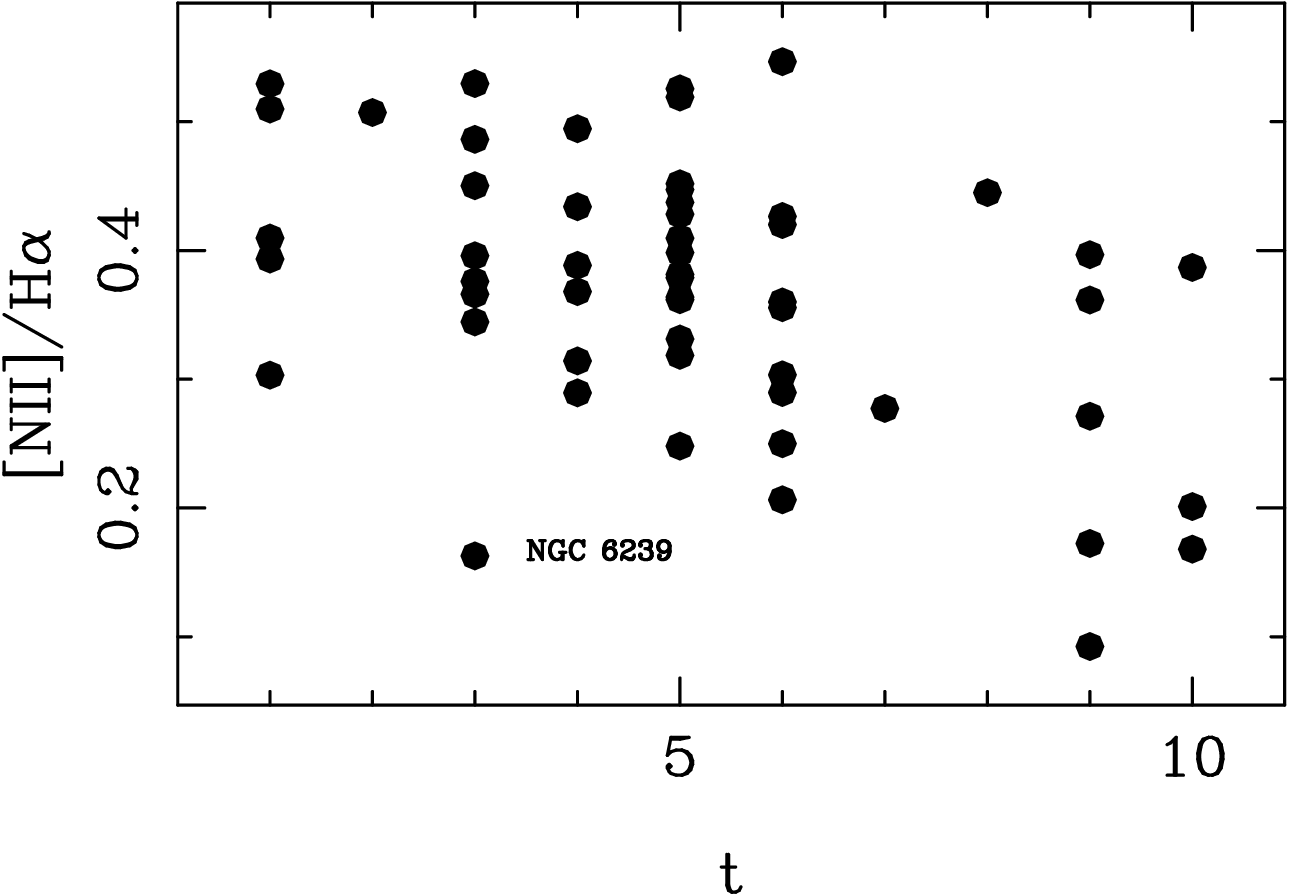,width=6truecm,angle=-90}
\caption{[NII]/H$\alpha$ ratios as a function of the morphological type.}
\label{niivst}
\end{figure}

In spite of the rough character of the estimator we use here, some
correlations are already hinted.  There is a relation between the
metallicity of the HII-like nuclei and the morphological type, the
early type spirals having more metallic nuclei than the later spirals
(Fig.  \ref{niivst}, and Table~\ref{metalicidad}). This result is
consistent with the suggestion by Oey \& Kennicutt (1993) that early
type spirals are more metal rich than later types.  That results rests
on the difference between Sa/Sab galaxies in one side, and the later
types in the other, since we don't find any difference between Sb/Sbc
and Sc/Scd objects. Indeed, the later than Scd types are still of
lower Z and look as a different family.  The results reported here are
in agreement with the work by Zaritsky et al. (1994) based on a
completely different data set, in the sense that a tendency does exist
for the metallicity to decrease when moving along the Hubble
sequence. We have investigated if such a trend could be due to a
systematically stronger H$\alpha$ absorption in early types.  The
absence of any appreciable trend between the H$\alpha$ EW and the
morphological type t argues against that explanation, and leaves the
relation as genuine.

Finally, the analysis of a possible Z enhancement produced by the
presence of instabilities like bars or by the interaction with nearby
neighbors has produced negative results. The range and median value of
Z does not seem to be altered when those aspects are taken into
account.

\begin{table}[h]
\caption[]{[NII]/H$\alpha$ ratios for the three spectral nuclear types, and for the
different morphological types.}
\label{metalicidad}
\begin{flushleft}
\begin{scriptsize}
\begin{tabular}{l  c }
\hline
\noalign{\smallskip}
\multicolumn {1}{l}{}
& \multicolumn {1}{c}{[NII]/H$\alpha$} \\
\noalign{\smallskip}
\hline\noalign{\smallskip}
SNT=1 & 0.38\\
SNT=2 & 0.98\\
SNT=3 & 0.76\\
t=1,2 & 0.46\\
t=3,4 & 0.38\\
t=5,6 & 0.38\\
t$>$6 & 0.27\\
\hline
\end{tabular}
\end{scriptsize}
\end{flushleft}
\end{table}

\begin{figure}
\psfig{file=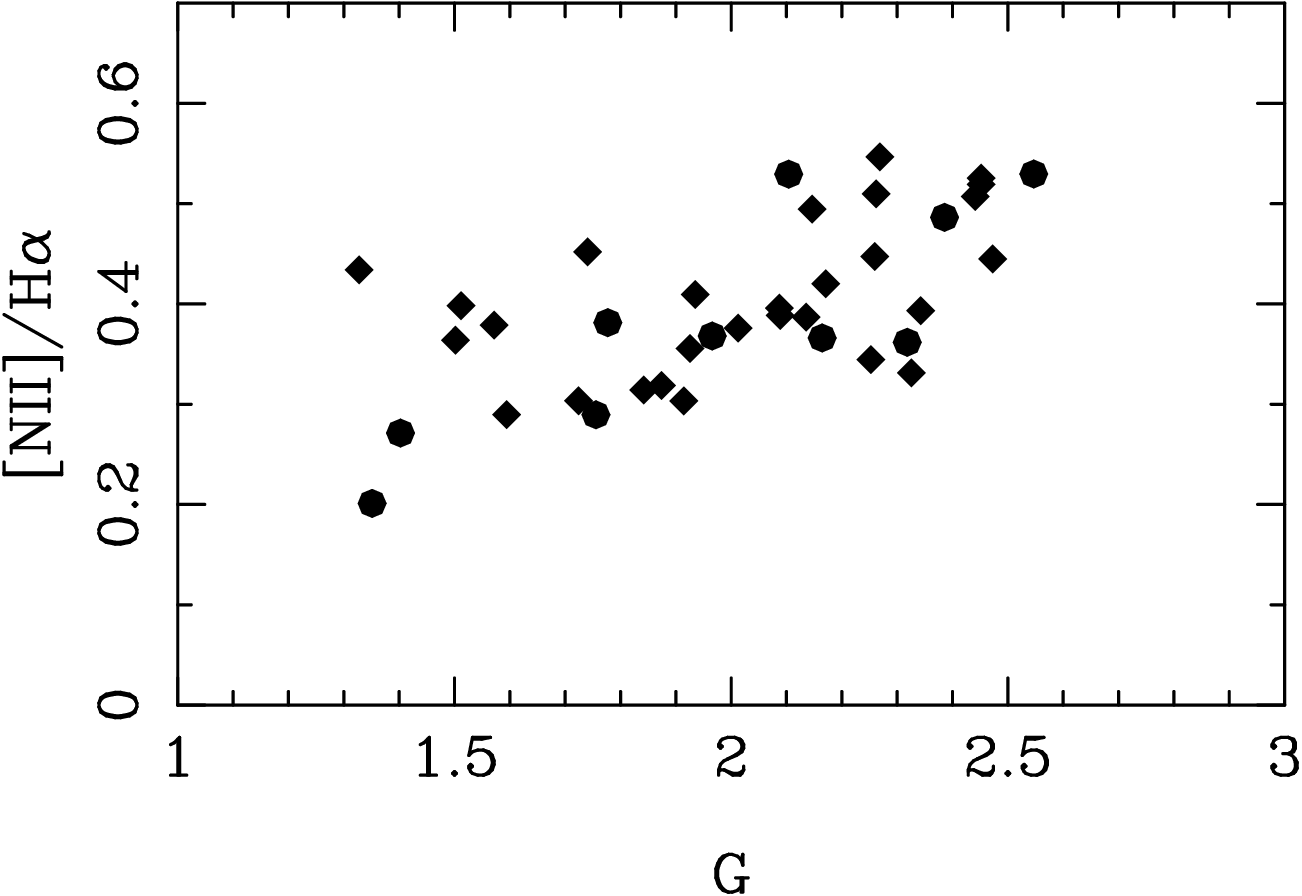,width=6truecm,angle=-90}
\caption{Nuclear [NII]/H$\alpha$ ratios as a function of the gradient of the solid-body
region of the rotation curve, G. (G is in logarithmic scale). Interacting galaxies are marked with solid circles.}
\label{niivsG}
\end{figure}

We have shown in Paper II that the relations are better defined if the type 
is replaced by a more quantitative
parameter such as B/D ratio, or the inner gradient G, with which it
is tightly correlated.
Here too,
we find a good correlatin between Z and the gradient G (see Fig. \ref{niivsG}),
statistically significant at 99.99\% confidence level (R = 0.63, for 38 objects).
This add to the previous findings about the quality of the parameter G to
characterize the global properties of spiral galaxies.

The other two global properties which appear to be related to the
metallicity are the absolute magnitude, M$_B$ and the maximum rotation
velocity.  We find that Z increases with both, the central velocity
and the luminosity, i.e, massive galaxies are more metal rich. This
agrees with the results reported by Zaritsky et al. (1994) and by
Dutil \& Roy (1999). Again no difference is found between isolated and
interacting galaxies. These results would suggest that the
instabilities produced by gravitational interaction, even if they can
drive gas to the center (Barnes \& Hernquist 1991; Mihos et al.), do
not have major effects in the central region for mild interaction as
the ones reported in this work.

\subsection{Extranuclear HII regions}

\begin{figure}
\psfig{file=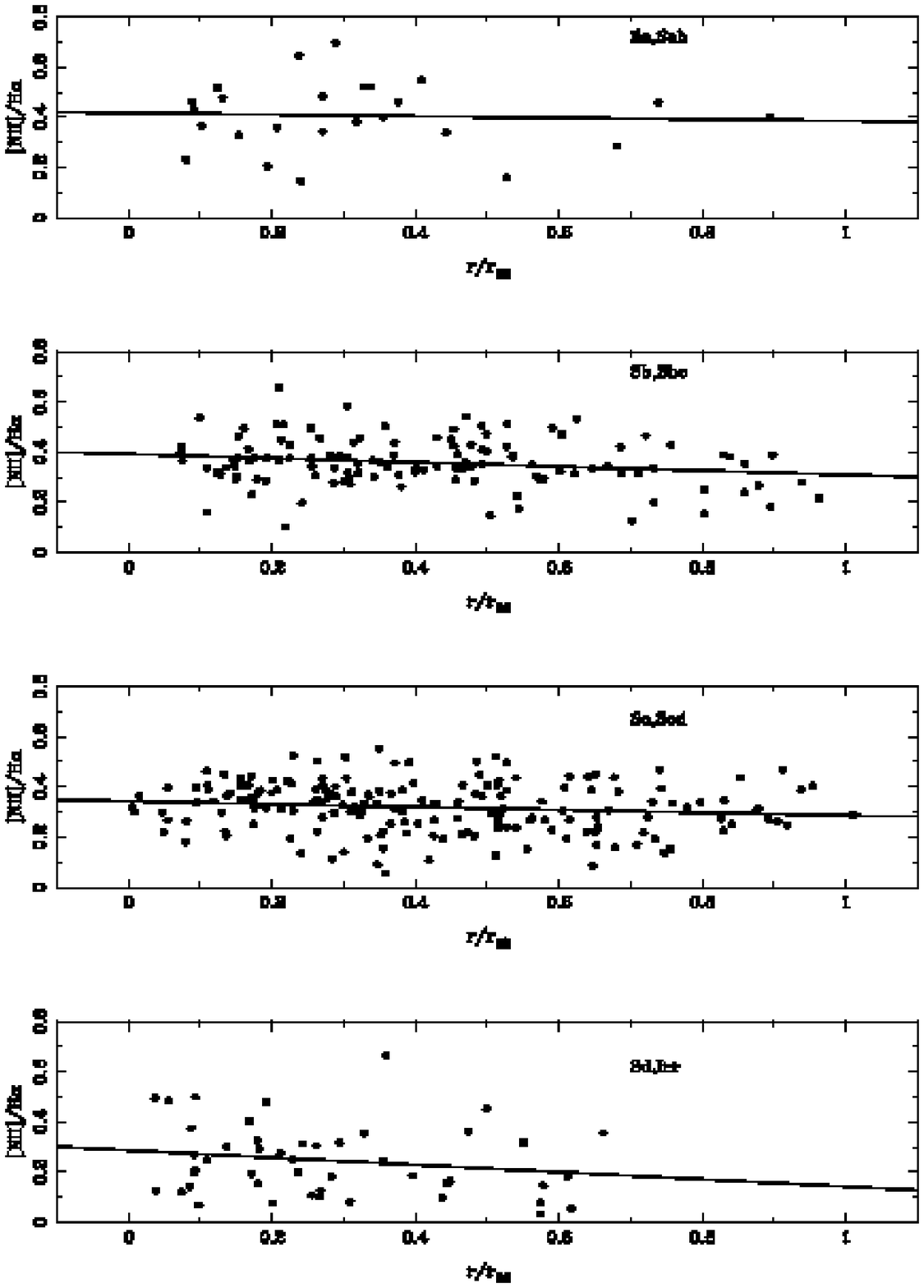,height=12truecm}
\caption{Disk [NII]/H$\alpha$ ratios as a function of the distance to the center
(normalized to R$_{25}$).}
\label{grad-tipo-mod}
\end{figure}

The extranuclear HII regions detected in all our 2D spectra were
measured and used to characterize the metallicity of the disks.  As
discussed before, our data does not allow to conclude on any
individual galaxy, but can be used to look for general trends when the
population of the disks is considered as a whole. To be able to
combine data from different galaxies, we have normalized to R$_{25}$
(as given in the RC3) the galactocentric distances of the HII regions.
In principle the choice of one or another radius to normalize could
have some effect on the results about Z gradients (Zaritsky et al.
1994). Since we are interested on general trends and not in a real
quantification of the gradient we consider that the choice of the
normalization is of minor importance and decided to use the isophotal
radius, accessible for most of the galaxies studied. From the HII
regions measured we have only selected all the data with H$\alpha$
equivalent width larger than 10~\AA.  In that way we select the better
S/N data, and avoid including regions with important Balmer
absorption, that could induce inconsistencies in the estimation of the
metallicity. The number of regions we consider here is 392, in 98
galaxies.  In the following we report the results obtained when the
general metallicity trends in the disk are analyzed in relation to the
morphology of the galaxy, the Nuclear Spectral Type and the effects
produced by interaction.  Different authors have claimed (see Vila
Costas \& Edmunds 1992 for a review and Zaritsky et al. 1994) that a Z
radial gradient does exist in disk like galaxies. In
Fig. \ref{grad-tipo-mod} the metallicity estimator [NII]/H$\alpha$ is
presented for the different morphological types. A slight tendency
seems to be present for the gradient to be steeper in later types,
whereas it is about zero for Sa and Sb spirals. This agrees with the
claim by Oey \& Kennicutt (1993) of a larger global metallicity and
almost flat gradients in early type spirals.

We have calculated from the disk data the expected central
[NII]/H$\alpha$ values, using the formal fitting to the data. The
values we find in that way ranges from 0.41 for the earlier types, to
0.28 for the latest types. They compare very well with what we have
found just measuring the line ratio of the nuclear regions, 0.44 for
the early types, and 0.27 for the latest types. That consistency adds
confidence to the reality of the trends we have found, and to the way
of estimating the metallicity form the line ratio.

\begin{figure}
\psfig{file=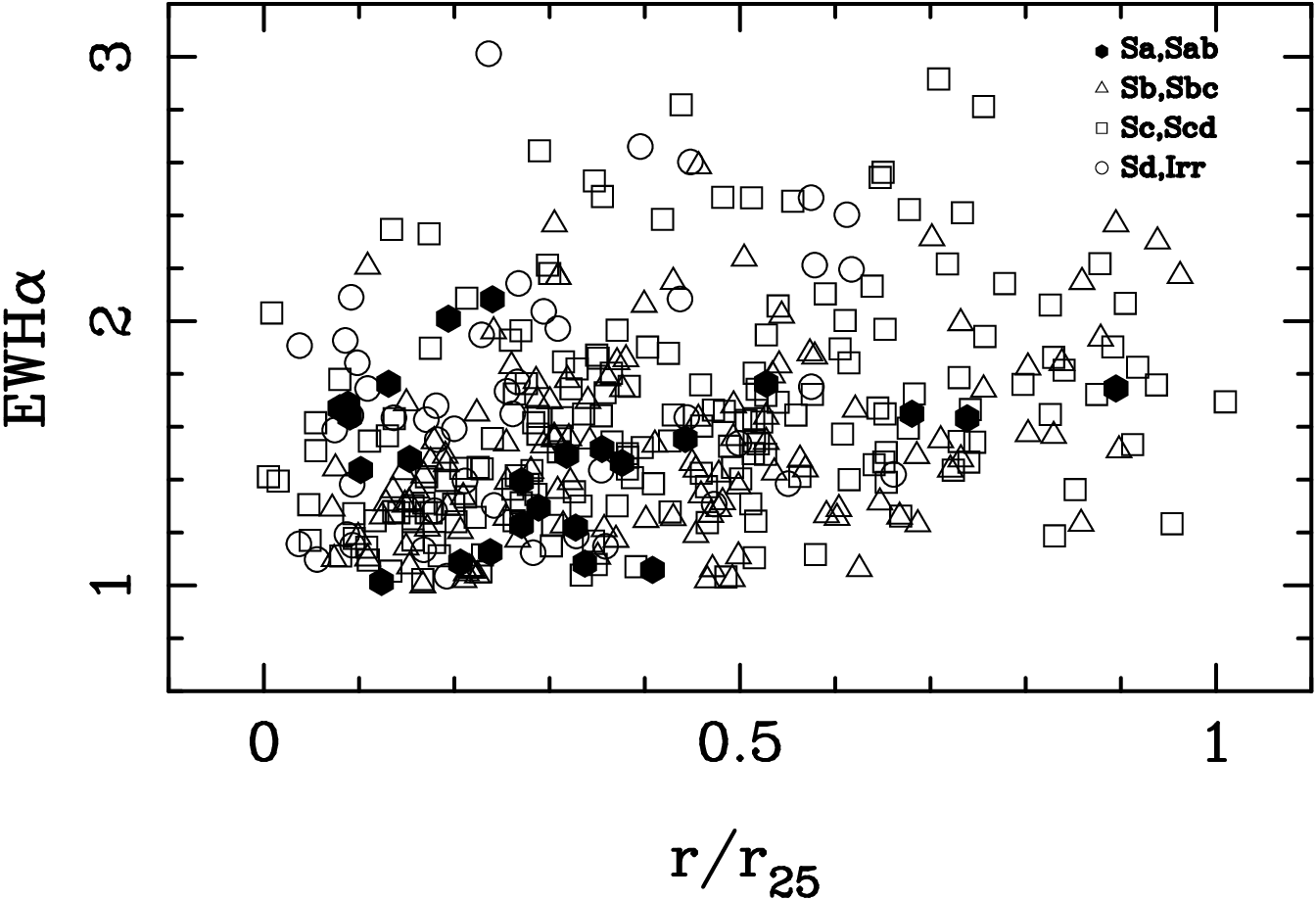,width=6truecm,angle=-90}
\caption{Disk equivalent widths of H$\alpha$ (in \AA) as a function of the morphological
type, t. EWs are plotted in logarithmic scale.}
\label{ew-ext-tipo}
\end{figure}

The H$\alpha$ Balmer line has been extensively used to measure the
ratio of the current to the average past Star Formation Rate in
Galaxies (see Kennicutt 1983, Kennicutt et al. 1994, and Stasinska \&
Sodr\'e 2001). Kennicutt (1994) found smooth progression in the Star
Formation History with the Hubble type, with a ratio of current to
past SFR increasing from 0.01-0.1 for Sa type to 0.5-2 for a typical
Sc disk. The data reported here are only barely consistent with such
claim. In Fig. \ref{ew-ext-tipo}, where the H$\alpha$ equivalent width
is plotted versus the radial distance, the only effect is a larger
dispersion on later types than in earlier spirals towards larger EW in
the later types but a clear separation between different morphologies
is not obvious. It has to be noticed that Kennicutt data are referred
to the integrated EW whereas here we are trying to get the trend based
on the distribution of HII regions crossed by the slit through the
disk of the galaxies.  We cannot extract a definitive conclusion from
our data and, therefore, we cannot say that our data are in
contradiction with Kennicutt's study, even if such a conclusion is
hinted by our results.

\begin{figure}
\psfig{file=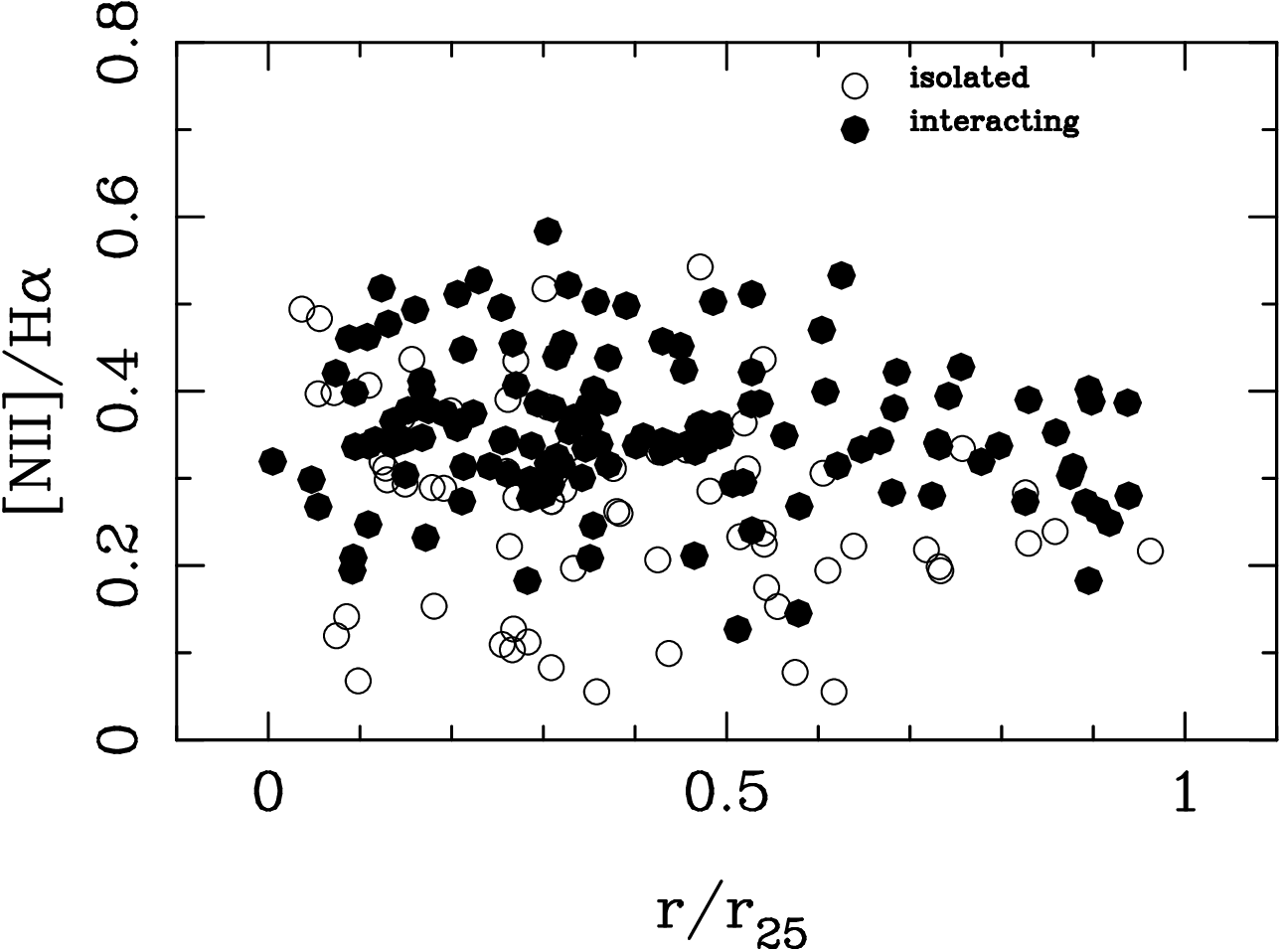,width=6truecm,angle=-90}
\caption{Metallicity estimator for the HII regions in the disks of isolated and interacting spirals.}
\label{grad-inter}
\end{figure}

Let's now compare isolated and interacting systems. Regarding the
metallicity, it appears that interacting galaxies tend to show a
larger [NII]/H$\alpha$ ratio in all the mapped regions (see
Fig. \ref{grad-inter}).  The median value of [NII]/H$\alpha$ for the disk of
isolated normal galaxies amount to 0.27, in contrast with a median
value of 0.35 for the interacting systems. 

It is usually accepted that the interaction process drives gas to the
central regions producing an enhancement in the star formation
events. Consequently, a larger Z in the bulges of those galaxies would
be predicted, and it's found in our data and other studies. The point
here is that we have also found a higher Z along the disks of the same
galaxies. It seems then that the interaction affects the whole galaxy,
producing star formation in all the disk, depending on the conditions
(see for example, M\'arquez \& Moles 1994).

However, no difference in the H$\alpha$ EW has been found between
interacting and isolated galaxies, as if the global star formation
rate now was essentially the same in both families. To understand this
result we have to take into account that only mildly interacting
systems are included in our sample, for which the effects of the
interaction are expected to be much less important than in stronger
interactions. In that sense, we notice that our results are compatible
with those found by Kennicutt et al. (1987) for a large fraction of
galaxies in their complete pairs sample.  Combining both results,
higher Z and normal present star formation rate, it seems that the
enrichment is only produced as a secular, accumulative effect along
the galaxy life, without marked episodes, in those mildly interacting
systems. This result is consistent with those by Bergvall et
al. (2001), who find reddest disks in interacting galaxies.  Nevertheless, 
we have
already noticed that the morphological types of interacting galaxies
tend to be earlier than for isolated ones, so the reported higher 
metallicities could be reflecting the difference in metallicities between 
early and late type galaxies. Larger samples
of isolated galaxies would be needed to further analyze the
metallicities of early types spirals as compared to those of
interacting spirals with the same morphologies. 

The situation for AGNs is somewhat similar, since active spirals are
known to mainly reside in early types spirals (see for instance Moles
et al. 1995). Given the limited sample we are considering (11 AGNs, 7
of them belonging to interacting systems) eventual differences in
metallicity cannot be addressed.

\section{Summary and conclusions}

We have obtained long slit spectra along the major axes of a sample of
spiral galaxies selected to be either isolated or in isolated pairs,
with similar intermediate-scale environment and with a
recognizable and well defined spiral morphology. We have further
investigated their environmental status and reclassified them, what
allows us to define a sample of isolated objects, to be later compared
with mildly interacting spirals (with small satellites and/or
companions of similar size). The main results we have obtained are the
following:

\begin{itemize}
\item{We have confirmed previous results (Paper II) that isolated galaxies tend to be of later 
Hubble types and lower luminosity than the interacting galaxies}
\item{The outer parts of the rotation curves of isolated galaxies tend to be flatter than in 
interacting galaxies. They show similar relations between global parameters. The scatter of the
Tully-Fisher relation defined by  isolated galaxies appears to be significantly lower than 
that of interacting galaxies}
\item{There is a clear trend between the metallicity of the HII-like nuclei and the morphological type of the galaxy, the earlier types showing larger Z values. Extrapolation of the Z-trend in 
the disk to the central position gives consistent results with the direct measurement of the 
nuclear HII region. No trend with the interaction status was found}
\item{We report here for the first time the existence of a tight correlation between Z and 
the gradient of the inner rigid solid rotation part of the rotation curve, G}
\item{The Z-gradient of the disks depends on the type, being almost flat for early spirals, and 
increasing for later types}
\item{[NII]/H$\alpha$ ratios appear to be larger for disk HII regions 
interacting galaxies. This could be simply due to the fact that early types 
are more frequent among them. (A similar result is obtained for AGNs, but 
we cannot further test it given 
the small size of our sample). On the other hand, the H$\alpha$ EW present 
similar values in all kind of galaxies. 
At face value those results would indicate that mildly interacting galaxies 
(as those in our sample) have different histories form normal galaxies. This 
difference however has no marked episodes (bursts) of star formation, but 
only small cumulative effects that result in more metallic (and redder) disks 
and nuclei.  
}
\end{itemize}


\begin{acknowledgements}

We are very grateful to the anonymous referee, whose comments and
suggestions helped us improving the presentation.  We thank
Prof. G. Paturel who kindly made available to us prior to publication
the Galaxy Catalog we have used to determine the interaction status of
our sample galaxies. We also thank L. Cariggi for her careful reading
of the manuscript and valuable comments. I.~M\'arquez acknowledges
financial support from the Spanish Ministerio de Ciencia y
Tecnolog\'{\i}a and the IAA.  D. Jes\'us Varela acknowledges a
scholarship from the Ministerio de Ciencia y Tecnolog\'{\i}a.  This
work is financed by DGICyT grants PB93-0139, PB96-0921, PB98-0521,
PB98-0684, ESP98-1351, AYA2001-2089 and the Junta de Andaluc\'{\i}a.
This research has made use of the NASA/IPAC extragalactic database
(NED), which is operated by the Jet Propulsion Laboratory under
contract with the National Aeronautics and Space Administration.

\end{acknowledgements}

\begin{figure*}[htp]
\psfig{figure=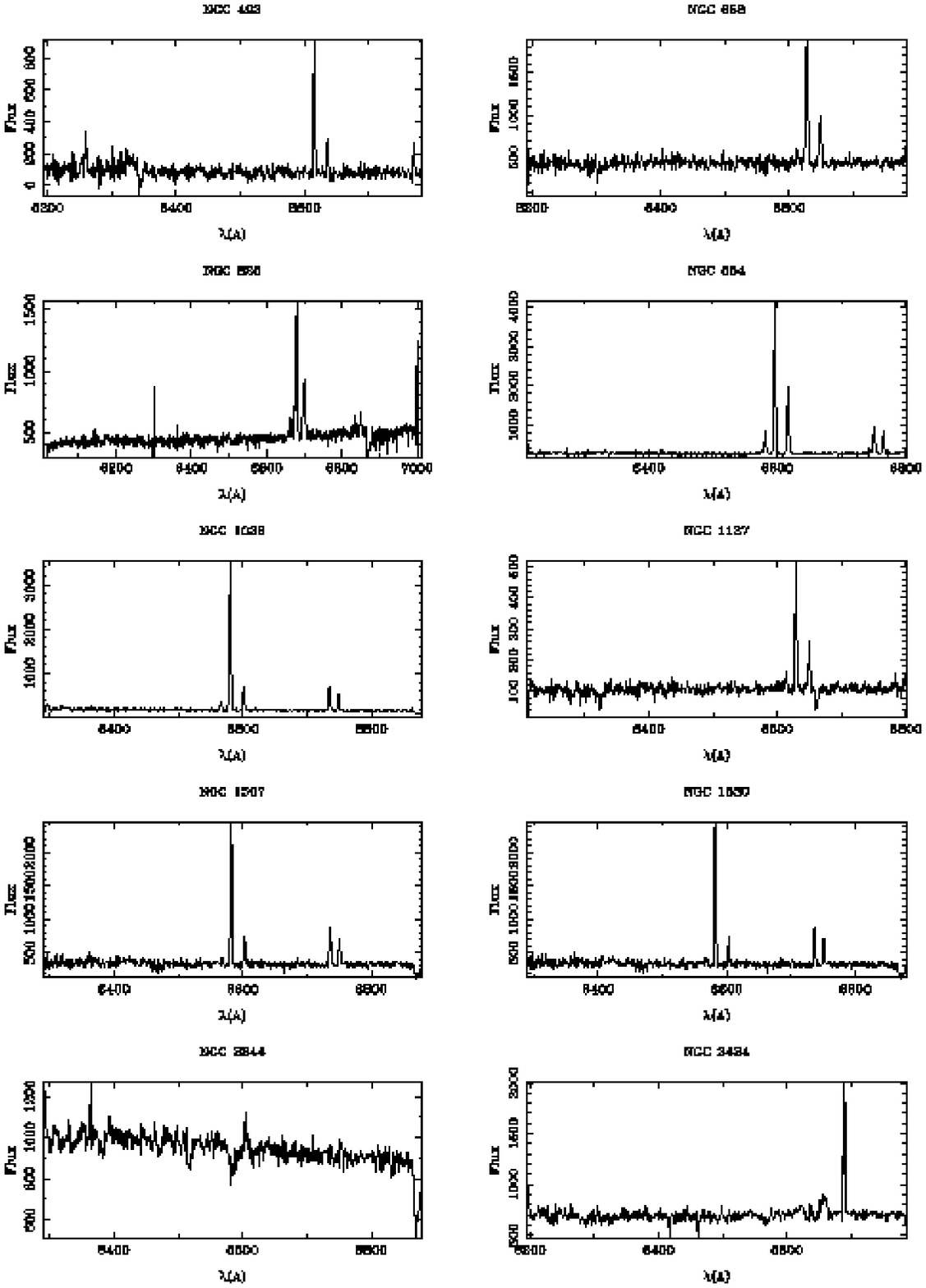,width=17truecm}
\caption[nucleo1]{Nuclear spectra for the whole sample.}
\label{nucleo}
\end{figure*}

\begin{figure*}[htp]
\psfig{figure=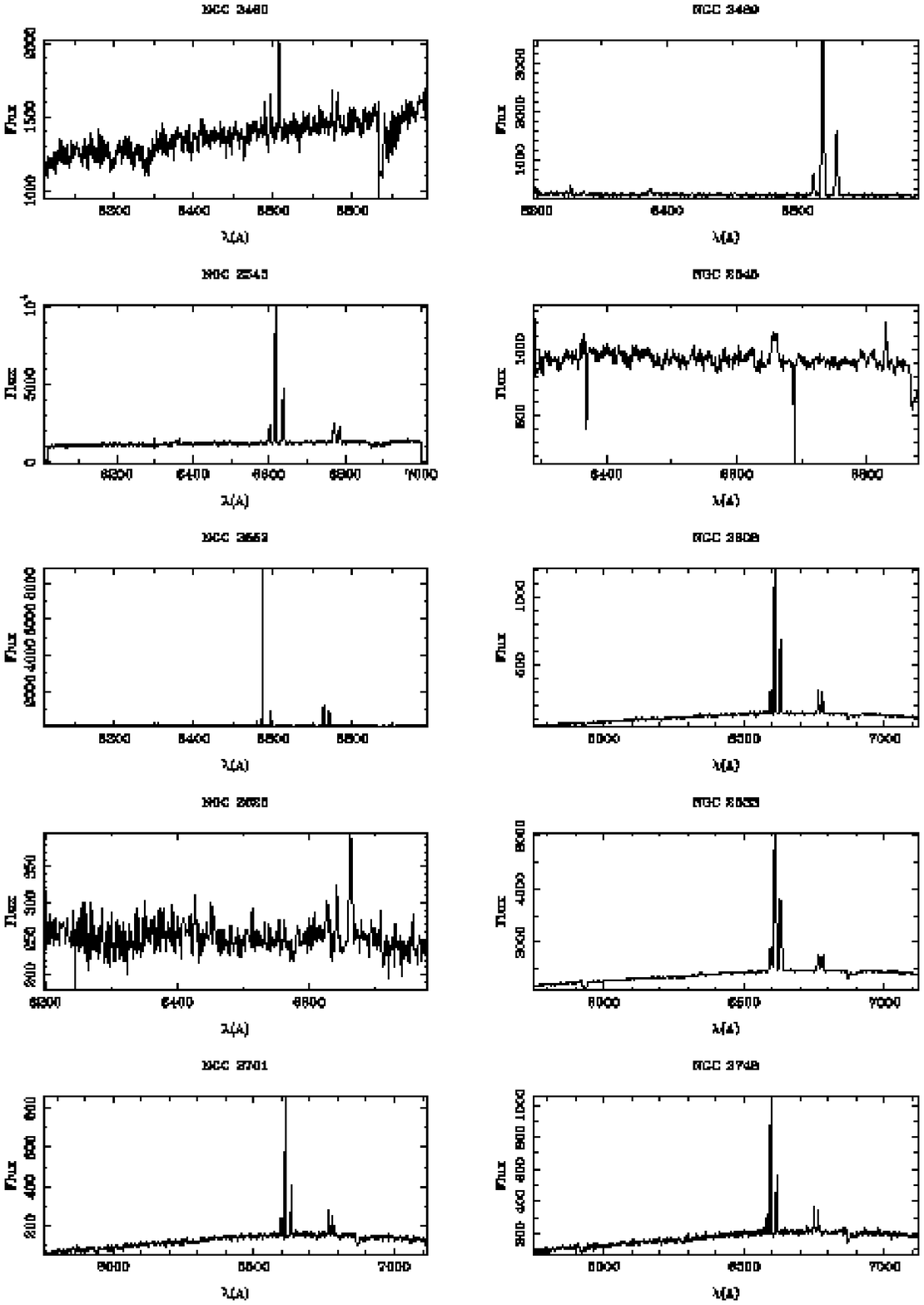,width=17truecm}
\addtocounter{figure}{-1}
\caption[nucleo2]{Nuclear spectra for the whole sample (cont.).}
\end{figure*}

\begin{figure*}[htp]
\psfig{figure=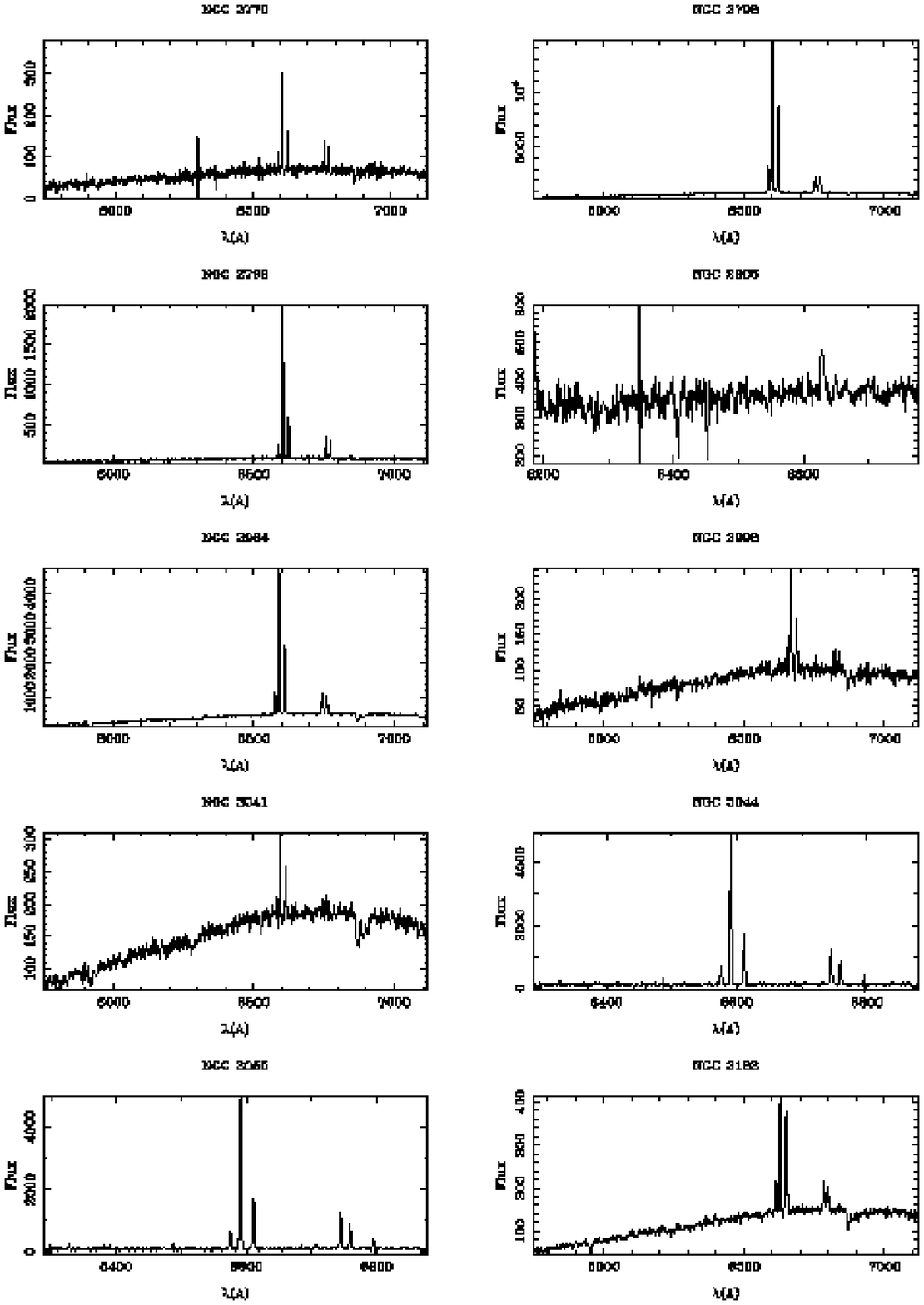,width=17truecm}
\addtocounter{figure}{-1}
\caption[nucleo3]{Nuclear spectra for the whole sample (cont.). }
\end{figure*}

\begin{figure*}[htp]
\psfig{figure=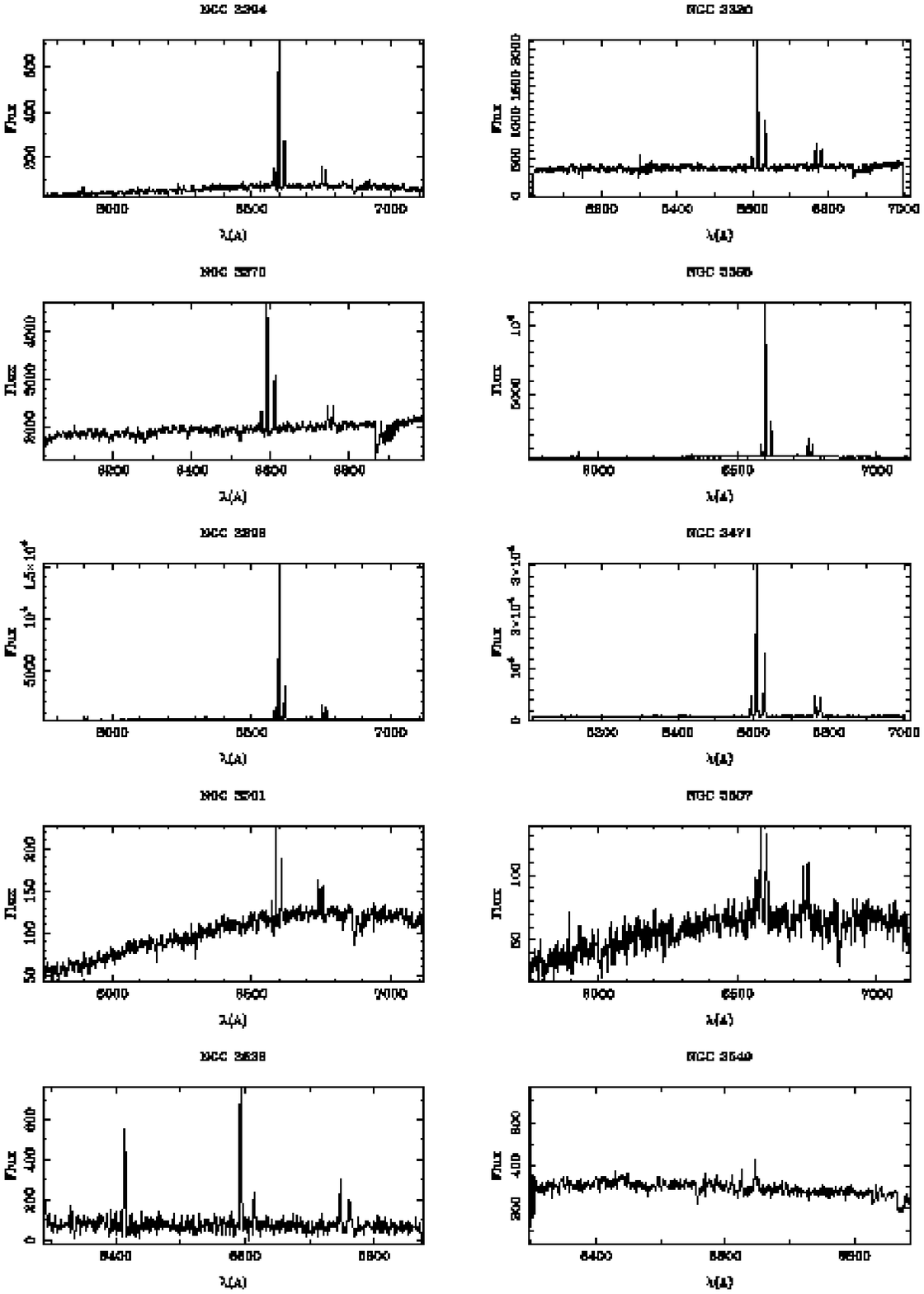,width=17truecm}
\addtocounter{figure}{-1}
\caption[nucleo4]{Nuclear spectra for the whole sample (cont.).}
\end{figure*}

\begin{figure*}[htp]
\psfig{figure=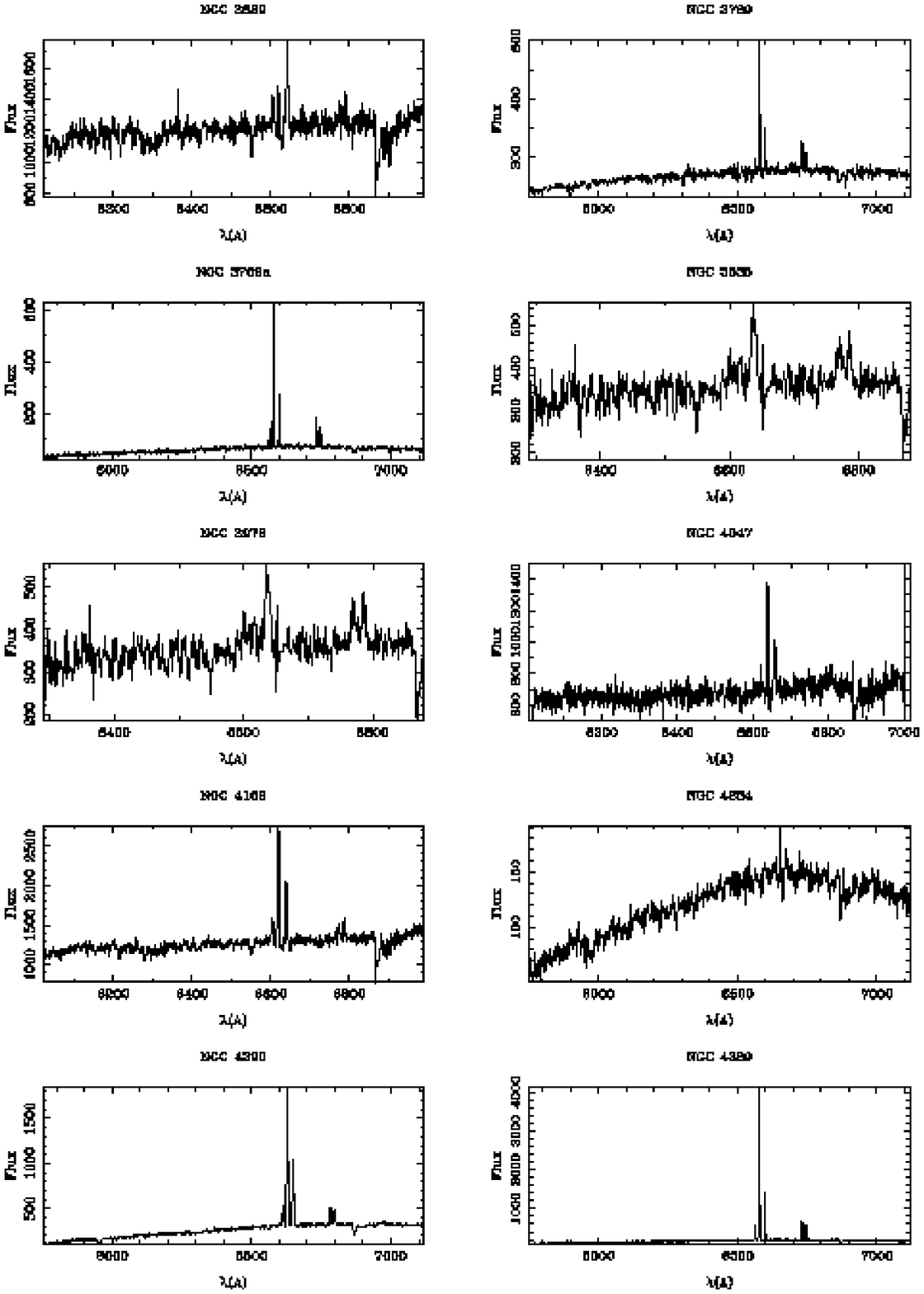,width=17truecm}
\addtocounter{figure}{-1}
\caption[nucleo5]{Nuclear spectra for the whole sample (cont.).}
\end{figure*}

\begin{figure*}[htp]
\psfig{figure=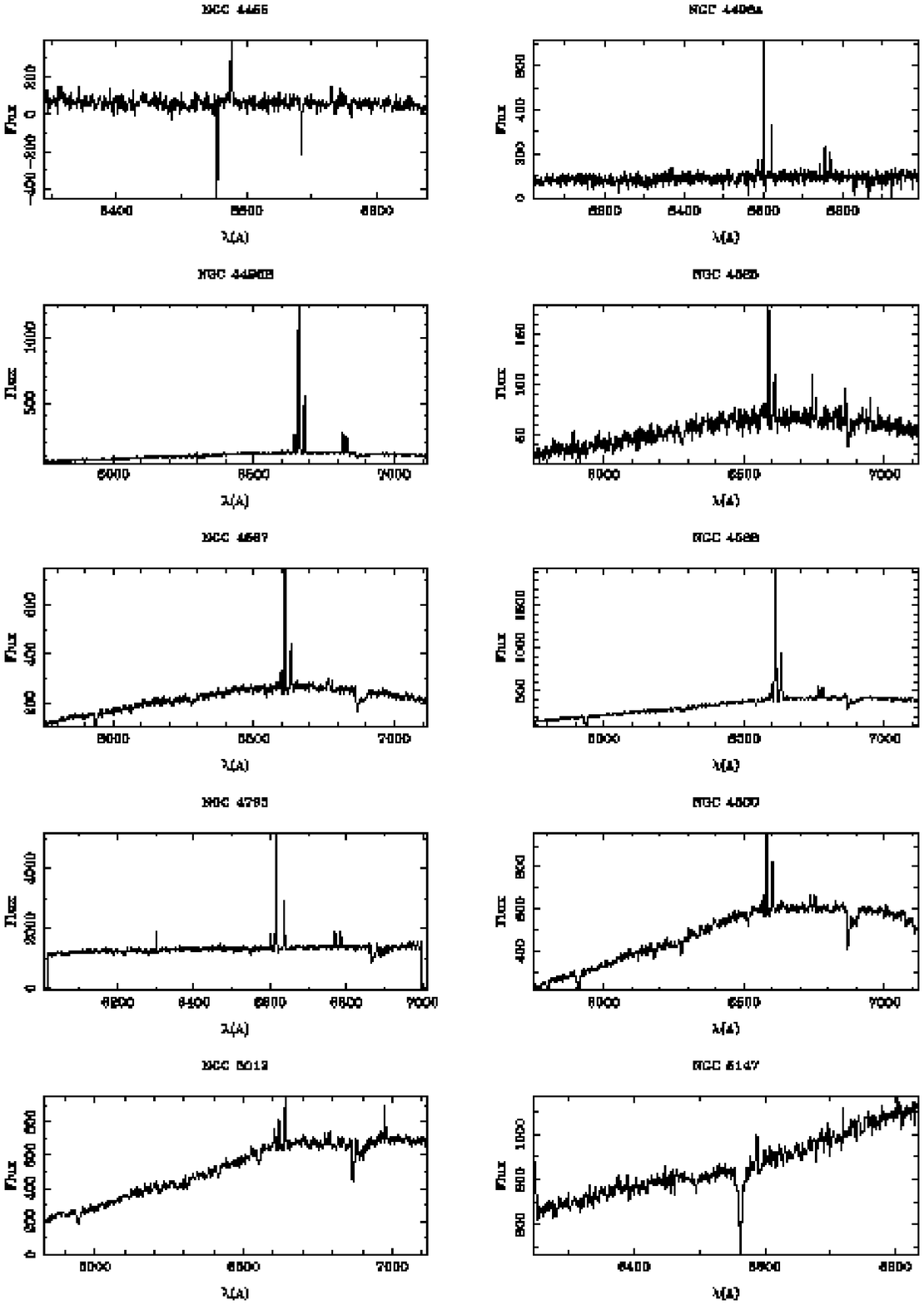,width=17truecm}
\addtocounter{figure}{-1}
\caption[nucleo6]{Nuclear spectra for the whole sample (cont.).}
\end{figure*}

\begin{figure*}[htp]
\psfig{figure=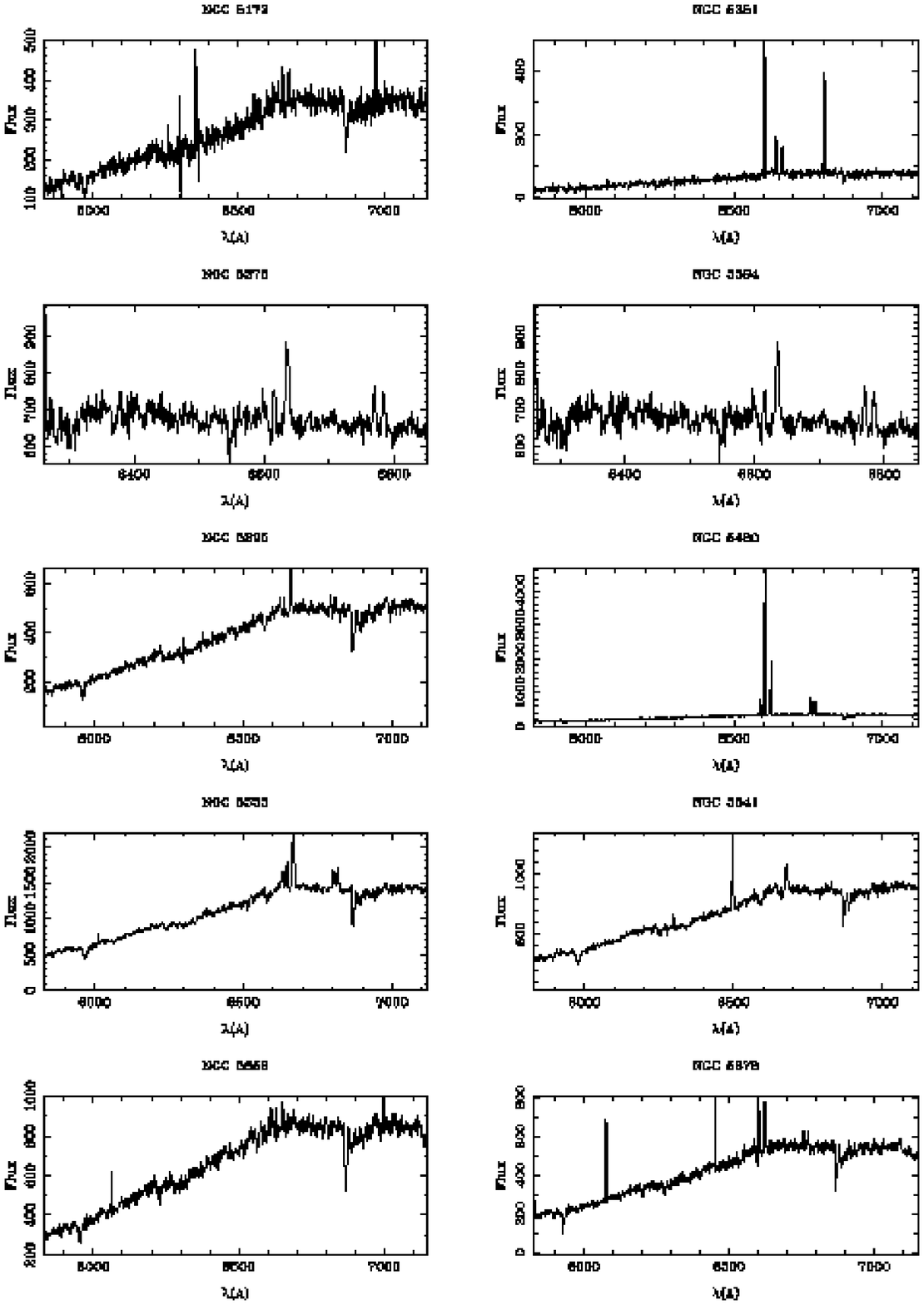,width=17truecm}
\addtocounter{figure}{-1}
\caption[nucleo7]{Nuclear spectra for the whole sample (cont.).}
\end{figure*}

\begin{figure*}[htp]
\psfig{figure=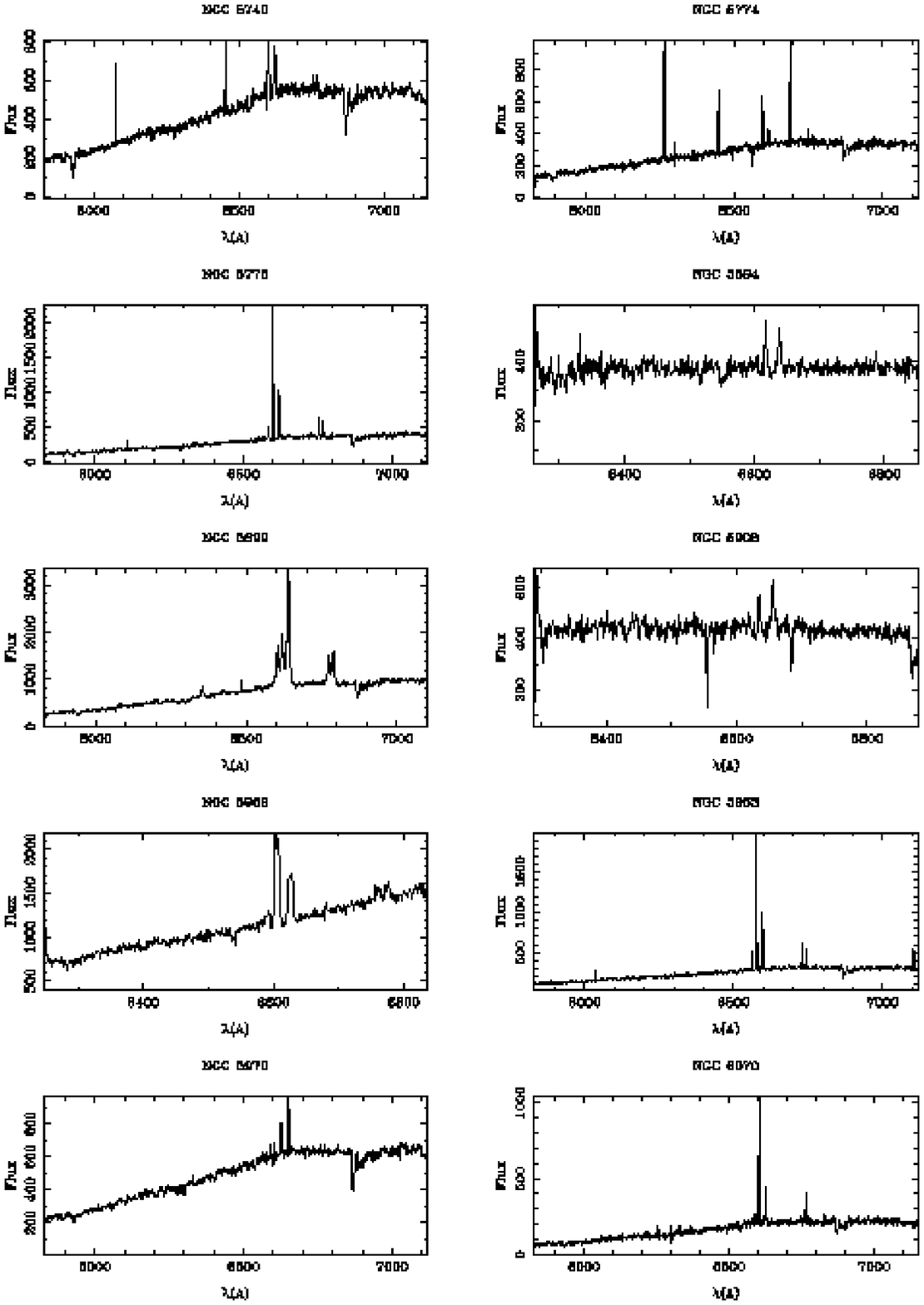,width=17truecm}
\addtocounter{figure}{-1}
\caption[nucleo8]{Nuclear spectra for the whole sample (cont.).}
\end{figure*}

\begin{figure*}[htp]
\psfig{figure=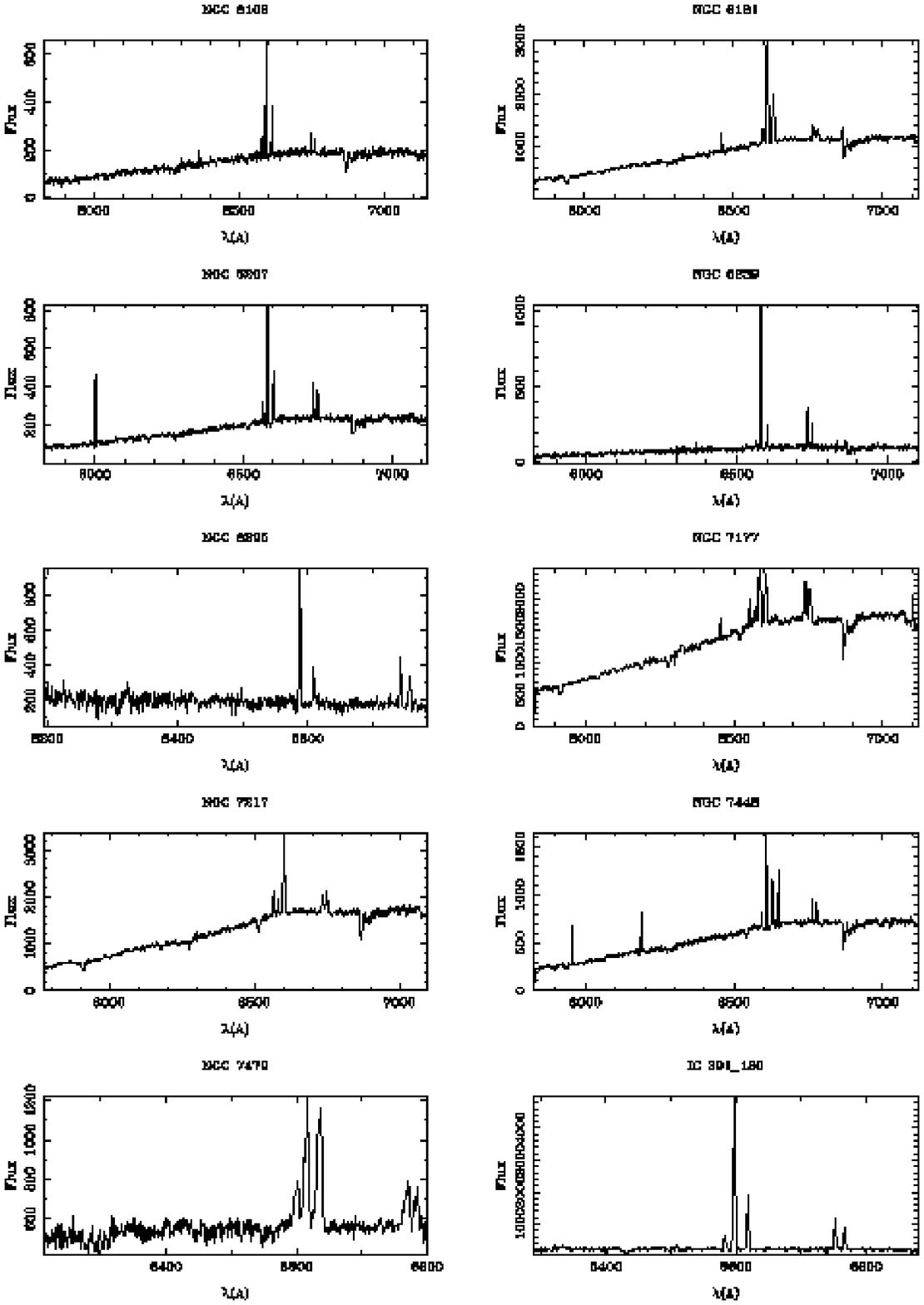,width=17truecm}
\addtocounter{figure}{-1}
\caption[nucleo9]{Nuclear spectra for the whole sample (cont.).}
\end{figure*}

\begin{figure*}[htp]
\psfig{figure=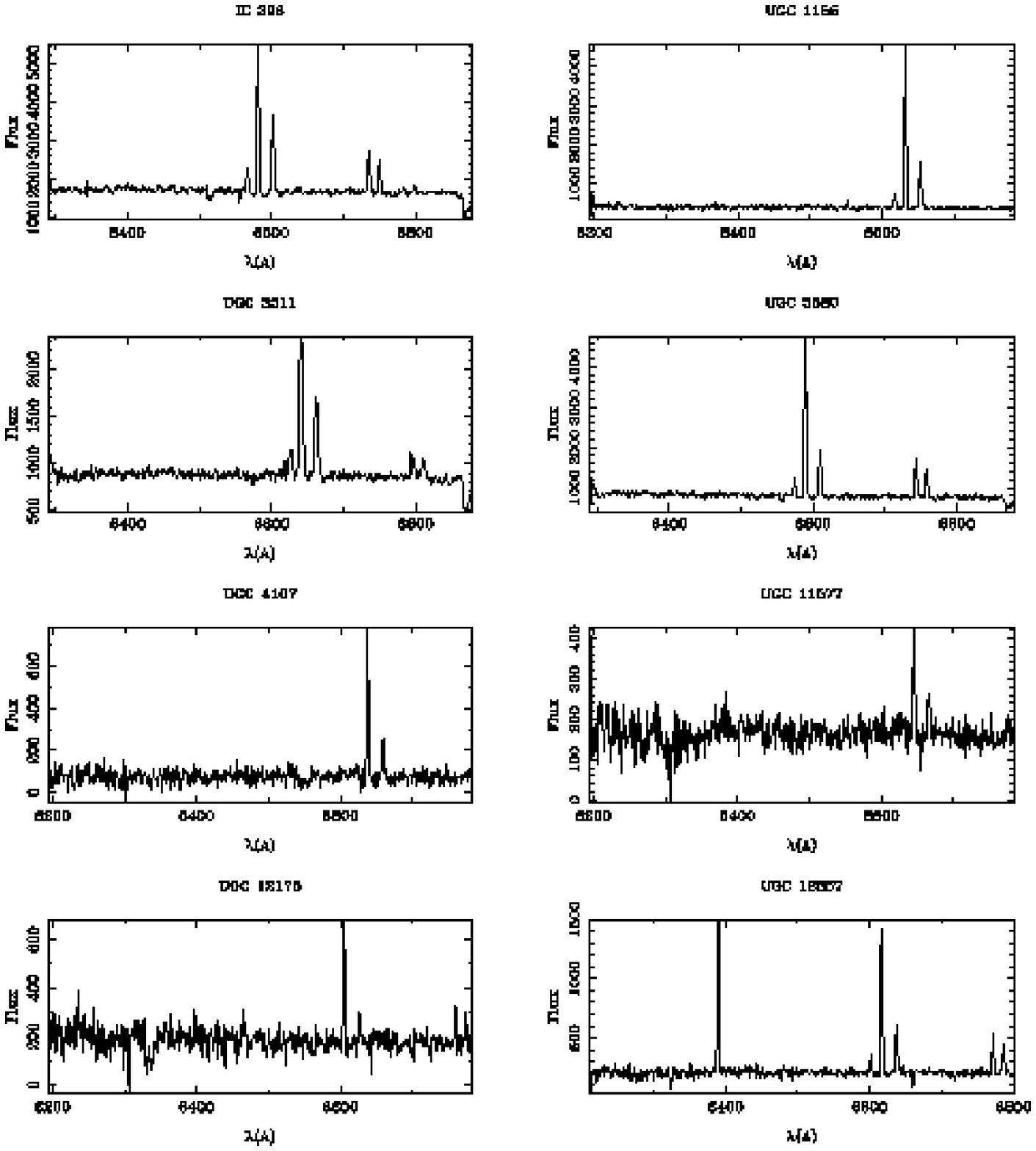,width=17truecm}
\addtocounter{figure}{-1}
\caption[nucleo10]{Nuclear spectra for the whole sample (cont.).}
\end{figure*}

\end{document}

%% file: tabla_lin1.tex
NGC~493    &    0.0 &     36.590 &  --  &  --  &      0.290 &      0.054 &  --  &  --  &  -- \\
NGC~493    &  -24.4 &     36.020 &  --  &  --  &      0.138 &      0.037 &  --  &  --  &  -- \\
NGC~493    &   -8.1 &     60.180 &  --  &  --  &      0.181 &      0.038 &  --  &  --  &  -- \\
NGC~493    &   13.9 &     42.810 &  --  &  --  &      0.206 &      0.075 &  --  &  --  &  -- \\
NGC~493    &   37.1 &     63.570 &  --  &  --  &      0.253 &      0.039 &  --  &  --  &  -- \\
NGC~493    &   56.8 &     44.140 &  --  &  --  &      0.269 &      0.059 &  --  &  --  &  -- \\
NGC~493    &   66.1 &    365.600 &  --  &  --  &      0.219 &      0.026 &  --  &  --  &  -- \\
NGC~658    &    0.0 &     15.300 &  --  &  --  &      0.396 &      0.042 &  --  &  --  &  -- \\
NGC~658    &  -11.6 &     20.460 &  --  &  --  &      0.312 &      0.056 &  --  &  --  &  -- \\
NGC~658    &   16.2 &     35.160 &  --  &  --  &      0.289 &      0.100 &  --  &  --  &  -- \\
NGC~828    &    0.0 &     13.830 &  --  &  --  &      0.510 &      0.029 &      0.225 &      0.052 &      0.863\\
NGC~828    &   -7.8 &     43.570 &  --  &  --  &      0.433 &      0.031 &      0.288 &      0.055 &      1.159\\
NGC~864    &    0.0 &     44.200 &  --  &  --  &      0.409 &      0.017 &      0.336 &      0.031 &      0.919\\
NGC~864    &  -95.1 &    264.400 &  --  &  --  &      0.158 &      2.057 &      0.375 &      4.113 &      1.566\\
NGC~864    &  -61.5 &    657.800 &  --  &  --  &      0.193 &      1.863 &  --  &  --  &  -- \\
NGC~864    &  -49.9 &    296.300 &  --  &  --  &      0.224 &      1.581 &  --  &  --  &  -- \\
NGC~864    &  -41.8 &    163.000 &  --  &  --  &      0.327 &      5.373 &  --  &  --  &  -- \\
NGC~864    &   40.6 &    440.800 &  --  &  --  &      0.361 &      2.021 &      0.323 &      4.043 &      1.485\\
NGC~1036   &    0.0 &     67.740 &      0.015 &      0.014 &      0.168 &      0.015 &      0.287 &      0.030 &      1.320\\
NGC~1036   &  -11.6 &    138.800 &      0.014 &      0.009 &      0.127 &      0.010 &      0.201 &      0.020 &      1.306\\
NGC~1137   &    0.0 &     11.450 &  --  &  --  &      0.376 &      0.061 &  --  &  --  &  -- \\
NGC~1137   &    4.6 &     19.530 &  --  &  --  &      0.398 &      0.087 &  --  &  --  &  -- \\
NGC~1507   &    0.0 &     20.320 &  --  &  --  &      0.172 &      0.036 &      0.478 &      0.072 &      1.365\\
NGC~1507   &  -27.8 &     54.320 &  --  &  --  &      0.110 &      0.021 &      0.247 &      0.043 &      1.453\\
NGC~1507   &  -19.7 &     47.910 &  --  &  --  &      0.153 &      0.030 &      0.391 &      0.061 &      1.310\\
NGC~1507   &   -8.1 &     38.710 &  --  &  --  &      0.119 &      0.029 &      0.308 &      0.058 &      1.501\\
NGC~1507   &    9.3 &     84.570 &  --  &  --  &      0.141 &      0.016 &      0.237 &      0.032 &      1.370\\
NGC~1507   &   47.6 &    121.200 &  --  &  --  &      0.099 &      0.020 &      0.304 &      0.041 &      1.461\\
NGC~1507   &   62.6 &    293.200 &  --  &  --  &      0.077 &      0.048 &      0.297 &      0.097 &      1.138\\
NGC~1507   &   85.8 &    -28.760 & -- & -- & -- & -- &      0.251 &      0.165 &      1.589\\
NGC~1530   &    0.0 &     36.740 &      0.010 &      0.009 &      0.345 &      0.011 &      0.181 &      0.019 &      1.260\\
NGC~1530   &  -74.2 &     68.040 &  --  &  --  &      0.225 &      0.053 &      0.285 &      0.103 &      1.526\\
NGC~1530   &  -62.6 &    385.400 &  --  &  --  &      0.333 &      0.081 &      0.323 &      0.159 &      1.415\\
NGC~2344   &    0.0 &      0.000 &  --  &  --  &  --  &  --  &  --  &  --  &  -- \\
NGC~2344   &  -31.3 &     69.610 &  --  &  --  &      0.269 &      0.108 &  --  &  --  &  -- \\
NGC~2344   &    8.1 &      4.079 &  --  &  --  &      0.381 &      0.161 &  --  &  --  &  -- \\
NGC~2424   &    0.0 &      0.000 &  --  &  --  &  --  &  --  &  --  &  --  &  -- \\
NGC~2424   &  -34.8 &    232.300 &  --  &  --  &      0.317 &      0.088 &  --  &  --  &  -- \\
NGC~2424   &   25.5 &     44.750 &  --  &  --  &      0.374 &      0.071 &  --  &  --  &  -- \\
NGC~2460   &    0.0 &      0.520 &      0.317 &      0.318 &      3.301 &      0.330 &      1.604 &      0.640 &      0.864\\
NGC~2460   &  -24.1 &     16.560 &      0.191 &      0.090 &      0.522 &      0.095 &      0.463 &      0.181 &      1.060\\
NGC~2460   &  -12.3 &      5.880 &  --  &  --  &      0.691 &      0.092 &      0.545 &      0.175 &      1.233\\
NGC~2460   &   -8.5 &      5.630 &  --  &  --  &      0.789 &      0.084 &      0.459 &      0.155 &      0.958\\
NGC~2460   &    3.3 &      1.310 &  --  &  --  &      1.748 &      0.107 &      0.953 &      0.204 &      0.851\\
NGC~2460   &    9.1 &     10.310 &  --  &  --  &      0.518 &      0.038 &      0.362 &      0.070 &      1.070\\
NGC~2460   &   20.2 &      8.630 &  --  &  --  &      0.569 &      0.088 &      0.412 &      0.167 &      2.004\\
NGC~2469   &    0.0 &     54.730 &      0.034 &      0.090 &      0.388 &      0.015 &  --  &  --  &  -- \\
NGC~2469   &  -12.8 &     71.730 &      0.029 &      0.090 &      0.262 &      0.026 &  --  &  --  &  -- \\
NGC~2469   &   10.4 &    146.900 &      0.015 &      0.090 &      0.274 &      0.011 &  --  &  --  &  -- \\
NGC~2543   &    0.0 &     31.560 &      0.009 &      0.010 &      0.450 &      0.011 &      0.273 &      0.020 &      1.267\\
NGC~2543   &  -61.8 &     85.830 &  --  &  --  &      0.267 &      0.126 &      0.395 &      0.251 &      1.646\\
NGC~2543   &  -40.0 &     27.490 &  --  &  --  &      0.301 &      0.153 &  --  &  --  &  -- \\
NGC~2543   &   31.9 &     15.520 &      0.036 &      0.070 &      0.492 &      0.076 &      0.433 &      0.145 &      1.389\\
NGC~2545   &    0.0 &      0.000 &  --  &  --  &  --  &  --  &  --  &  --  &  -- \\
NGC~2545   &  -16.2 &     24.810 &  --  &  --  &      0.342 &      0.073 &      0.267 &      0.142 &      1.551\\
NGC~2545   &   16.2 &     16.950 &  --  &  --  &      0.483 &      0.125 &  --  &  --  &  -- \\
NGC~2552   &    0.0 &    236.900 &      0.013 &      0.010 &      0.092 &      0.011 &      0.234 &      0.022 &      1.397\\
NGC~2552   &  -46.2 &     43.210 &  --  &  --  &      0.155 &      0.152 &      0.278 &      0.304 &      0.805\\
NGC~2552   &  -20.8 &     39.180 &  --  &  --  &      0.076 &      0.203 &  --  &  --  &  -- \\
NGC~2552   &   -3.9 &     80.840 &  --  &  --  &      0.125 &      3.789 &      0.264 &      7.578 &      1.495\\
NGC~2552   &   59.8 &     56.460 &  --  &  --  &      0.034 &      0.077 &      0.365 &      0.160 &      1.243\\
NGC~2608   &    0.0 &     36.150 &      0.013 &      0.019 &      0.495 &      0.025 &      0.334 &      0.043 &      1.083\\
NGC~2608   &   -6.8 &     15.250 &      0.017 &      0.053 &      0.535 &      0.067 &      0.540 &      0.121 &      1.075\\
NGC~2608   &   40.6 &     18.460 &  --  &  --  &      0.498 &      0.103 &      0.259 &      0.186 &      1.223\\
NGC~2628   &    0.0 &      0.810 &  --  &  --  &      3.398 &      0.461 &  --  &  --  &  -- \\
NGC~2628   &  -18.6 &     49.330 &  --  &  --  &      0.436 &      0.060 &  --  &  --  &  -- \\
NGC~2628   &   -9.3 &     17.320 &  --  &  --  &      0.434 &      0.069 &  --  &  --  &  -- \\
NGC~2628   &   10.4 &     14.130 &  --  &  --  &      0.518 &      0.065 &  --  &  --  &  -- \\
NGC~2633   &    0.0 &     44.370 &      0.012 &      0.006 &      0.530 &      0.009 &      0.256 &      0.014 &      0.961\\
NGC~2633   &  -23.7 &     24.610 &      0.028 &      0.041 &      0.454 &      0.050 &      0.360 &      0.090 &      1.366\\
NGC~2633   &  -11.8 &     22.890 &      0.033 &      0.035 &      0.493 &      0.045 &      0.373 &      0.078 &      1.208\\
NGC~2633   &   15.2 &     16.070 &      0.029 &      0.043 &      0.511 &      0.053 &      0.415 &      0.096 &      0.647\\
NGC~2633   &   65.9 &    232.500 &      0.016 &      0.020 &      0.183 &      0.024 &      0.235 &      0.045 &      1.223\\

%% file: tabla_lin2.tex
NGC~2701   &    0.0 &     17.030 &  --  &  --  &      0.398 &      0.038 &      0.344 &      0.069 &      1.379\\
NGC~2701   &  -49.0 &     34.820 &  --  &  --  &      0.137 &      0.087 &      0.583 &      0.180 &      1.484\\
NGC~2701   &  -20.3 &     32.040 &  --  &  --  &      0.303 &      0.043 &      0.435 &      0.083 &      1.901\\
NGC~2701   &   25.4 &     25.580 &  --  &  --  &      0.216 &      0.076 &      0.274 &      0.147 &      1.365\\
NGC~2748   &    0.0 &     17.570 &  --  &  --  &      0.434 &      0.037 &      0.363 &      0.070 &      1.226\\
NGC~2748   &  -25.4 &     27.100 &  --  &  --  &      0.389 &      0.058 &      0.422 &      0.110 &      1.160\\
NGC~2748   &   -6.8 &     27.740 &  --  &  --  &      0.365 &      0.027 &      0.333 &      0.048 &      1.399\\
NGC~2748   &   15.2 &     26.080 &      0.042 &      0.034 &      0.366 &      0.039 &      0.339 &      0.073 &      1.303\\
NGC~2748   &   37.2 &     34.110 &      0.018 &      0.035 &      0.328 &      0.040 &      0.367 &      0.076 &      1.176\\
NGC~2748   &   52.4 &     73.520 &      0.026 &      0.020 &      0.295 &      0.025 &      0.294 &      0.045 &      1.326\\
NGC~2748   &   72.7 &     66.760 &      0.029 &      0.038 &      0.153 &      0.041 &      0.154 &      0.078 &      1.409\\
NGC~2770   &    0.0 &     13.200 &  --  &  --  &      0.424 &      0.072 &      0.670 &      0.140 &      1.206\\
NGC~2770   &  -86.2 &    647.700 &  --  &  --  &      0.155 &      0.086 &      0.504 &      0.176 &      1.088\\
NGC~2770   &  -64.2 &     25.870 &  --  &  --  &      0.277 &      0.205 &  --  &  --  &  -- \\
NGC~2770   &  -37.2 &     22.630 &  --  &  --  &      0.304 &      0.053 &      0.522 &      0.103 &      1.376\\
NGC~2770   &  -21.0 &     12.950 &  --  &  --  &      0.386 &      0.084 &      0.499 &      0.160 &      1.435\\
NGC~2770   &   -1.7 &     24.830 &  --  &  --  &      0.365 &      0.107 &      0.502 &      0.207 &      1.360\\
NGC~2770   &   15.2 &     11.340 &  --  &  --  &      0.450 &      0.107 &      0.641 &      0.204 &      0.917\\
NGC~2770   &   25.3 &     18.120 &  --  &  --  &      0.423 &      0.093 &      0.523 &      0.175 &      1.404\\
NGC~2770   &   32.1 &     27.200 &  --  &  --  &      0.372 &      0.060 &      0.483 &      0.113 &      1.193\\
NGC~2770   &   52.4 &     26.600 &  --  &  --  &      0.276 &      0.072 &      0.506 &      0.142 &      1.444\\
NGC~2770   &   74.4 &     44.580 &  --  &  --  &      0.169 &      0.091 &      0.336 &      0.183 &      1.576\\
NGC~2798   &    0.0 &     96.710 &      0.009 &      0.004 &      0.529 &      0.006 &      0.228 &      0.010 &      1.016\\
NGC~2798   &  -10.1 &     57.940 &      0.035 &      0.017 &      0.477 &      0.022 &      0.408 &      0.038 &      1.201\\
NGC~2798   &    6.8 &     48.390 &      0.030 &      0.016 &      0.460 &      0.021 &      0.376 &      0.036 &      1.323\\
NGC~2799   &    0.0 &     75.460 &      0.009 &      0.016 &      0.271 &      0.020 &      0.318 &      0.038 &      1.335\\
NGC~2799   &  -13.5 &     20.090 &  --  &  --  &      0.314 &      0.062 &      0.409 &      0.120 &      0.929\\
NGC~2799   &   11.8 &     24.870 &  --  &  --  &      0.274 &      0.066 &      0.314 &      0.126 &      1.046\\
NGC~2906   &    0.0 &      1.750 &  --  &  --  &      0.250 &      0.157 &  --  &  --  &  -- \\
NGC~2906   &  -13.9 &     10.980 &  --  &  --  &  --  &  --  &  --  &  --  &  -- \\
NGC~2906   &    9.3 &     12.990 &  --  &  --  &  --  &  --  &  --  &  --  &  -- \\
NGC~2964   &    0.0 &     45.350 &      0.018 &      0.008 &      0.486 &      0.011 &      0.045 &      0.017 &      1.118\\
NGC~2964   &  -25.4 &     33.850 &  --  &  --  &      0.386 &      0.044 &      0.308 &      0.079 &      1.750\\
NGC~2964   &   22.0 &     18.040 &      0.051 &      0.065 &      0.496 &      0.077 &      0.405 &      0.140 &      1.200\\
NGC~2964   &   37.2 &    140.900 &      0.009 &      0.012 &      0.342 &      0.016 &      0.286 &      0.028 &      1.361\\
NGC~2998   &    0.0 &      4.290 &  --  &  --  &      0.579 &      0.081 &      0.294 &      0.145 &      0.961\\
NGC~2998   &  -32.1 &     92.400 &      0.030 &      0.029 &      0.316 &      0.036 &      0.276 &      0.064 &      1.465\\
NGC~2998   &  -11.8 &     19.050 &  --  &  --  &      0.365 &      0.049 &      0.289 &      0.090 &      1.332\\
NGC~2998   &   10.1 &     17.580 &      0.053 &      0.046 &      0.344 &      0.052 &      0.220 &      0.095 &      1.585\\
NGC~3041   &    0.0 &      3.250 &  --  &  --  &      0.703 &      0.110 &      0.261 &      0.200 &      1.047\\
NGC~3041   &  -64.2 &     52.830 &  --  &  --  &      0.343 &      0.094 &      0.252 &      0.176 &      1.167\\
NGC~3041   &    6.8 &      8.500 &  --  &  --  &      0.456 &      0.078 &      0.205 &      0.136 &      1.334\\
NGC~3041   &   33.8 &     16.890 &      0.107 &      0.084 &      0.435 &      0.093 &      0.301 &      0.170 &      1.735\\
NGC~3041   &   72.7 &     93.160 &      0.029 &      0.040 &      0.279 &      0.048 &      0.292 &      0.088 &      1.227\\
NGC~3044   &    0.0 &     89.120 &      0.026 &      0.013 &      0.319 &      0.015 &      0.391 &      0.029 &      1.531\\
NGC~3044   &  -95.1 &    349.800 &  --  &  --  &      0.087 &      1.825 &      0.201 &      3.649 &      1.012\\
NGC~3044   &  -61.5 &    242.400 &  --  &  --  &      0.107 &      0.026 &      0.178 &      0.052 &      1.206\\
NGC~3044   &  -51.0 &    339.300 &  --  &  --  &      0.090 &      0.038 &      0.151 &      0.076 &      2.285\\
NGC~3044   &  -44.1 &    151.400 &  --  &  --  &      0.141 &      0.036 &      0.280 &      0.072 &      1.508\\
NGC~3044   &  -19.7 &    222.300 &  --  &  --  &      0.223 &      0.021 &      0.328 &      0.040 &      1.436\\
NGC~3044   &    1.2 &    107.400 &      0.033 &      0.014 &      0.300 &      0.016 &      0.399 &      0.031 &      1.505\\
NGC~3044   &   25.5 &    214.000 &      0.027 &      0.024 &      0.252 &      0.025 &      0.361 &      0.050 &      1.457\\
NGC~3044   &   59.2 &     79.780 &  --  &  --  &      0.254 &      0.051 &      0.354 &      0.100 &      1.414\\
NGC~3055   &    0.0 &     95.100 &      0.021 &      0.007 &      0.452 &      0.009 &      0.289 &      0.016 &      1.207\\
NGC~3055   &  -17.4 &     24.120 &  --  &  --  &      0.357 &      0.109 &      0.468 &      0.211 &      1.459\\
NGC~3055   &   30.2 &    293.900 &  --  &  --  &      0.204 &      1.117 &      0.270 &      2.234 &      1.313\\
NGC~3183   &    0.0 &      9.800 &  --  &  --  &      1.092 &      0.061 &      0.547 &      0.099 &      1.233\\
NGC~3183   &  -15.2 &     11.580 &  --  &  --  &      0.509 &      0.100 &      0.384 &      0.181 &      1.565\\
NGC~3183   &   11.8 &      6.950 &  --  &  --  &      0.619 &      0.094 &      0.551 &      0.173 &      1.947\\
NGC~3294   &    0.0 &     32.150 &      0.019 &      0.032 &      0.379 &      0.039 &      0.338 &      0.071 &      1.439\\
NGC~3294   &  -69.3 &     29.330 &      0.028 &      0.048 &      0.452 &      0.059 &      0.434 &      0.107 &      1.709\\
NGC~3294   &  -42.3 &     33.270 &      0.028 &      0.031 &      0.415 &      0.038 &      0.382 &      0.069 &      1.563\\
NGC~3294   &  -22.0 &     27.730 &      0.040 &      0.042 &      0.387 &      0.051 &      0.362 &      0.093 &      1.365\\
NGC~3294   &  -11.8 &     35.170 &  --  &  --  &      0.388 &      0.050 &      0.297 &      0.088 &      1.466\\
NGC~3294   &   18.6 &     78.980 &      0.011 &      0.023 &      0.356 &      0.030 &      0.316 &      0.053 &      1.507\\
NGC~3320   &  -62.6 &     17.110 &  --  &  --  &      0.404 &      1.367 &  --  &  --  &  -- \\
NGC~3320   &  -54.5 &     15.420 &  --  &  --  &      0.345 &      0.275 &  --  &  --  &  -- \\
NGC~3320   &  -28.0 &     18.960 &  --  &  --  &      0.268 &      0.081 &      0.402 &      0.160 &      2.033\\
NGC~3320   &  -13.9 &     21.630 &  --  &  --  &      0.323 &      0.028 &      0.287 &      0.052 &      1.418\\
NGC~3320   &   42.9 &     31.900 &  --  &  --  &      0.244 &      0.227 &      0.489 &      0.454 &      1.259\\
NGC~3320   &    0.0 &     11.690 &  --  &  --  &      0.426 &      0.037 &      0.305 &      0.069 &      1.008\\
NGC~3320   &    9.3 &     18.840 &  --  &  --  &      0.375 &      0.049 &      0.312 &      0.094 &      1.169\\
NGC~3320   &   17.4 &     26.020 &  --  &  --  &      0.344 &      0.058 &      0.295 &      0.111 &      1.190\\
NGC~3320   &   26.7 &      9.440 &  --  &  --  &      0.484 &      0.158 &      0.836 &      0.315 &      1.045\\
NGC~3320   &   42.9 &     24.560 &  --  &  --  &      0.232 &      0.117 &      0.435 &      0.234 &      1.130\\
NGC~3320   &   74.2 &     21.560 &  --  &  --  &      0.313 &      0.008 &      0.363 &      0.016 &      1.210\\
NGC~3320   &   80.0 &     27.250 &  --  &  --  &      0.295 &      0.196 &      0.252 &      0.387 &      1.580\\

%% file: tabla_lin3.tex
NGC~3370   &    0.0 &      6.060 &  --  &  --  &      0.461 &      0.026 &      0.276 &      0.050 &      1.038\\
NGC~3370   &  -69.6 &    256.900 &  --  &  --  &      0.194 &      1.657 &      0.294 &      3.315 &      1.483\\
NGC~3370   &  -57.9 &    100.600 &      0.023 &      0.071 &      0.194 &      0.074 &      0.601 &      0.151 &      1.329\\
NGC~3370   &  -48.8 &     42.730 &  --  &  --  &      0.233 &      0.080 &      0.491 &      0.161 &      1.266\\
NGC~3370   &  -40.3 &     75.550 &  --  &  --  &      0.207 &      0.023 &      0.403 &      0.046 &      1.551\\
NGC~3370   &   18.9 &     20.300 &      0.048 &      0.030 &      0.378 &      0.033 &      0.435 &      0.064 &      1.381\\
NGC~3370   &  -30.6 &     55.460 &      0.032 &      0.034 &      0.287 &      0.037 &      0.470 &      0.072 &      1.371\\
NGC~3370   &  -12.3 &     36.860 &      0.015 &      0.014 &      0.298 &      0.017 &      0.291 &      0.031 &      1.458\\
NGC~3370   &    2.0 &      6.220 &      0.029 &      0.016 &      0.476 &      0.019 &      0.288 &      0.034 &      1.084\\
NGC~3370   &   10.4 &     12.480 &      0.024 &      0.047 &      0.407 &      0.049 &      0.319 &      0.095 &      0.988\\
NGC~3370   &   27.3 &     37.380 &      0.052 &      0.053 &      0.297 &      0.055 &      0.465 &      0.109 &      1.526\\
NGC~3370   &   31.2 &     65.960 &      0.049 &      0.022 &      0.334 &      0.026 &      0.414 &      0.048 &      1.116\\
NGC~3370   &   36.4 &     56.680 &      0.070 &      0.034 &      0.260 &      0.036 &      0.435 &      0.072 &      1.217\\
NGC~3370   &   52.7 &    284.300 &      0.022 &      0.016 &      0.153 &      0.018 &      0.258 &      0.035 &      1.430\\
NGC~3395   &    0.0 &    143.800 &      0.012 &      0.006 &      0.206 &      0.007 &      0.202 &      0.013 &      1.298\\
NGC~3395   &  -94.6 &     61.140 &  --  &  --  &      0.254 &      0.073 &      0.437 &      0.143 &      1.801\\
NGC~3395   &  -32.1 &    292.700 &      0.018 &      0.027 &      0.127 &      0.028 &      0.246 &      0.057 &      0.979\\
NGC~3395   &  -22.0 &     72.440 &      0.057 &      0.046 &      0.208 &      0.050 &      0.442 &      0.100 &      1.110\\
NGC~3396   &    0.0 &    155.400 &      0.012 &      0.005 &      0.201 &      0.006 &      0.171 &      0.012 &      1.249\\
NGC~3396   &  -10.1 &     55.610 &      0.026 &      0.019 &      0.247 &      0.022 &      0.369 &      0.043 &      1.357\\
NGC~3396   &    8.5 &    123.100 &      0.023 &      0.013 &      0.194 &      0.015 &      0.253 &      0.029 &      1.368\\
NGC~3471   &    0.0 &     93.080 &      0.012 &      0.004 &      0.410 &      0.006 &      0.267 &      0.010 &      1.175\\
NGC~3501   &    0.0 &      3.360 &  --  &  --  &      0.589 &      0.132 &      0.666 &      0.252 &      0.904\\
NGC~3501   &  -84.5 &     27.300 &  --  &  --  &      0.280 &      0.111 &      0.409 &      0.215 &      2.554\\
NGC~3501   &  -67.6 &     13.100 &  --  &  --  &      0.268 &      0.103 &      0.606 &      0.208 &      1.430\\
NGC~3501   &  -45.6 &     11.780 &      0.073 &      0.078 &      0.498 &      0.089 &      0.604 &      0.168 &      1.951\\
NGC~3501   &  -30.4 &      9.830 &      0.075 &      0.075 &      0.435 &      0.084 &      0.414 &      0.157 &      1.381\\
NGC~3501   &  -10.1 &      6.790 &  --  &  --  &      0.443 &      0.090 &      0.442 &      0.170 &      1.309\\
NGC~3501   &    8.4 &      3.400 &  --  &  --  &      0.665 &      0.119 &      0.654 &      0.226 &      0.954\\
NGC~3501   &   38.9 &     10.960 &  --  &  --  &      0.370 &      0.098 &      0.594 &      0.191 &      1.308\\
NGC~3501   &   59.1 &     42.080 &  --  &  --  &      0.294 &      0.051 &      0.363 &      0.096 &      1.354\\
NGC~3507   &    0.0 &      6.900 &  --  &  --  &      0.890 &      0.120 &      0.849 &      0.220 &      0.711\\
NGC~3526   &    0.0 &     35.420 &  --  &  --  &      0.248 &      0.061 &      0.609 &      0.124 &      1.157\\
NGC~3526   &  -16.2 &     41.290 &  --  &  --  &      0.112 &      0.060 &      0.325 &      0.121 &      1.499\\
NGC~3549   &    0.0 &      0.730 &  --  &  --  &      2.197 &      0.432 &  --  &  --  &  -- \\
NGC~3549   &  -62.6 &     46.920 &  --  &  --  &      0.387 &      0.092 &  --  &  --  &  -- \\
NGC~3549   &  -49.9 &     31.550 &  --  &  --  &      0.405 &      0.119 &  --  &  --  &  -- \\
NGC~3549   &  -25.5 &     17.590 &  --  &  --  &      0.504 &      0.066 &      0.296 &      0.123 &      0.884\\
NGC~3549   &  -17.4 &     14.560 &  --  &  --  &      0.373 &      0.058 &      0.220 &      0.108 &      1.621\\
NGC~3549   &   16.2 &     10.530 &  --  &  --  &      0.405 &      0.067 &      0.264 &      0.127 &      1.205\\
NGC~3549   &   37.1 &     30.560 &  --  &  --  &      0.386 &      0.058 &      0.275 &      0.111 &      2.156\\
NGC~3549   &   45.2 &     17.300 &  --  &  --  &      0.408 &      0.128 &      0.332 &      0.249 &      1.845\\
NGC~3689   &    0.0 &      0.510 &      0.698 &      0.332 &      4.393 &      0.350 &      1.649 &      0.666 &      0.884\\
NGC~3689   &  -18.2 &     34.900 &      0.017 &      0.024 &      0.375 &      0.028 &      0.395 &      0.052 &      1.420\\
NGC~3689   &   -7.8 &     20.030 &      0.025 &      0.018 &      0.352 &      0.021 &      0.286 &      0.039 &      1.094\\
NGC~3689   &    8.5 &     25.780 &      0.036 &      0.021 &      0.358 &      0.024 &      0.301 &      0.044 &      1.367\\
NGC~3689   &   30.6 &     25.120 &      0.091 &      0.078 &      0.443 &      0.083 &      0.472 &      0.160 &      1.409\\
NGC~3769A  &    0.0 &     30.300 &      0.020 &      0.032 &      0.362 &      0.039 &      0.388 &      0.072 &      1.478\\
NGC~3769A  &   -8.4 &     44.570 &  --  &  --  &      0.305 &      0.036 &      0.351 &      0.066 &      1.462\\
NGC~3769A  &   15.2 &     20.330 &  --  &  --  &      0.362 &      0.057 &      0.586 &      0.111 &      1.222\\
NGC~3769A  &   79.4 &     64.740 &      0.020 &      0.045 &      0.050 &      0.046 &      0.312 &      0.101 &      1.451\\
NGC~3769A  &   87.9 &    155.400 &      0.021 &      0.114 &      0.043 &      0.114 &      0.232 &      0.229 &      1.167\\
NGC~3769   &    0.0 &      9.840 &      0.052 &      0.051 &      0.366 &      0.056 &      0.404 &      0.107 &      1.898\\
NGC~3769   &  -37.2 &     17.710 &  --  &  --  &      0.338 &      0.128 &      0.843 &      0.261 &      1.014\\
NGC~3769   &  -23.7 &     24.700 &      0.011 &      0.050 &      0.343 &      0.060 &      0.541 &      0.117 &      1.442\\
NGC~3769   &  -10.1 &     12.660 &  --  &  --  &      0.338 &      0.068 &      0.451 &      0.131 &      1.743\\
NGC~3769   &   15.2 &      5.150 &  --  &  --  &      0.601 &      0.131 &      0.846 &      0.253 &      0.672\\
NGC~3835   &    0.0 &      0.000 &  --  &  --  &  --  &  --  &  --  &  --  &  -- \\
NGC~3835   &  -22.0 &     29.080 &  --  &  --  &      0.460 &      0.100 &  --  &  --  &  -- \\
NGC~3835   &  -13.9 &     13.370 &  --  &  --  &      0.645 &      0.121 &  --  &  --  &  -- \\
NGC~3835   &   18.6 &     31.210 &  --  &  --  &      0.380 &      0.084 &      0.254 &      0.162 &      1.709\\
NGC~3976   &    0.0 &      0.000 &  --  &  --  &  --  &  --  &  --  &  --  &  -- \\
NGC~3976   &  -86.2 &     55.110 &  --  &  --  &      0.428 &      0.369 &      0.255 &      0.733 &      1.354\\
NGC~3976   &  -42.3 &     69.950 &  --  &  --  &      0.438 &      0.039 &      0.241 &      0.064 &      1.788\\
NGC~3976   &  -30.4 &     14.910 &  --  &  --  &      0.455 &      0.084 &      0.265 &      0.149 &      1.402\\
NGC~3976   &  -20.3 &      7.200 &      0.092 &      0.079 &      0.508 &      0.089 &      0.253 &      0.161 &      0.822\\
NGC~3976   &   20.3 &      8.620 &      0.038 &      0.054 &      0.491 &      0.063 &      0.328 &      0.113 &      1.282\\
NGC~3976   &   98.0 &    140.700 &      0.031 &      0.276 &      0.353 &      0.277 &      0.287 &      0.554 &      1.473\\
NGC~4047   &    0.0 &      4.700 &  --  &  --  &      0.494 &      0.062 &  --  &  --  &  -- \\
NGC~4047   &  -27.0 &      4.470 &  --  &  --  &      0.471 &      0.063 &  --  &  --  &  -- \\
NGC~4047   &  -14.5 &     35.760 &      0.034 &      0.038 &      0.382 &      0.044 &      0.370 &      0.082 &      1.179\\
NGC~4047   &   -9.1 &     28.740 &  --  &  --  &      0.289 &      0.027 &      0.199 &      0.050 &      0.948\\
NGC~4047   &   -5.9 &     18.250 &  --  &  --  &      0.319 &      0.040 &      0.201 &      0.073 &      1.226\\
NGC~4047   &    7.1 &     13.910 &  --  &  --  &      0.293 &      0.037 &      0.214 &      0.070 &      1.238\\
NGC~4047   &   22.4 &     11.530 &  --  &  --  &      0.543 &      0.149 &      0.561 &      0.287 &      1.619\\
NGC~4047   &   24.7 &     19.860 &  --  &  --  &      0.391 &      0.128 &      0.418 &      0.248 &      1.996\\

%% file: tabla_lin4.tex
NGC~4162   &    0.0 &      5.720 &      0.014 &      0.033 &      0.660 &      0.036 &      0.320  &     0.067  &     0.770\\
NGC~4162   &  -66.0 &    200.300 &      0.024 &      0.035 &      0.280 &      0.039 &      0.331 &      0.074 &      1.098\\
NGC~4162   &  -45.5 &     20.610 &      0.106 &      0.088 &      0.333 &      0.091 &      0.474 &      0.180 &      1.541\\
NGC~4162   &  -39.6 &     29.920 &      0.035 &      0.048 &      0.349 &      0.052 &      0.408 &      0.101 &      1.566\\
NGC~4162   &  -31.9 &     27.430 &      0.025 &      0.040 &      0.424 &      0.045 &      0.416 &      0.085 &      1.409\\
NGC~4162   &  -22.1 &     16.880 &      0.017 &      0.032 &      0.440 &      0.036 &      0.299 &      0.067 &      1.663\\
NGC~4162   &    9.1 &     11.320 &      0.028 &      0.041 &      0.425 &      0.045 &      0.259 &      0.084 &      1.241\\
NGC~4162   &   12.3 &     29.460 &      0.021 &      0.031 &      0.378 &      0.035 &      0.336 &      0.065 &      1.407\\
NGC~4162   &   25.4 &     60.790 &      0.032 &      0.028 &      0.339 &      0.032 &      0.266 &      0.059 &      1.154\\
NGC~4284   &    0.0 &      1.200 &      0.096 &      0.028 &      0.817 &      0.290 &  --  &  --  &  -- \\
NGC~4284   &  -20.3 &      9.840 &  --  &  --  &      0.676 &      0.224 &  --  &  --  &  -- \\
NGC~4284   &   16.9 &     10.930 &  --  &  --  &      0.429 &      0.114 &      0.416 &      0.216 &      1.315\\
NGC~4290   &    0.0 &     31.000 &      0.008 &      0.014 &      0.507 &      0.019 &      0.266 &      0.032 &      1.188\\
NGC~4290   &   28.7 &     11.460 &  --  &  --  &      0.547 &      0.141 &  --  &  --  &  -- \\
NGC~4290   &  -23.7 &     12.100 &      2.283 &      0.014 &      0.521 &      0.141 &  --  &  --  &  -- \\
NGC~4389   &    0.0 &     64.520 &      0.008 &      0.011 &      0.289 &      0.014 &      0.247 &      0.024 &      1.326\\
NGC~4389   &  -49.0 &     46.280 &  --  &  --  &      0.314 &      0.067 &      0.387 &      0.126 &      1.438\\
NGC~4389   &  -42.3 &     26.770 &  --  &  --  &      0.386 &      0.095 &      0.417 &      0.177 &      1.732\\
NGC~4389   &  -23.7 &     49.770 &      0.009 &      0.022 &      0.281 &      0.027 &      0.305 &      0.050 &      1.350\\
NGC~4389   &  -11.8 &     48.860 &      0.018 &      0.023 &      0.304 &      0.028 &      0.279 &      0.051 &      1.218\\
NGC~4389   &   15.2 &     19.710 &  --  &  --  &      0.375 &      0.031 &      0.393 &      0.057 &      1.335\\
NGC~4389   &   27.0 &     35.030 &      0.022 &      0.028 &      0.301 &      0.034 &      0.325 &      0.062 &      1.485\\
NGC~4389   &   54.1 &     30.860 &  --  &  --  &      0.422 &      0.060 &      0.406 &      0.110 &      1.271\\
NGC~4455   &    0.0 &     14.190 &  --  &  --  &      0.075 &      0.158 &  --  &  --  &  -- \\
NGC~4455   &  -51.0 &    157.000 &  --  &  --  &      0.055 &      0.032 &      0.166 &      0.065 &      1.239\\
NGC~4455   &  -22.0 &     58.890 &  --  &  --  &      0.103 &      0.073 &      0.301 &      0.147 &      2.004\\
NGC~4455   &   -8.1 &     69.550 &  --  &  --  &      0.068 &      0.033 &      0.202 &      0.067 &      1.266\\
NGC~4455   &   25.5 &     93.720 &  --  &  --  &      0.083 &      0.036 &      0.174 &      0.072 &      1.870\\
NGC~4496A  &    0.0 &     12.230 &  --  &  --  &      0.397 &      0.079 &      0.455 &      0.153 &      1.274\\
NGC~4496A  & -112.5 &    100.800 & -- & -- & -- & -- &      0.371 &      0.180 &      0.991\\
NGC~4496A  &  -35.1 &    108.900 &      0.058 &      0.028 &      0.316 &      0.106 &  --  &  --  &  -- \\
NGC~4496A  &  -16.3 &     43.100 &  --  &  --  &      0.300 &      0.051 &      0.326 &      0.099 &      1.522\\
NGC~4496A  &  -10.4 &     15.590 &  --  &  --  &      0.373 &      0.066 &      0.393 &      0.128 &      2.025\\
NGC~4496A  &   -4.5 &      7.610 &  --  &  --  &      0.432 &      0.160 &      0.768 &      0.318 &      1.685\\
NGC~4496A  &   11.1 &     14.210 &  --  &  --  &      0.499 &      0.092 &      0.630 &      0.179 &      1.304\\
NGC~4496A  &   27.3 &     88.680 &  --  &  --  &      0.251 &      0.044 &      0.348 &      0.087 &      1.632\\
NGC~4496A  &   53.5 &    400.000 &      0.030 &      0.762 &      0.164 &      0.762 &      0.150 &      1.524 &      1.641\\
NGC~4496B  &    0.0 &     49.360 &      0.013 &      0.017 &      0.387 &      0.022 &      0.261 &      0.038 &      1.342\\
NGC~4496B  &  -76.0 &     71.400 &  --  &  --  &      0.222 &      0.144 &      0.255 &      0.286 &      1.397\\
NGC~4496B  &  -60.8 &     61.260 &  --  &  --  &      0.226 &      0.053 &      0.276 &      0.101 &      1.389\\
NGC~4496B  &  -20.3 &     26.320 &  --  &  --  &      0.356 &      0.066 &      0.341 &      0.121 &      1.410\\
NGC~4496B  &   16.9 &     24.270 &  --  &  --  &      0.316 &      0.070 &      0.348 &      0.131 &      1.321\\
NGC~4525   &    0.0 &      4.600 &  --  &  --  &      0.372 &      0.137 &      0.644 &      0.271 &      1.364\\
NGC~4525   &  -49.0 &      9.940 &  --  &  --  &      0.328 &      0.155 &      0.447 &      0.303 &      0.857\\
NGC~4525   &  -11.8 &      4.810 &  --  &  --  &      0.373 &      0.159 &      0.643 &      0.316 &      0.848\\
NGC~4525   &   21.8 &     13.300 &  --  &  --  &      0.183 &      0.095 &      0.424 &      0.192 &      1.203\\
NGC~4567   &    0.0 &      6.930 &  --  &  --  &      0.377 &      0.053 &  --  &  --  &  -- \\
NGC~4567   &  -42.3 &     26.270 &  --  &  --  &      0.343 &      0.054 &      0.282 &      0.098 &      1.101\\
NGC~4567   &  -15.2 &     16.410 &  --  &  --  &      0.232 &      0.057 &      0.159 &      0.105 &      0.893\\
NGC~4567   &   13.5 &     20.650 &  --  &  --  &      0.378 &      0.053 &      0.199 &      0.093 &      1.435\\
NGC~4567   &   47.3 &      9.130 &  --  &  --  &      0.474 &      0.101 &      0.417 &      0.188 &      1.444\\
NGC~4568   &    0.0 &     18.930 &  --  &  --  &      0.368 &      0.022 &      0.168 &      0.040 &      1.141\\
NGC~4568   &  -42.3 &     22.790 &  --  &  --  &      0.293 &      0.054 &      0.272 &      0.099 &      0.803\\
NGC~4568   &  -20.3 &     20.230 &  --  &  --  &      0.344 &      0.041 &      0.213 &      0.074 &      0.916\\
NGC~4568   &  -10.1 &     12.600 &  --  &  --  &      0.421 &      0.060 &      0.205 &      0.107 &      1.006\\
NGC~4568   &   18.6 &     22.920 &  --  &  --  &      0.339 &      0.035 &      0.210 &      0.064 &      1.036\\
NGC~4568   &   43.9 &     36.050 &  --  &  --  &      0.315 &      0.047 &      0.158 &      0.082 &      1.005\\
NGC~4568   &   67.6 &     36.080 &  --  &  --  &      0.349 &      0.063 &      0.278 &      0.113 &      1.092\\
NGC~4793   &  -43.6 &     88.780 &  --  &  --  &      0.240 &      0.060 &      0.284 &      0.118 &      0.973\\
NGC~4793   &  -33.8 &     24.280 &  --  &  --  &      0.348 &      0.059 &      0.418 &      0.116 &      1.252\\
NGC~4793   &  -28.6 &     14.720 &  --  &  --  &      0.334 &      0.114 &  --  &  --  &  -- \\
NGC~4793   &  -17.6 &    122.200 &      0.005 &      0.007 &      0.313 &      0.009 &      0.216 &      0.016 &      1.215\\
NGC~4793   &   21.4 &     84.770 &      0.018 &      0.013 &      0.344 &      0.015 &      0.340 &      0.028 &      1.210\\
NGC~4793   &   -3.9 &     20.210 &  --  &  --  &      0.299 &      0.017 &      0.222 &      0.032 &      1.340\\
NGC~4793   &    0.0 &      8.940 &  --  &  --  &      0.428 &      0.027 &      0.258 &      0.051 &      1.050\\
NGC~4793   &    4.5 &     32.390 &  --  &  --  &      0.267 &      0.019 &      0.193 &      0.034 &      1.150\\
NGC~4793   &    0.4 &     25.770 &      0.020 &      0.022 &      0.319 &      0.025 &      0.299 &      0.047 &      1.167\\
NGC~4793   &   26.0 &     70.110 &      0.033 &      0.024 &      0.324 &      0.027 &      0.342 &      0.050 &      1.586\\
NGC~4793   &   38.4 &     23.670 &  --  &  --  &      0.211 &      0.104 &  --  &  --  &  -- \\
NGC~4800   &    0.0 &      4.090 &  --  &  --  &      0.634 &      0.064 &      0.314 &      0.120 &      0.780\\
NGC~4800   &  -23.7 &     12.890 &  --  &  --  &      0.473 &      0.064 &      0.339 &      0.115 &      1.016\\
NGC~4800   &  -10.1 &      8.320 &  --  &  --  &      0.494 &      0.063 &      0.255 &      0.115 &      1.098\\
NGC~4800   &    5.1 &      7.140 &  --  &  --  &      0.467 &      0.043 &      0.248 &      0.082 &      0.951\\
NGC~4800   &   23.7 &     23.660 &  --  &  --  &      0.403 &      0.038 &      0.237 &      0.065 &      1.342\\